\documentclass[10pt]{article}
\PassOptionsToPackage{hyphens}{url}   % long model URLs may break after a hyphen

\usepackage{amsmath}
\usepackage{scicom}
\usepackage[authoryear]{natbib}

\usepackage{hyperref}
\usepackage{bookmark}
\usepackage{url}
\usepackage{nicefrac}
\usepackage{microtype}
\usepackage{tabularx}
\usepackage{multirow}
\usepackage{float}
\usepackage{adjustbox}
\usepackage{longtable}
\usepackage{array}
\usepackage[section]{placeins}

\newcolumntype{L}[1]{>{\raggedright\arraybackslash}p{#1}}
\input{numbers}

\newcommand{\pending}[1]{\textcolor{orange}{[pending: #1]}}

\hypersetup{
pdftitle={The Voice on the Line},
pdfauthor={Husein Zolkepli},
pdfsubject={Audio and Speech Processing},
pdfkeywords={voice isolation, target speaker extraction, speech enhancement, speech recognition, streaming, GPU serving},
colorlinks=true,
linkcolor=scLink,
citecolor=scLink,
urlcolor=scLink
}
\title{The Voice on the Line}
\scshorttitle{The Voice on the Line}
\scauthor{Husein Zolkepli}{husein.zolkepli@scicom.com.my}
\scaffiliation{Scicom (MSC) Berhad, Malaysia}
\scdate{November 2026}

\begin{document}

\maketitle

\begin{abstract}
A transcriber listening through one microphone also hears everyone near it. We train causal voice isolators that keep
only the caller for a speech recogniser: masking transformers with 20 ms of latency that enroll the caller from the
call itself. Short evaluation windows hide two failures that real calls expose: forgetting the caller when another voice
talks alone for a while, and damaging a caller who is alone. Long calls with turns, memory beyond the attention window
and real headset crosstalk fix the first, at a cost to the second. WavLM-Large's own weights, made causal and trained
inside the model, move that trade, and training windows cut from longer calls remove most of a lone caller's cold
start. The release (\val{model/MW/params}M parameters) makes \val{main/rel_selfl/ami_test_long/wer}\% word errors on
60 s AMI windows, against \val{main/voxlock_live/ami_test_long/wer}\% for VoxLock, the strongest open isolator, and
\val{nine/rel_selfl/wer}\% on average over nine test sets, against \val{nine/voxlock_live/wer}\%; it is level on Krisp's
benchmark, and a lone caller still loses more words than with VoxLock. Distilled without WavLM into
\val{cpu/C22/params-m}M and \val{small/c10/params-m}M parameters, it runs live on one CPU thread inside a LiveKit agent,
at \val{vint/c10/core/cpu8}\% and \val{vint/c10/core/c108}\% of a core per call, with \val{r/C22_ftlong_kd_selfl/dev4}\%
and \val{r/C10n_ftlong_kd_selfl/dev4}\% error on our development sets (the release \val{r/rel_selfl/dev4}\%). In an
agent that talks back, noise interrupts it \val{vint/gate/agent/all} times a minute of its speech behind our agent model,
which gates its output on a caller-presence head, against \val{vint/voxlock/fi} for VoxLock. We open-source the models,
with a LiveKit plugin, at \url{https://huggingface.co/collections/Scicom-intl/voice-isolator}.
\end{abstract}

\section{Introduction}
\label{sec:intro}

\begin{figure}[H]
\centering
\includegraphics[width=0.92\linewidth]{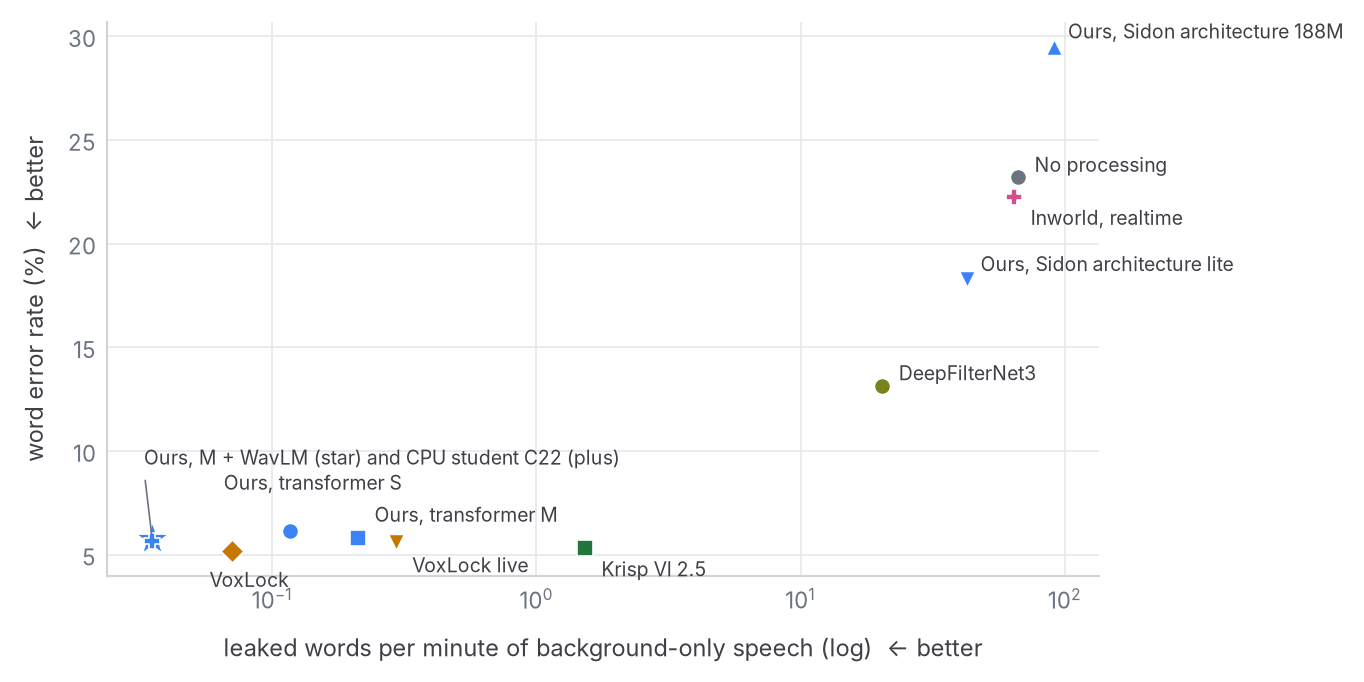}
\caption{Krisp: word error rate against leaked words per minute of background-only speech, one point per system, ours
live. Noise suppression (DeepFilterNet3) keeps the other voices; isolators remove them.}
\label{fig:krisptrade}
\end{figure}

A call-centre transcriber listens through one microphone, and so does everyone near it. On Krisp's benchmark of 265 real
recordings with a second voice \citep{krisp2026stt}, Parakeet TDT 0.6B v2 \citep{nvidia2025parakeettdt06bv2} gets
\val{krisp/raw/wer}\% of the words wrong, almost all of them other people's words (Figure~\ref{fig:krisptrade}). Noise
suppressors keep those words, because they are speech. VoxLock \citep{siddiqui2026voxlock}, the strongest open isolator,
removes them with a sample of the caller taken from the call.

We train causal isolators for GPU serving that start with no sample, enroll the caller from the loudest 3 s heard so
far, and add no look-ahead to a 20 ms window \citep{dubey2024icassp}. Our contributions:
\begin{itemize}
\item an open evaluation that reproduces VoxLock's and adds long meeting windows and real call-centre calls;
\item two failures that short windows hide, forgetting the caller over a long call and damaging a lone caller, and the
training stages that fix them;
\item WavLM-Large made causal and trained inside the isolator, which improves both failures at once, with ablations of
the objective, enrollment, data, model size and resynthesis;
\item a batched GPU server with stress tests, and \val{cpu/C22/params-m}M and \val{small/c10/params-m}M students that
run on one CPU thread and serve ten times the calls on a GPU;
\item a LiveKit plugin measured inside running agents that talk back: a voicebot mode for agents that speak first,
agent models whose caller-presence gate puts them ahead of NVIDIA's RE-USE, and a phone-line fine-tune.
\end{itemize}
The models are open at \url{https://huggingface.co/collections/Scicom-intl/voice-isolator}, the plugin at
\url{https://github.com/Scicom-AI-Enterprise-Organization/STT-API/tree/main/stt_api/livekit_plugin/voice_isolator}.

\section{Background}
\label{sec:background}

\textbf{Three problems.} Noise suppression removes everything that is not speech and keeps every voice
\citep{reddy2021icassp,dubey2024icassp,schroter2023deepfilternet}. Target speaker extraction keeps one voice given a
clue, usually an enrollment clip \citep{zmolikova2023neural,wang2019voicefilter,zmolikova2019speakerbeam,xu2020spex,ge2020spex,zeng2025useftse,zhang2025multilevel}.
Primary-speaker isolation keeps whoever is on the microphone; nearness alone is a weak cue
\citep{dubey2024icassp,patterson2022distance,parnamaa2024personalized}, so VoxLock cuts an enrollment from the call.

\textbf{Two failures for a recogniser.} Over-suppression deletes the caller's words
\citep{wang2020voicefilterlite,iwamoto2022how,iwamoto2024how,sato2021should}; leakage inserts other people's
\citep{dubey2024icassp,microsoft2022icassp}. Following VoxLock, we report word error rate with its parts: \emph{missed}
words (substitutions and deletions), \emph{inserted} words, and \emph{leaked} words, recognised words whose midpoint
falls where only the background talks, per minute of that time.

\textbf{Real recordings.} Simulated mixtures overstate extraction on real conversations \citep{li2025realt,wang2026slt}.
Krisp records the second voice for real and labels who talks when \citep{krisp2026stt}; in AMI and NOTSOFAR every
participant's head-worn microphone also hears the others
\citep{carletta2006ami,vinnikov2024notsofar,microsoft2024notsofardata}. Intervals resample speakers, not files
\citep{bisani2004bootstrap,liu2020statistical}.

\textbf{Latency.} An overlap-add STFT model holds each sample for its window plus any look-ahead
\citep{wang2023stftdomain}; the 2023 DNS challenge caps the sum at 20 ms \citep{microsoft2022icassp,dubey2024icassp}.
VoxLock's is 24 ms, DeepFilterNet3's 40 ms \citep{schroter2023deepfilternet}. On GPUs a call also waits for its chunk
and the batched step.

\section{Data}
\label{sec:data}

\begin{figure}[H]
\centering
\includegraphics[width=\linewidth]{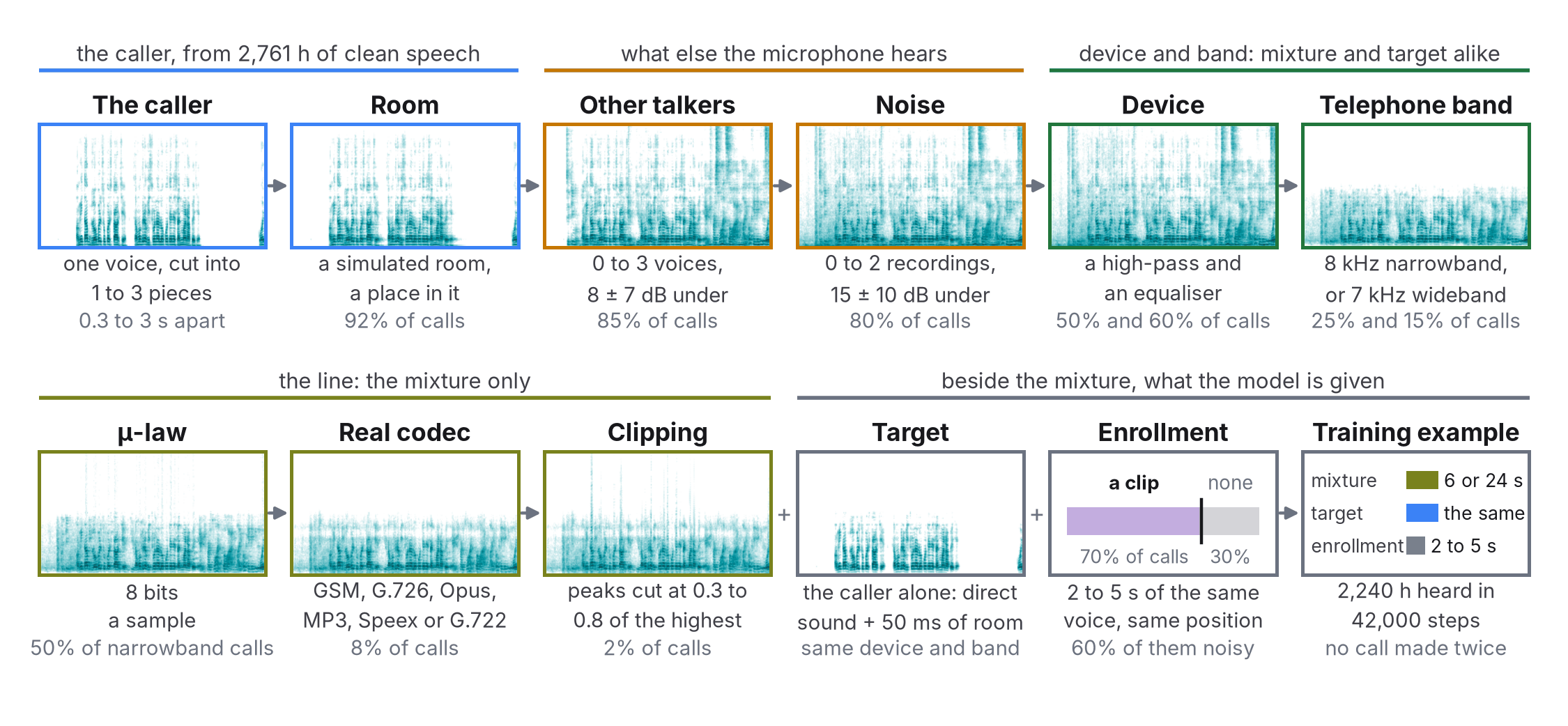}
\caption{How a training example is made: one simulated call, 8 s across and 0 to 8 kHz up, after each stage of the
simulator, with how often a call gets the stage. Voice, noise and room are held out; the colour scale is shared.
Every probability is in Appendix~\ref{app:sim}.}
\label{fig:pipeline}
\end{figure}

Every training example is a call simulated on the fly from clean speech (Figure~\ref{fig:pipeline}). The target is the
person on the microphone alone, through the same room, device and band as the mixture, so the model never invents
bandwidth and never keeps anyone else.

\subsection{Speech, noise and rooms}

\begin{figure}[H]
\centering
\includegraphics[width=0.92\linewidth]{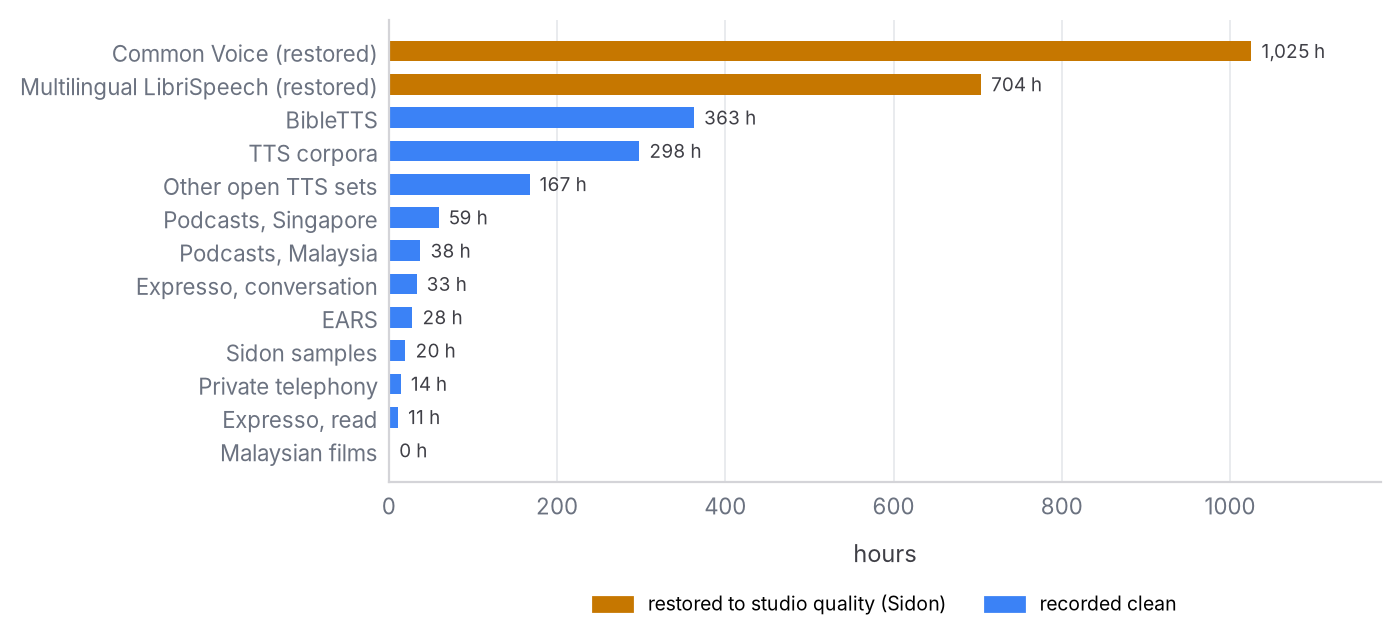}
\caption{Training speech by source. Common Voice \citep{ardila2020common} and Multilingual LibriSpeech
\citep{pratap2020mls} are restored to studio quality with Sidon \citep{nakata2025sidon}; the rest is recorded clean.}
\label{fig:pool}
\end{figure}

\begin{figure}[H]
\centering
\includegraphics[width=0.92\linewidth]{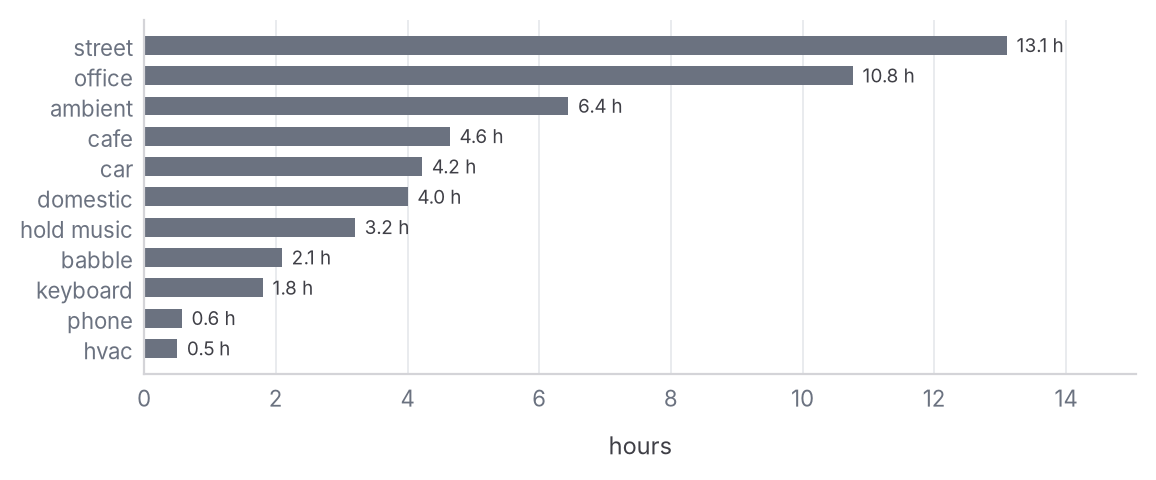}
\caption{Noise by kind. One clip in 20 is held out for development and tests.}
\label{fig:noise}
\end{figure}

\textbf{Speech and noise.} The speech is the \val{data/pool-hours} h clean pool of our restoration work
(Figure~\ref{fig:pool}), in \val{data/cv-langs} Common Voice languages and more, \val{data/english-hours} h of it
English. Speakers are split 199 to 1 into training and development. Noise is \val{data/noise-hours} h from five corpora
\citep{snyder2015musan,fonseca2022fsd50k,gemmeke2017audio,thiemann2013diverse,piczak2015esc} (Figure~\ref{fig:noise}).

\begin{figure}[H]
\centering
\includegraphics[width=0.92\linewidth]{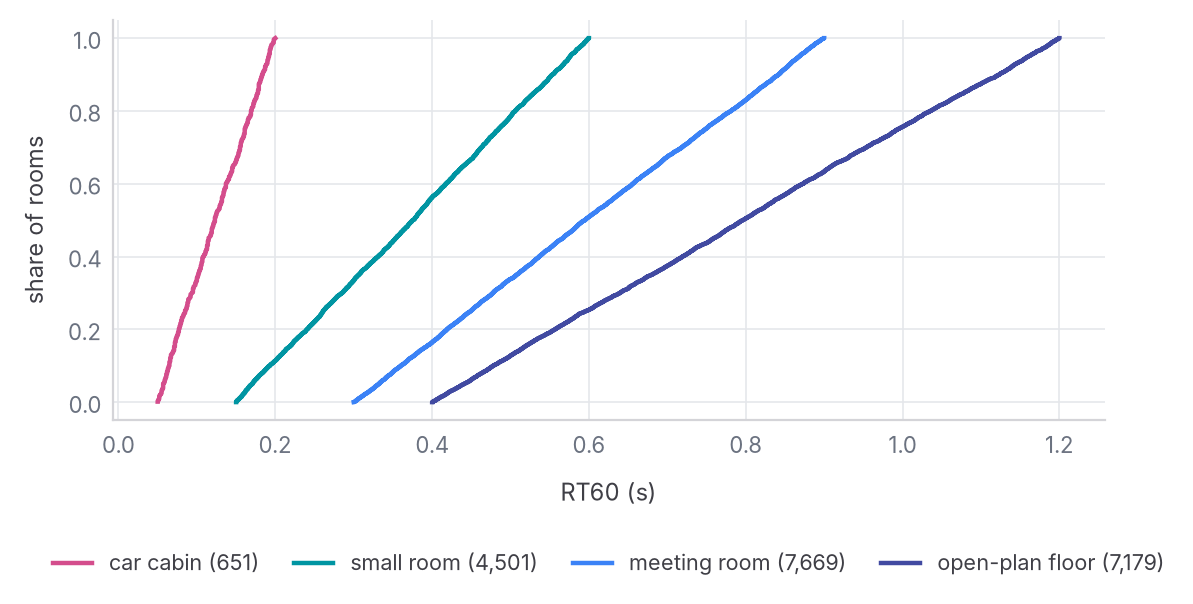}
\caption{Reverberation time of the simulated rooms, by type. Car cabins are \val{data/car-share}\% of rooms, since
talker placement fails more often there and the room is redrawn.}
\label{fig:rooms}
\end{figure}

\begin{figure}[H]
\centering
\includegraphics[width=0.92\linewidth]{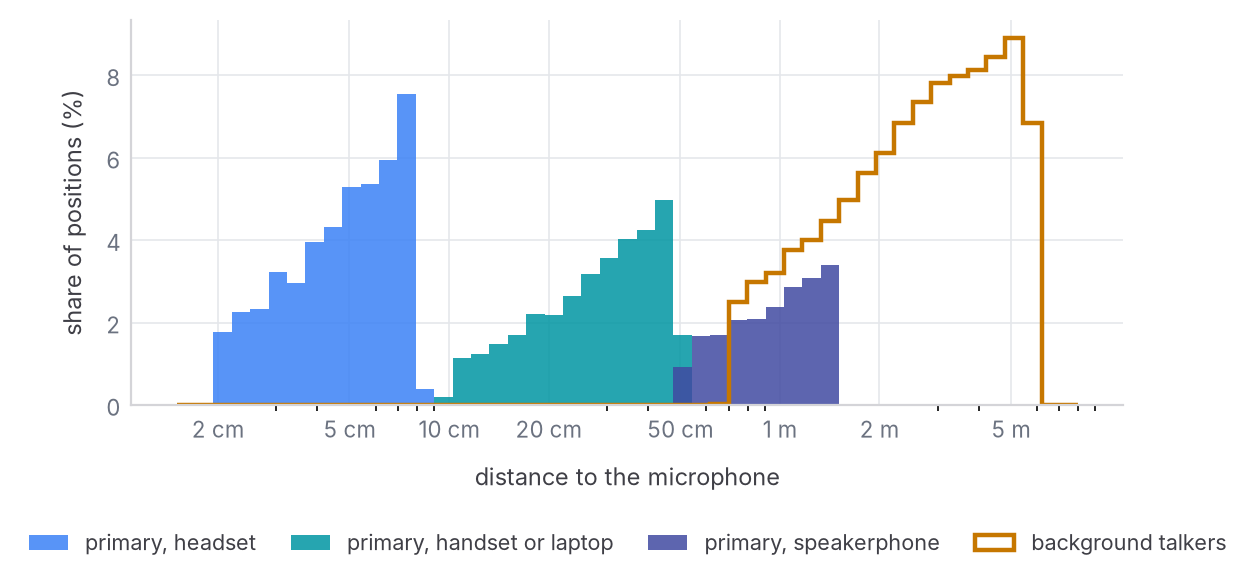}
\caption{Distance from each talker to the microphone. The primary sits at a headset, a handset or a speakerphone;
background talkers are usually further away, but a tenth are rendered dry, as close as the primary.}
\label{fig:distances}
\end{figure}

\textbf{Rooms.} \val{data/rooms} shoebox rooms are simulated with the image method
\citep{allen1979image,scheibler2018pyroomacoustics}, each with one microphone and four talker positions
(Figures~\ref{fig:rooms} and~\ref{fig:distances}); the last 1,000 are held out.

\subsection{Synthetic crosstalk, noise and the line}

\begin{figure}[H]
\centering
\includegraphics[width=0.92\linewidth]{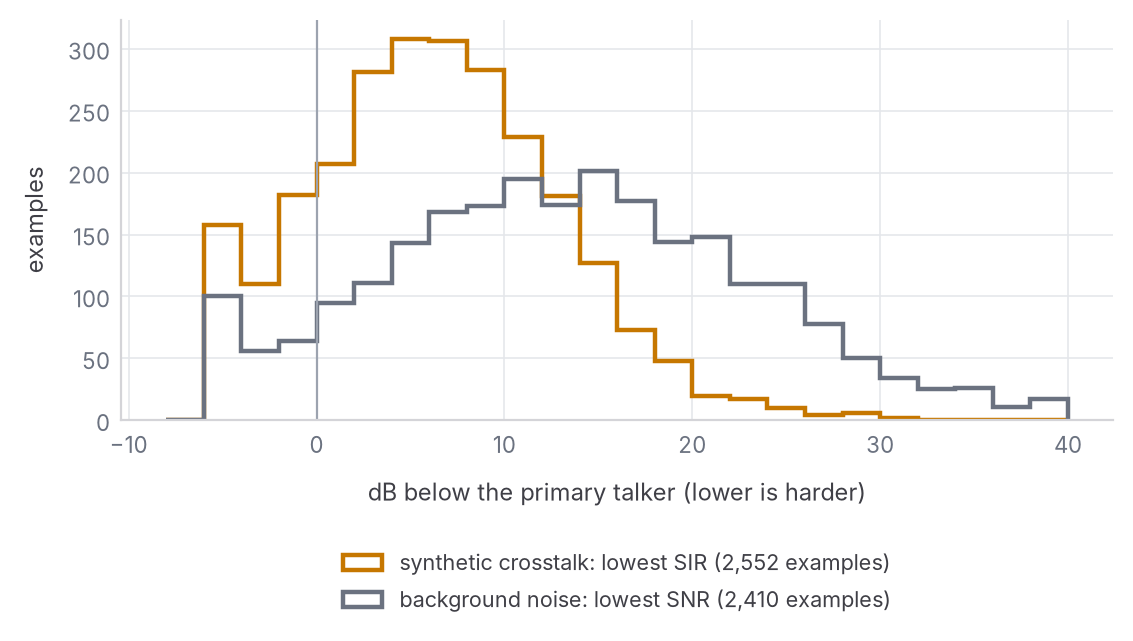}
\caption{Difficulty drawn by the simulator over 3,000 examples: the lowest signal-to-interference ratio among an
example's background talkers and the lowest signal-to-noise ratio among its noises, relative to the primary.}
\label{fig:sirsnr}
\end{figure}

\begin{figure}[H]
\centering
\includegraphics[width=0.92\linewidth]{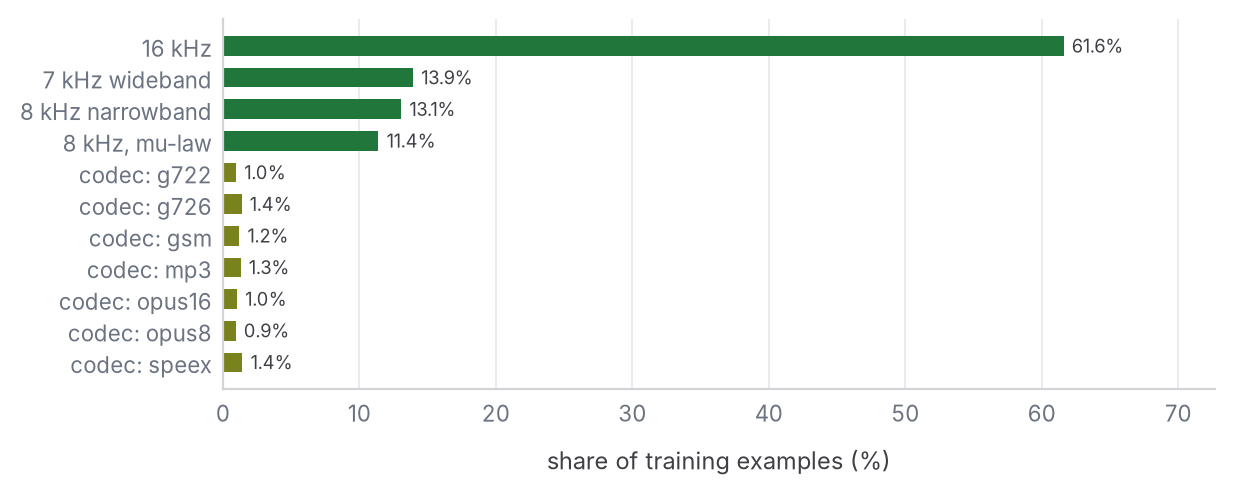}
\caption{The line: band limits (green) and real codecs run through ffmpeg on the mixture (olive). G.723.1 and Codec2
are held out for testing.}
\label{fig:channels}
\end{figure}

\textbf{The mixture.} The primary's speech is cut into pieces with pauses. Up to three background talkers are added,
mostly inside those pauses, so \val{data/bgonly-share}\% of examples hold more than half a second where only someone
else talks. Up to two noises follow, then the device and the line (Figures~\ref{fig:sirsnr} and~\ref{fig:channels}).

\textbf{Enrollment.} Seven examples in ten carry 2 to 5 s of the same voice through the same position and device,
noisy in 60\% of cases. The rest carry none, so one model also learns to find the primary from the audio alone.

\subsection{Long calls}
\label{sec:longdata}

\begin{figure}[H]
\centering
\includegraphics[width=\linewidth]{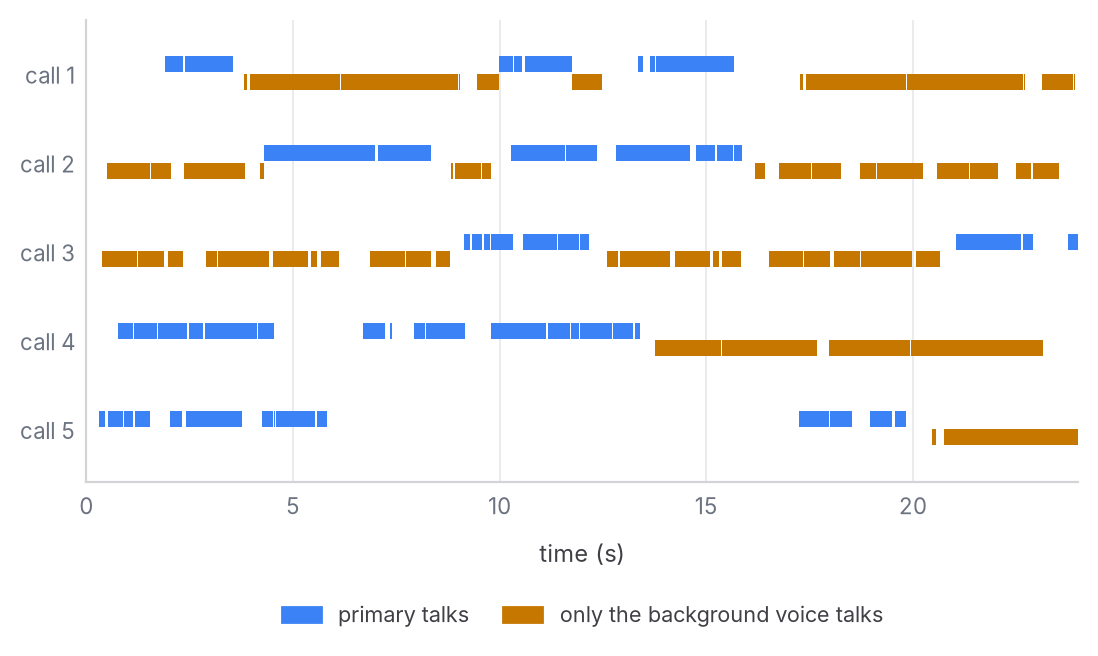}
\caption{Five simulated calls of the long-call stage (development speakers). Blue: the caller talks; amber: only the
background voice does. The caller often returns after the other voice has talked alone for seconds.}
\label{fig:turns}
\end{figure}

\textbf{Turns.} Six-second examples teach separation, not memory: the caller is never silent for more than 3 s. The
long-call stage uses 24 s calls; in six of ten the caller talks in turns and one background voice takes most of the
pauses (Figure~\ref{fig:turns}). Background-only stretches over 8 s grow from \val{long/stage1/over8} s to
\val{long/turns/over8} s per call.

\subsection{Real callers}
\label{sec:callerdata}

\textbf{Four changes} come from real callers (Section~\ref{sec:caller}). The mixture level spans $-65$ to $-10$ dBFS,
since below $-45$ dBFS every model removed most of the caller. In half the long calls the caller's level drifts, by
gains drawn every 1 to 4 s with a 4 dB spread. A third of the long calls have nobody else. In the release's long-call
stage, three examples in ten are cut from a call 0.5 to 3 s longer, so whoever talks at the cut is already talking at
the first sample.

\subsection{Real crosstalk}

\textbf{Meeting channels.} In each AMI, NOTSOFAR and AliMeeting wearer's channel
\citep{carletta2006ami,vinnikov2024notsofar,yu2022m2met}, a window where nobody else speaks is an exact target, and a
window where only others speak is real crosstalk at its natural level. We use only meetings outside every evaluation
set: \val{realbank/all/meetings} meetings, \val{realbank/all/hours} h.

\subsection{Evaluation sets}

\textbf{VoxLock's sets and more.} Table~\ref{tab:evalsets} lists the sets; only the development sets choose a model.
Krisp and VoxLock's AMI windows (a wearer's turn and the 4 s after it) follow VoxLock
\citep{krisp2026stt,siddiqui2026voxlock}; we build the same windows on more AMI meetings, NOTSOFAR and AliMeeting
(Mandarin, scored by characters).

\begin{figure}[H]
\centering
\includegraphics[width=0.92\linewidth]{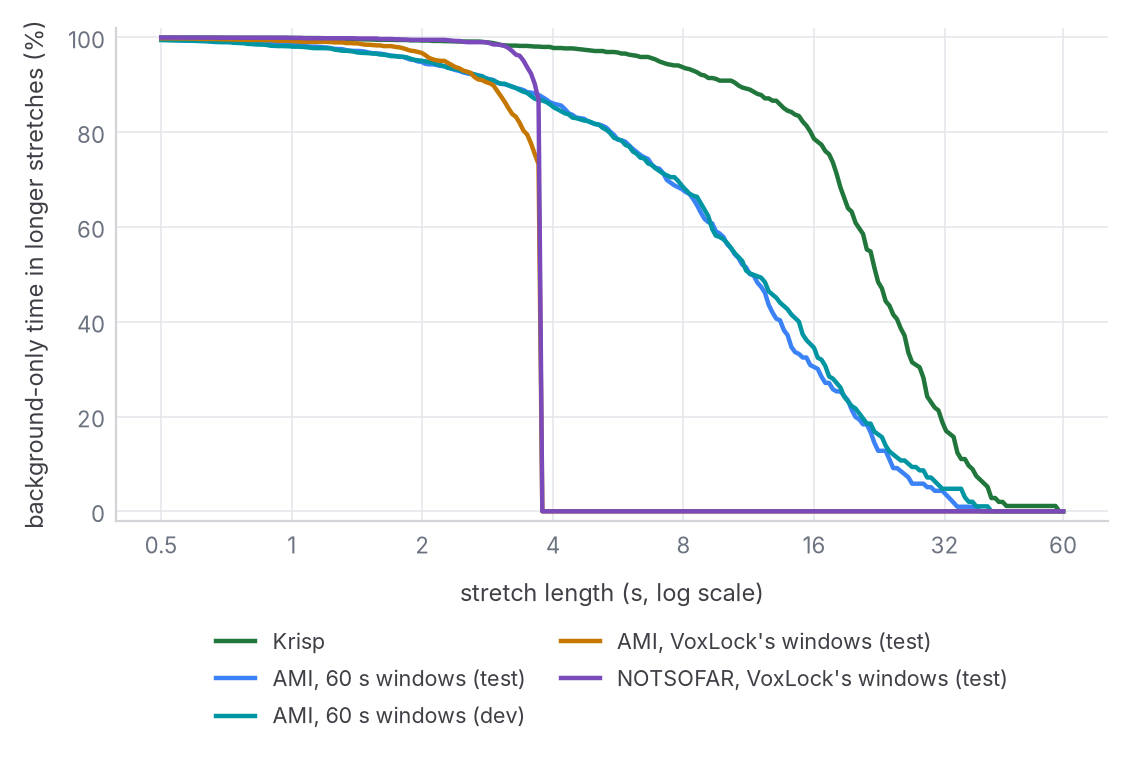}
\caption{Share of each set's background-only time in stretches longer than a given length (gaps under 1 s closed).
Every VoxLock-style window set stops at 4 s.}
\label{fig:stretches}
\end{figure}

\textbf{Long windows and calls.} Short windows never hold more than 4 s of background-only speech, while Krisp's comes
in stretches with a median of \val{evalbg/krisp/median} s (Figure~\ref{fig:stretches}). We add 60 s AMI windows that
each hold at least 8 s where only others talk, followed by the wearer. For calls we add AppTek's call-centre recordings
\citep{apptek2025callcenter}: 60 s of one speaker scored alone (any change is damage to a lone caller) and with another
call's speaker 0 to 12 dB below. ai-coustics' 83 real test calls \citep{aicoustics2025calls} are reported only, like
Krisp.

\begin{table}[H]
\centering
\caption{Evaluation sets. \emph{Callers}: distinct people on the microphone, the unit the bootstrap resamples.
\emph{Bg-only min}: minutes in which only someone else speaks, where every recognised word is a leaked word.}
\label{tab:evalsets}
\scriptsize
\begin{adjustbox}{max width=\linewidth}
\begin{tabular}{llrrrrrl}
\toprule
\schead \scth{Set} & \scth{Use} & \scth{Items} & \scth{Callers} & \scth{Minutes} & \scth{Ref. words} & \scth{Bg-only min} & \scth{ASR} \\
\midrule
Krisp VI benchmark & test & 265 & 47 & 337.5 & 31,418 & 85.3 & Parakeet v2 \\
AMI, VoxLock's windows & test & 134 & 19 & 25.5 & 2,874 & 7.5 & Parakeet v3 \\
AMI, evaluation meetings & test & 599 & 61 & 109.1 & 11,964 & 34.2 & Parakeet v3 \\
NOTSOFAR, evaluation split & test & 800 & 413 & 140.2 & 16,747 & 47.2 & Parakeet v3 \\
AMI, 60 s windows, evaluation meetings & test & 120 & 44 & 123.8 & 8,197 & 63.2 & Parakeet v3 \\
AMI, development meetings & selection & 400 & 65 & 67.5 & 7,148 & 22.4 & Parakeet v3 \\
AMI, 60 s windows, development meetings & selection & 120 & 50 & 122.5 & 7,867 & 62.0 & Parakeet v3 \\
Calls, one speaker, development split & selection & 120 & 120 & 120.0 & 8,050 & 0.0 & Parakeet v2 \\
Calls + another call's speaker, development split & selection & 120 & 120 & 120.0 & 8,050 & 4.9 & Parakeet v2 \\
Calls, one speaker, evaluation split & test & 120 & 120 & 120.0 & 8,008 & 0.0 & Parakeet v2 \\
Calls + another call's speaker, evaluation split & test & 120 & 120 & 120.0 & 8,008 & 19.4 & Parakeet v2 \\
ai-coustics test calls & test & 83 & 4 & 35.5 & 2,657 & 15.8 & Parakeet v2 \\
NOTSOFAR, development split & selection & 400 & 156 & 68.8 & 8,343 & 23.5 & Parakeet v3 \\
Simulated 60 s calls & analysis & 200 & 40 & 200.0 & 23,643 & 10.6 & Parakeet v2 \\
\bottomrule
\end{tabular}
\end{adjustbox}

\end{table}

\section{Model}
\label{sec:model}

We build two kinds of model, both for GPUs: a causal masking transformer in three sizes, our release, and Sidon's
restoration architecture at 16 kHz for comparison.

\subsection{Realtime: a causal transformer over frames}

\begin{figure}[H]
\centering
\includegraphics[width=1.0\linewidth]{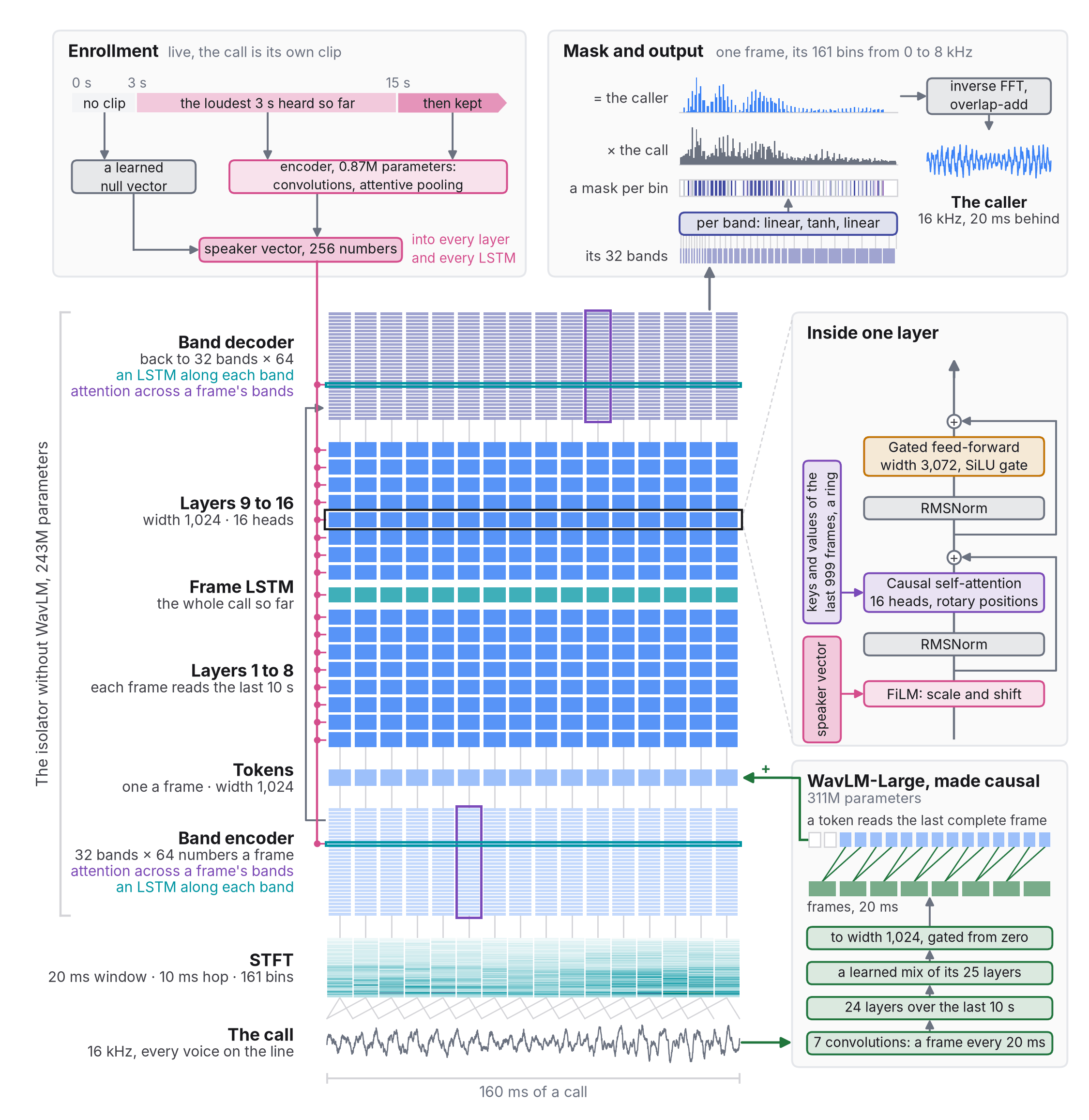}
\caption{The release as its frames: every row one stage or layer, every cell one 10 ms frame, over 160 ms of a call;
the audio is drawn. The speaker vector (pink) enters every layer and LSTM, and WavLM's 20 ms frames join the tokens
one tick late, adding no latency.}
\label{fig:former}
\end{figure}

\textbf{The FrameFormer.} Each 20 ms frame (10 ms hop) is split into 32 bands \citep{luo2023music,yu2023high},
encoded with attention across bands and a small LSTM along time in each band, and folded into one token
(Figure~\ref{fig:former}). A causal transformer attends over the last 10 s of tokens, with one frame LSTM after its
middle layer; a band decoder predicts a complex mask. The sizes differ only in width (Table~\ref{tab:sizes}).

\textbf{WavLM inside.} The release adds WavLM-Large \citep{chen2022wavlm}, pre-trained on 94k hours with overlapped
speech, and changes only what causality needs: a running level normalisation, a positional convolution over the past
only (it looked 1.28 s ahead), and causal attention over the last 10 s. Token $t$ reads WavLM frame
$\lfloor t/2 \rfloor - 1$, the last complete one, so no latency is added. A learned mix of its 25 layers enters the
tokens through a gate that starts at zero. For comparison, w2v-BERT 2.0 \citep{barrault2023seamless} and Whisper
large-v3's encoder \citep{radford2023robust} are made causal the same way.

\begin{table}[H]
\centering
\caption{The models. GMAC: multiply-accumulates per second of audio. State: what one live call keeps on the GPU between
frames.}
\label{tab:sizes}
\scriptsize
\begin{adjustbox}{max width=\linewidth}
\begin{tabular}{lllrrr}
\toprule
\schead \scth{Model} & \scth{Kind} & \scth{Shape} & \scth{Params (M)} & \scth{GMAC / s} & \scth{State / call} \\
\midrule
Transformer S & realtime, 20 ms & 16 layers, width 768 & 140 & 17.4 & 49.2 MB \\
Transformer M & realtime, 20 ms & 16 layers, width 1024 & 243 & 28.5 & 65.6 MB \\
Transformer L & realtime, 20 ms & 16 layers, width 1280 & 374 & 42.5 & 82.0 MB \\
S + WavLM & realtime, 20 ms & 16 layers, width 768 + WavLM 24 layers & 452 & 36.5 & 98.7 MB \\
M + WavLM (release) & realtime, 20 ms & 16 layers, width 1024 + WavLM 24 layers & 555 & 47.7 & 115.1 MB \\
L + WavLM & realtime, 20 ms & 16 layers, width 1280 + WavLM 24 layers & 687 & 61.6 & 131.5 MB \\
\midrule
Sidon-arch. 188M & offline, 4 s windows & w2v-BERT 2.0 + LoRA, 3072 ch & 580 + 187 & 139 & -- \\
Sidon-arch. lite & offline, 4 s windows & DistilHuBERT 2L + head, 1536 ch & 37 + 52 & 32 & -- \\
\bottomrule
\end{tabular}
\end{adjustbox}

\end{table}

\subsection{Enrollment and self-enrollment}

\textbf{A speaker vector} enters every layer through FiLM \citep{perez2018film}: a clip of the caller through a small
encoder (\val{model/enc/params}M parameters), or a learned null vector. Live, a call starts with the null vector, from
3 s in enrolls the loudest 3 s heard so far, and fixes it at 15 s: VoxLock's rule without its look-ahead
\citep{siddiqui2026voxlock}.

\subsection{Offline: Sidon's architecture at 16 kHz}

\begin{figure}[H]
\centering
\includegraphics[width=1.0\linewidth]{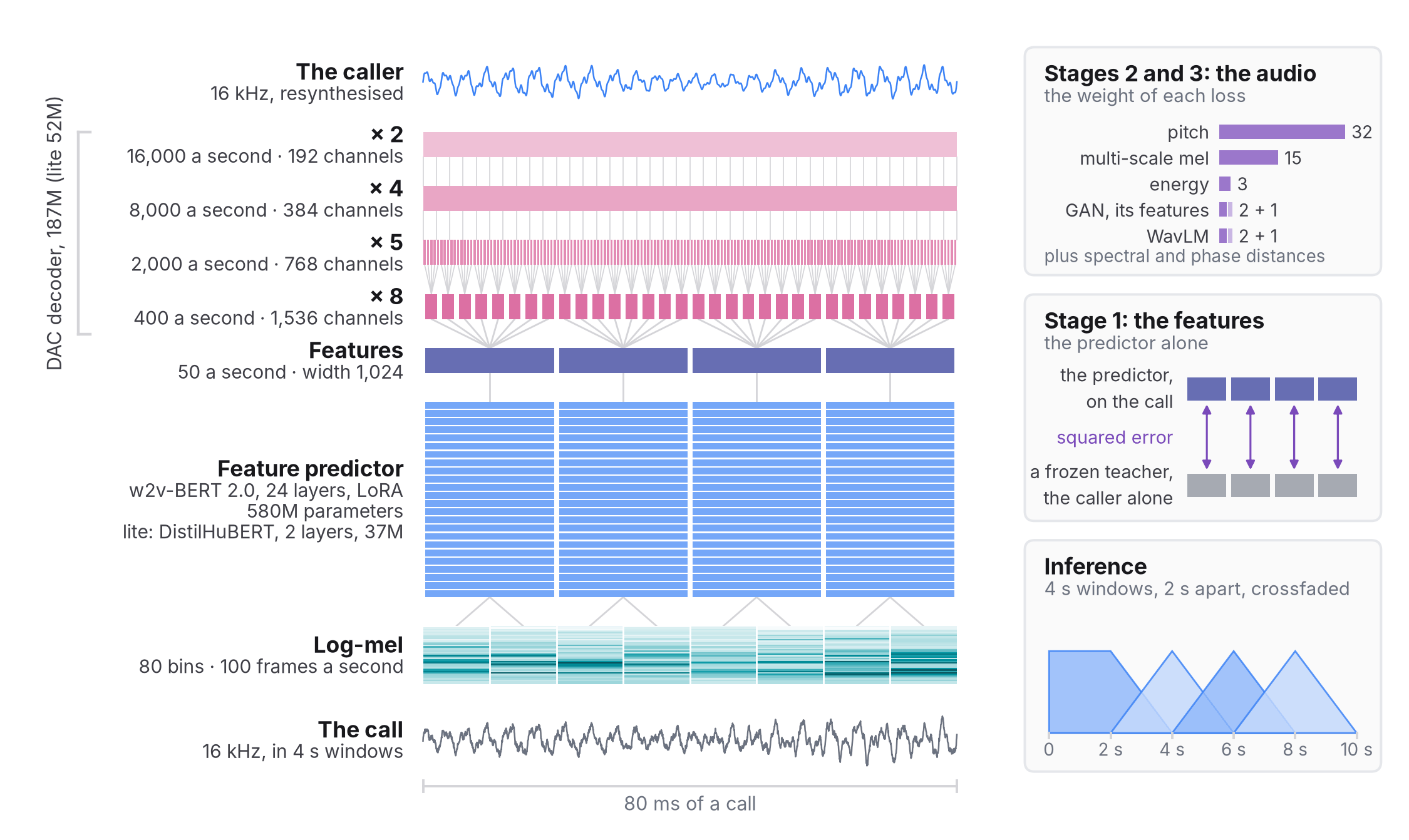}
\caption{The Sidon-architecture models as their frames over 80 ms: predicted features at 50 a second, upsampled to
16 kHz by the decoder. Right: inference in crossfaded windows and the two training signals.}
\label{fig:sidon}
\end{figure}

\textbf{Resynthesis.} A feature predictor maps the call to the w2v-BERT 2.0 features \citep{barrault2023seamless} of
the caller alone, and a DAC decoder \citep{kumar2023high} resynthesises them (Figure~\ref{fig:sidon}). We keep
CallEnhancer's two sizes, trained with Sidon's own scripts on our calls, ending at 16 kHz.

\subsection{Training}

\begin{figure}[H]
\centering
\includegraphics[width=1.0\linewidth]{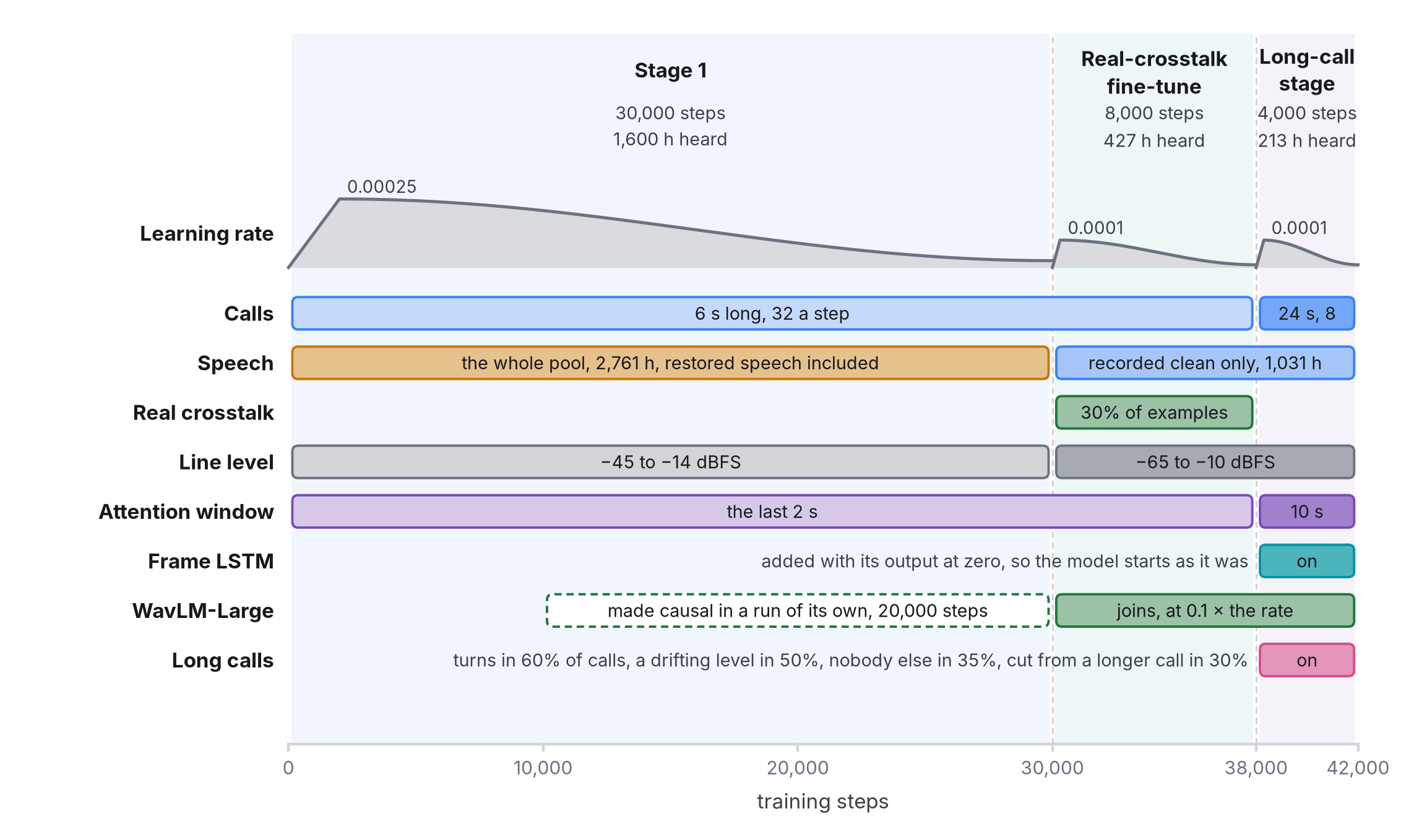}
\caption{The release recipe along its 42,000 steps, each bar as long as its stage. AdamW, bfloat16, weight averaging
(0.999). The long-call stage widens the attention window and adds the frame LSTM with its output at zero, so it
starts as the model before it.}
\label{fig:stages}
\end{figure}

\textbf{Three stages.} The transformer minimises negative signal-to-noise ratio \citep{wisdom2020unsupervised}, a
multi-resolution spectral distance \citep{wilson2018exploring,braun2021consolidated} and the output energy where the
caller is silent \citep{wisdom2021whats}, in three stages (Figure~\ref{fig:stages}). The WavLM branch joins at the
second. Every choice is made on the development sets.

\section{Evaluation}
\label{sec:eval}

\textbf{Systems.} We run every open system on the same 16 kHz files: VoxLock 1.1.1 with its published protocol (the
loudest 3 s of the first 15 s, which looks ahead) and in its live mode, its PyTorch script, DeepFilterNet3
\citep{schroter2023deepfilternet}, and Krisp VI 2.5 through Krisp's published outputs \citep{krisp2026stt}. Our models
run live (the loudest 3 s so far), with VoxLock's look-ahead sample, with no enrollment, or with an oracle clip.

\textbf{Scoring.} Krisp, the call sets and ai-coustics' calls use Parakeet TDT 0.6B v2 with Krisp's normaliser; the
meetings use Parakeet v3, AliMeeting Whisper large-v3
\citep{nvidia2025parakeettdt06bv2,nvidia2025parakeettdt06bv3,vaessen2025jiwer,zhang2021nemo}. Intervals come from
2,000 resamples of callers.

\section{Results}
\label{sec:results}

\subsection{Reproducing the baselines}

\textbf{Our evaluation reproduces VoxLock's.} Unprocessed Krisp scores \val{krisp/raw/wer}\% (VoxLock: 23.21\%), and
Krisp's published outputs reproduce VoxLock's rows exactly. VoxLock's PyTorch script gives
\val{krisp/voxlock_pt/wer}\% on Krisp and \val{ami_headset2/voxlock_pt/wer}\% on its AMI windows (published 4.95\% and
10.12\%); its streaming ONNX package scores worse. We report both.

\subsection{Against existing systems}

\begin{figure}[H]
\centering
\includegraphics[width=0.92\linewidth]{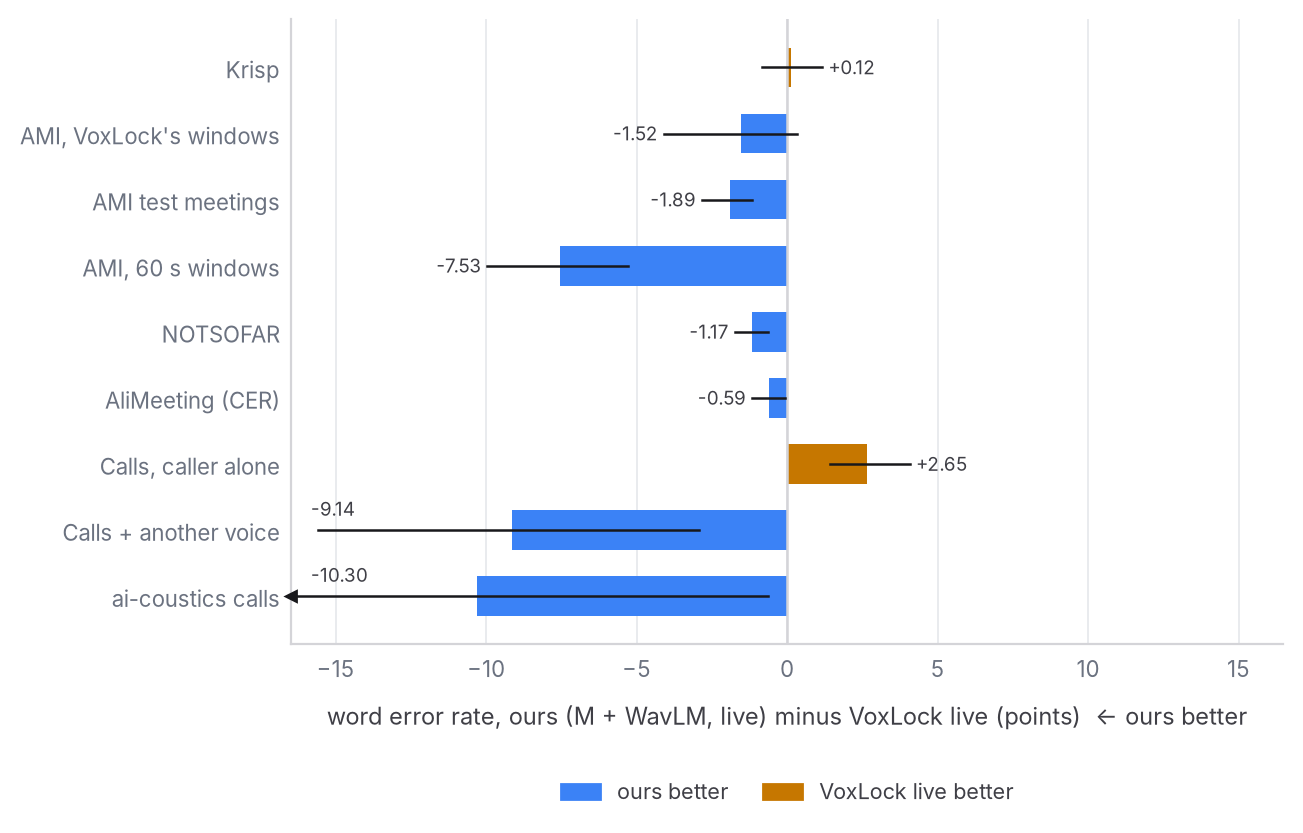}
\caption{Every test set: word error rate of the release, live, minus VoxLock live, with 95\% intervals over callers
(arrows: the interval runs past the axis). AliMeeting: character error rate; caller alone: change against plain
recognition. The release leads on every set with a second voice.}
\label{fig:vsvox}
\end{figure}

\textbf{Ahead wherever a second voice talks.} The release makes fewer errors than VoxLock live on every meeting and
call set with a second voice, is level on Krisp, and loses only on the lone caller (Figure~\ref{fig:vsvox};
Appendix~\ref{app:tables}): \val{main/rel_selfl/call_test_clean/wer}\% against
\val{main/voxlock_live/call_test_clean/wer}\% on clean calls. WavLM also fixes AliMeeting, where the loudest 3 s is
often someone else: \val{main/rel_selfl/ali_test/wer}\% character errors against VoxLock's
\val{main/voxlock_live/ali_test/wer}\%.

\begin{figure}[H]
\centering
\includegraphics[width=0.92\linewidth]{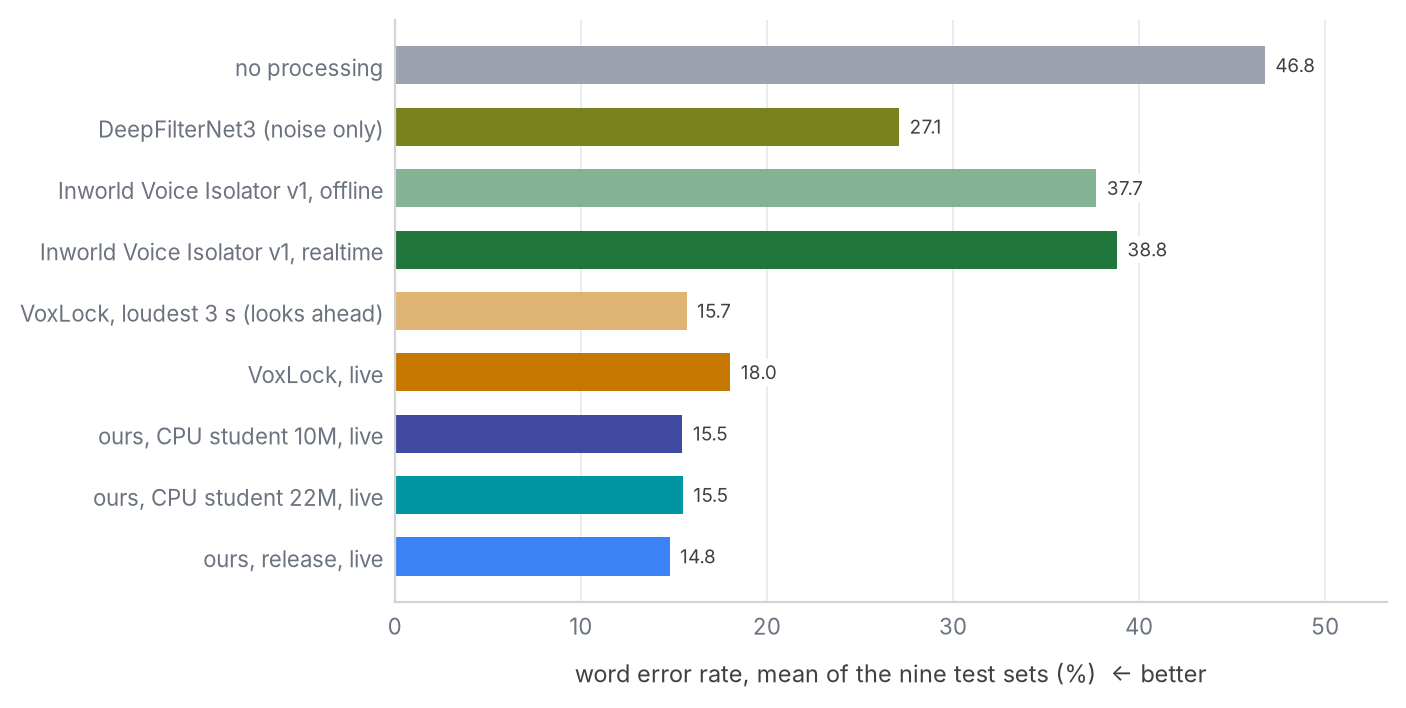}
\caption{Mean word error rate over the nine test sets (AliMeeting by characters), every system run the same way. Each
set is in Appendix~\ref{app:tables}.}
\label{fig:nine}
\end{figure}

\textbf{A noise remover keeps every voice.} Inworld Voice Isolator v1 \citep{inworld2026voiceisolator} has no
enrollment (Figures~\ref{fig:krisptrade} and~\ref{fig:nine}). Its realtime model averages \val{nine/inworld_rt/wer}\%
word errors over the nine test sets, against \val{nine/raw/wer}\% with no processing and \val{nine/rel_selfl/wer}\%
for the release, and passes \val{iw/inworld_rt/krisp/leak} words a minute of other people's speech on Krisp. It is the
gentlest on a lone caller: \val{iw/inworld_rt/call_test_clean/wer}\% on clean test calls.

\subsection{Long calls}
\label{sec:long}

\begin{figure}[H]
\centering
\includegraphics[width=0.92\linewidth]{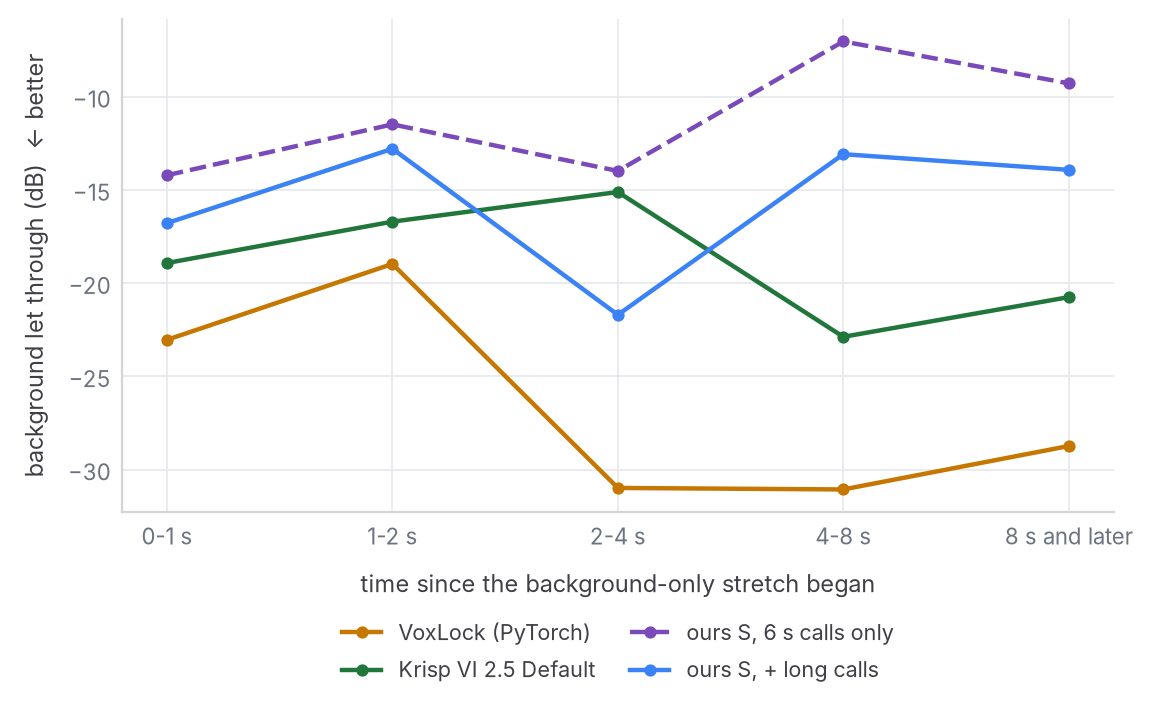}
\caption{Krisp: background let through by how long the background-only stretch has lasted, relative to each system's
gain while the caller talks. Our S forgets before the long-call stage and holds after it. Krisp never chose anything.}
\label{fig:bgtime}
\end{figure}

\begin{figure}[H]
\centering
\includegraphics[width=0.92\linewidth]{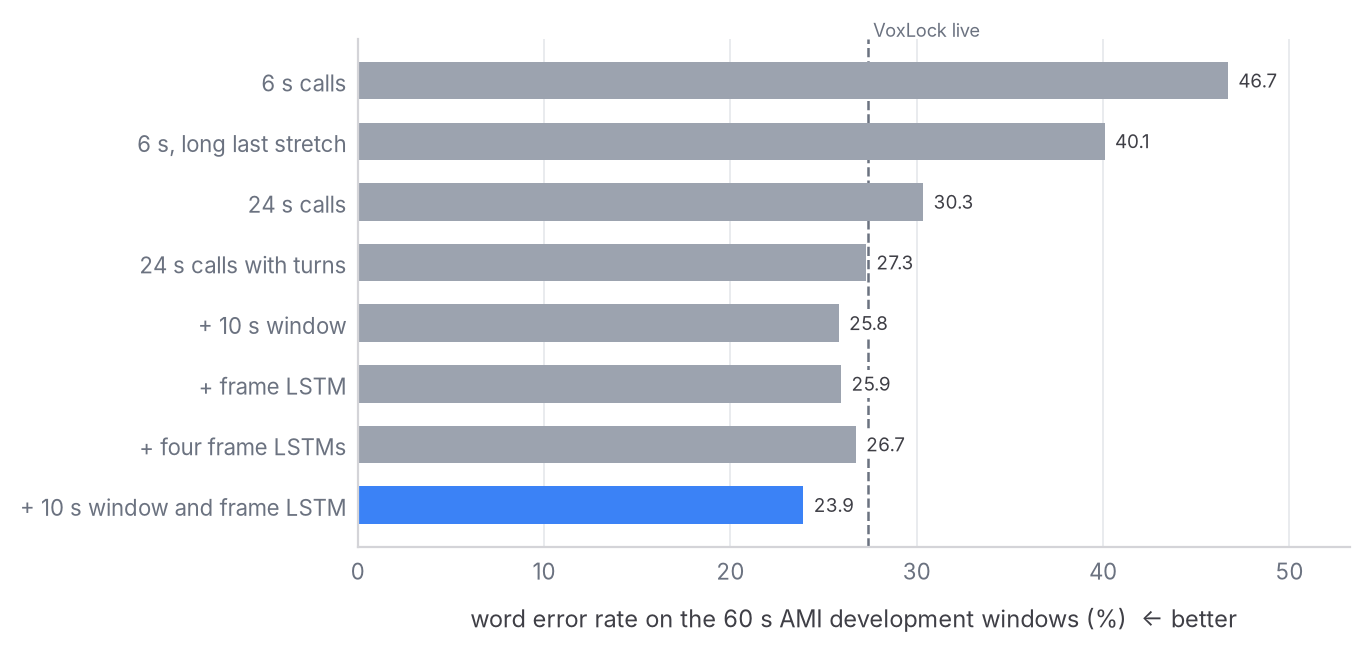}
\caption{The long-call stage at S, 4,000 steps from one checkpoint, one change per bar. Examples that end with the
other voice alone (second bar) also cut the returning caller.}
\label{fig:longcall}
\end{figure}

\textbf{A windowed transformer forgets the caller.} Once another voice has talked alone for a few seconds, the model
lets it through; VoxLock, a recurrent network, does not (Figure~\ref{fig:bgtime}). Only long windows expose this
(Figure~\ref{fig:stretches}). Calls of 24 s with turns, a 10 s window and one frame LSTM fix most of it
(Figure~\ref{fig:longcall}).

\subsection{A lone caller}
\label{sec:caller}

\begin{figure}[H]
\centering
\includegraphics[width=0.92\linewidth]{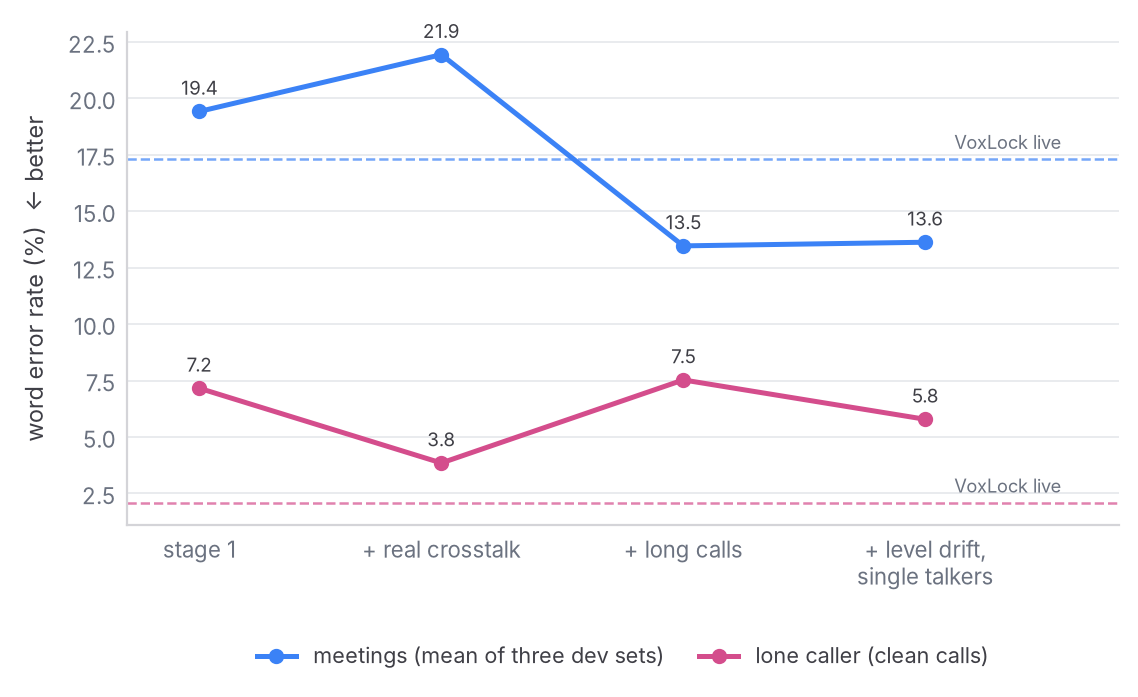}
\caption{S through the training stages, live enrollment: mean of the three meeting development sets, and clean real
calls, where any change is damage to a lone caller. Dashed: VoxLock live.}
\label{fig:recipe}
\end{figure}

\begin{figure}[H]
\centering
\includegraphics[width=0.92\linewidth]{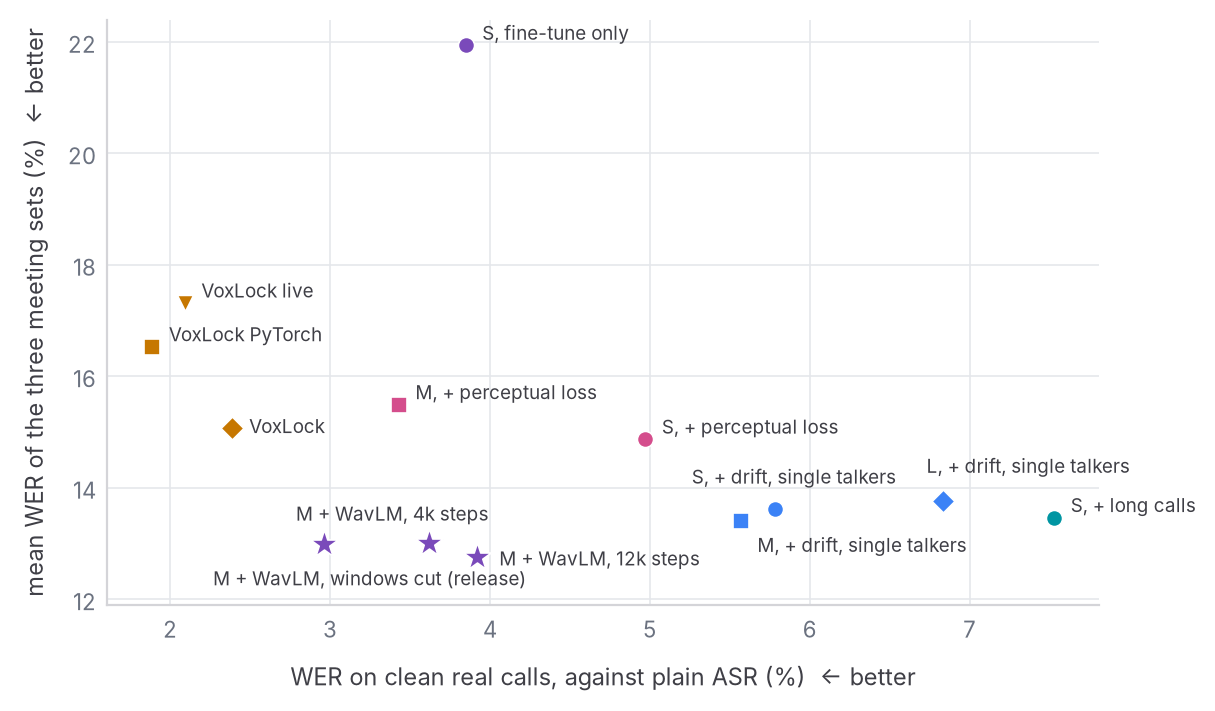}
\caption{Development sets, every system live. Down: less of the other voices in meetings. Left: less damage to a lone
caller. Only WavLM inside (stars) moves both.}
\label{fig:tradeoff}
\end{figure}

\textbf{Fixing memory costs the lone caller.} The long-call stage teaches the model to keep suppressing a lone voice,
and so to doubt a lone caller (Figure~\ref{fig:recipe}). Figure~\ref{fig:tradeoff} places every variant on this trade:
WavLM inside the model is the first change that moves both at once. The release recipe is the best compromise on
dev-4, the mean of the four development sets.

\begin{figure}[H]
\centering
\includegraphics[width=0.92\linewidth]{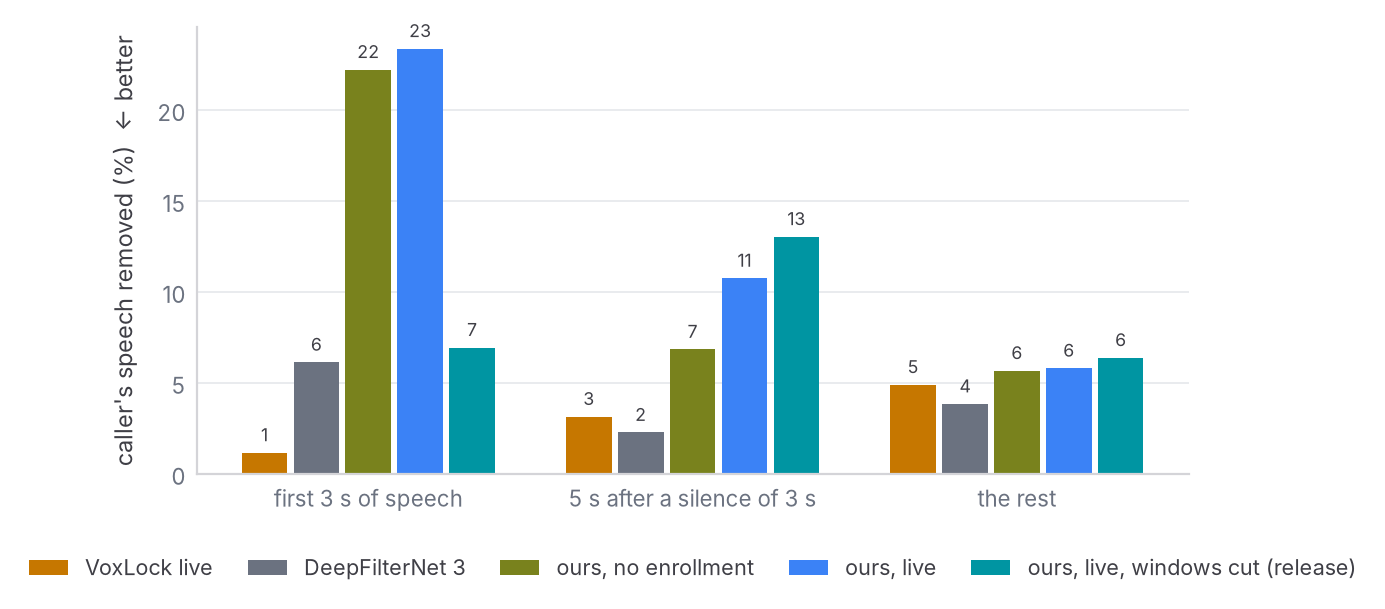}
\caption{Development calls, caller alone: share of the caller's speech removed (output 20 dB or more under the input)
in the first 3 s of speech, in the 5 s after a silence of 3 s or more, and elsewhere, before and after the window cut.}
\label{fig:callerwhere}
\end{figure}

\begin{figure}[H]
\centering
\includegraphics[width=0.92\linewidth]{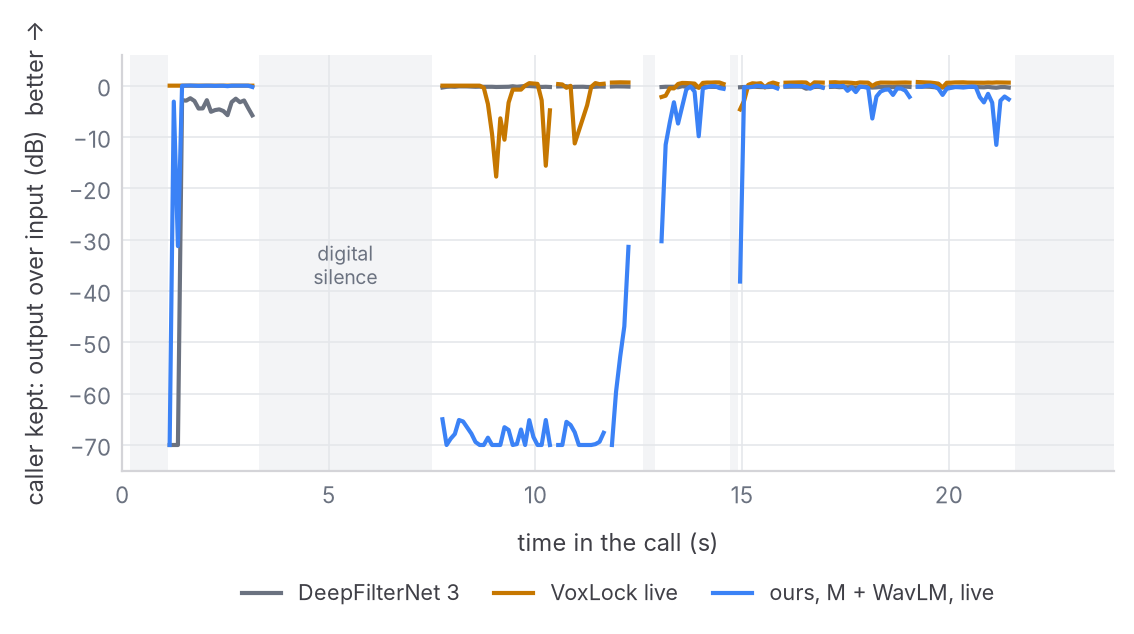}
\caption{One development call, caller alone: how much of the caller each system keeps per 100 ms of speech. After
digital silence (grey) the release removes the returning caller for seconds, as the model before the window cut does.}
\label{fig:callertrace}
\end{figure}

\textbf{Where the caller is lost.} A recorded call-centre line carries digital silence between turns
(\val{calls/call_dev_clean/silent}\% of 10 ms frames), unlike any simulated pause. M + WavLM loses the caller in two
places (Figures~\ref{fig:callerwhere} and~\ref{fig:callertrace}): \val{calls/call_dev_clean/pre_selfl/start}\% of the
first 3 s of speech, and \val{calls/call_dev_clean/pre_selfl/return}\% of the caller returning after a silence. Every
fix below was chosen on dev-4 alone.

\begin{figure}[H]
\centering
\includegraphics[width=0.92\linewidth]{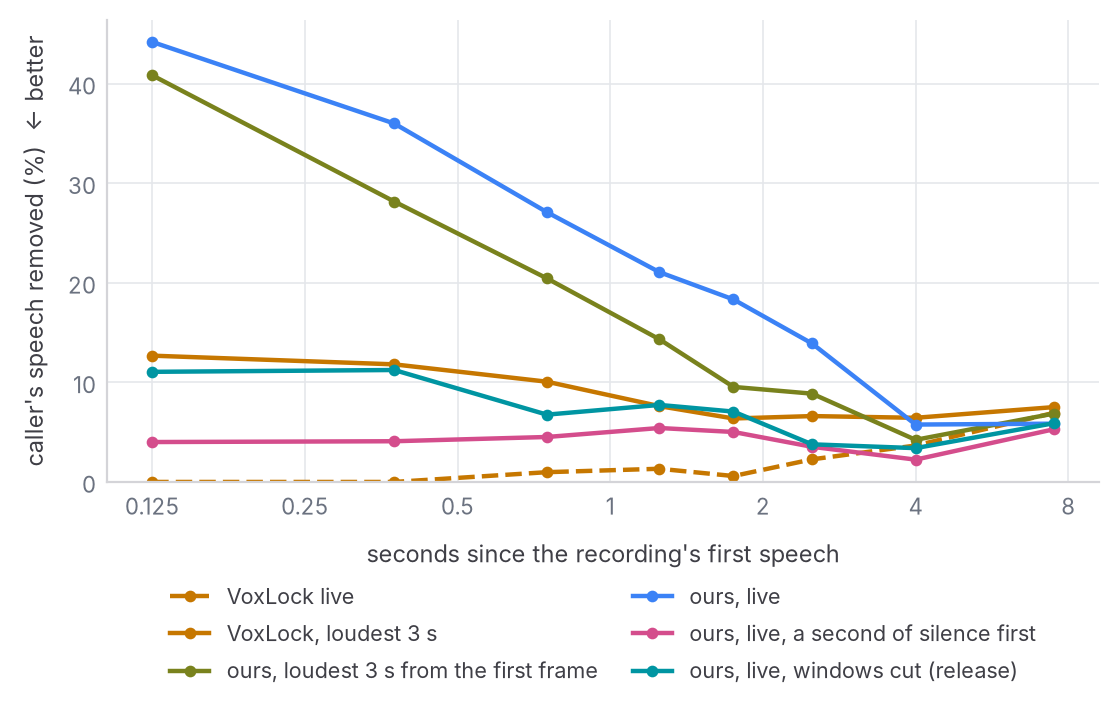}
\caption{Development calls, caller alone: share of the caller's speech removed against time since the recording's
first speech. M + WavLM from a cold start, with enrollment from the first frame, and after a second of digital silence;
the release, trained on windows cut from longer calls, removes least.}
\label{fig:calleronset}
\end{figure}

\textbf{The start is a cold start} (Figure~\ref{fig:calleronset}). Even enrolled from the first frame, the model
removes \val{calls/call_dev_clean/pre_loud/start}\% of the caller's first 3 s. Windows that open mid-speech lose
\val{calls/call_dev_clean/pre_selfl/start_t0}\%; windows where the caller starts after 2 s lose
\val{calls/call_dev_clean/pre_selfl/start_late}\%. No training example has anyone talking at its first sample.

\begin{figure}[H]
\centering
\includegraphics[width=0.92\linewidth]{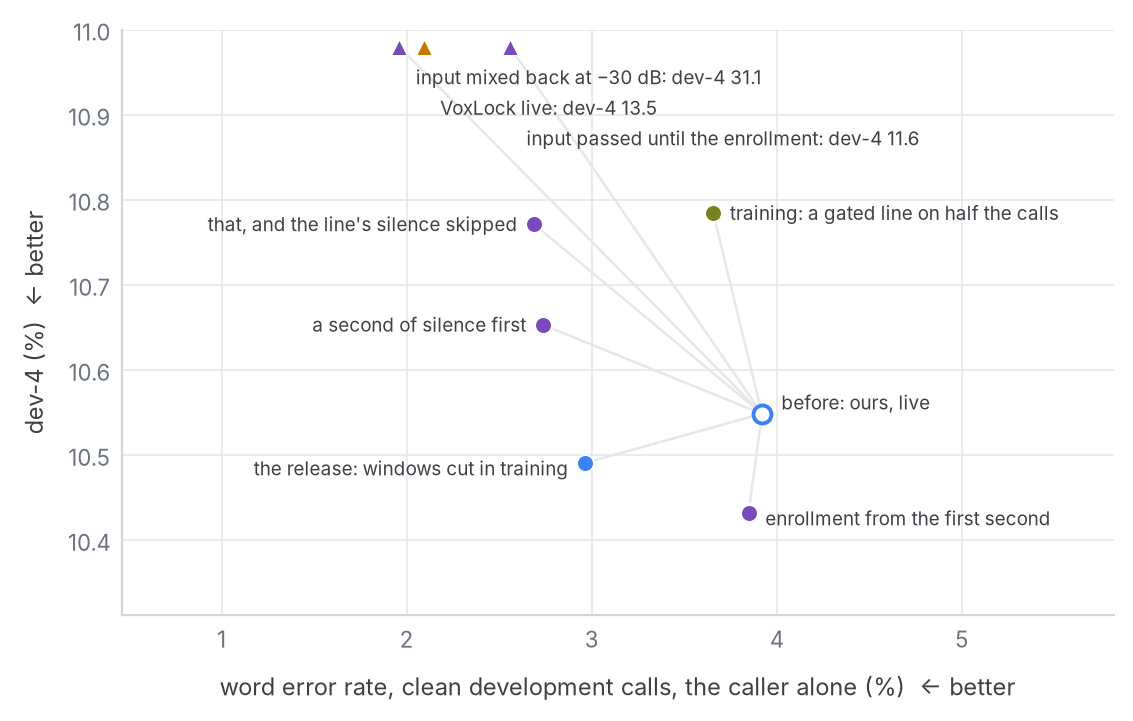}
\caption{Every fix tried for the lone caller, from M + WavLM before the window cut: word error rate on clean
development calls against dev-4. Fixes whose dev-4 lies above the axis sit on its top edge, labelled with their dev-4.}
\label{fig:rules}
\end{figure}

\begin{figure}[H]
\centering
\includegraphics[width=0.92\linewidth]{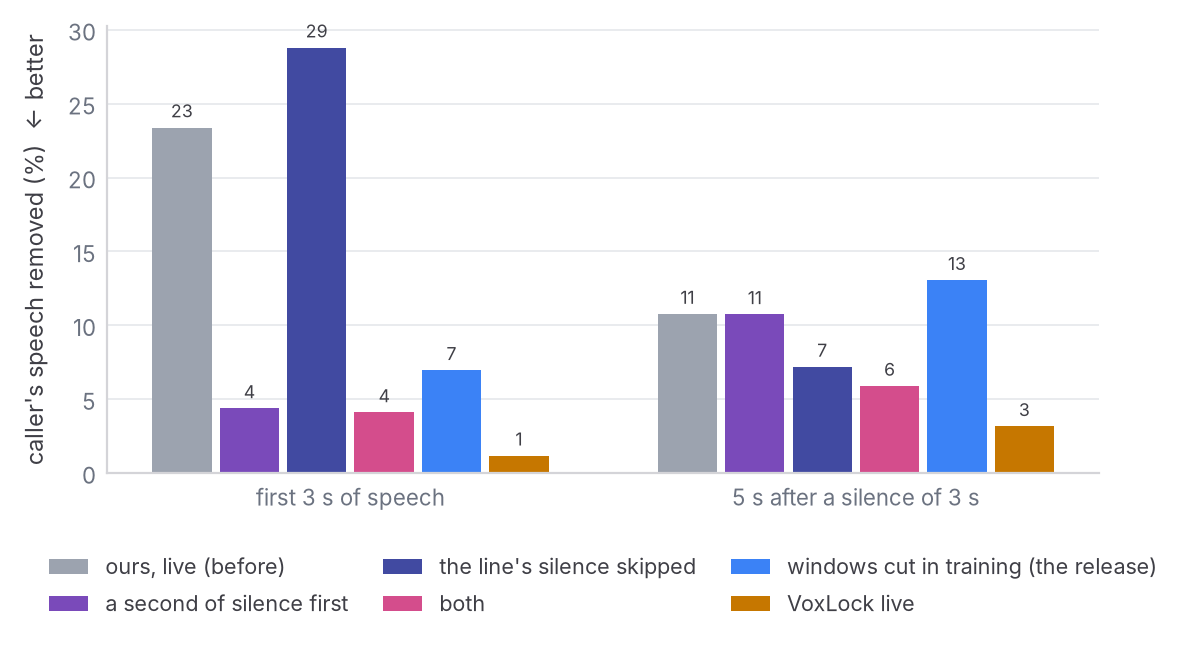}
\caption{Where each fix keeps the lone caller: share of speech removed in the first 3 s and in the 5 s after a silence
of 3 s, development calls.}
\label{fig:ruleswhere}
\end{figure}

\textbf{Fixes we did not keep} (Figures~\ref{fig:rules} and~\ref{fig:ruleswhere}). A second of digital silence before
each call saves the start but raises dev-4: the cold start also removed other voices that open meeting windows. Mixing
the input back lets every voice back. Earlier enrollment, skipping digital silence and a gated line in training do not
help dev-4.

\begin{figure}[H]
\centering
\includegraphics[width=0.92\linewidth]{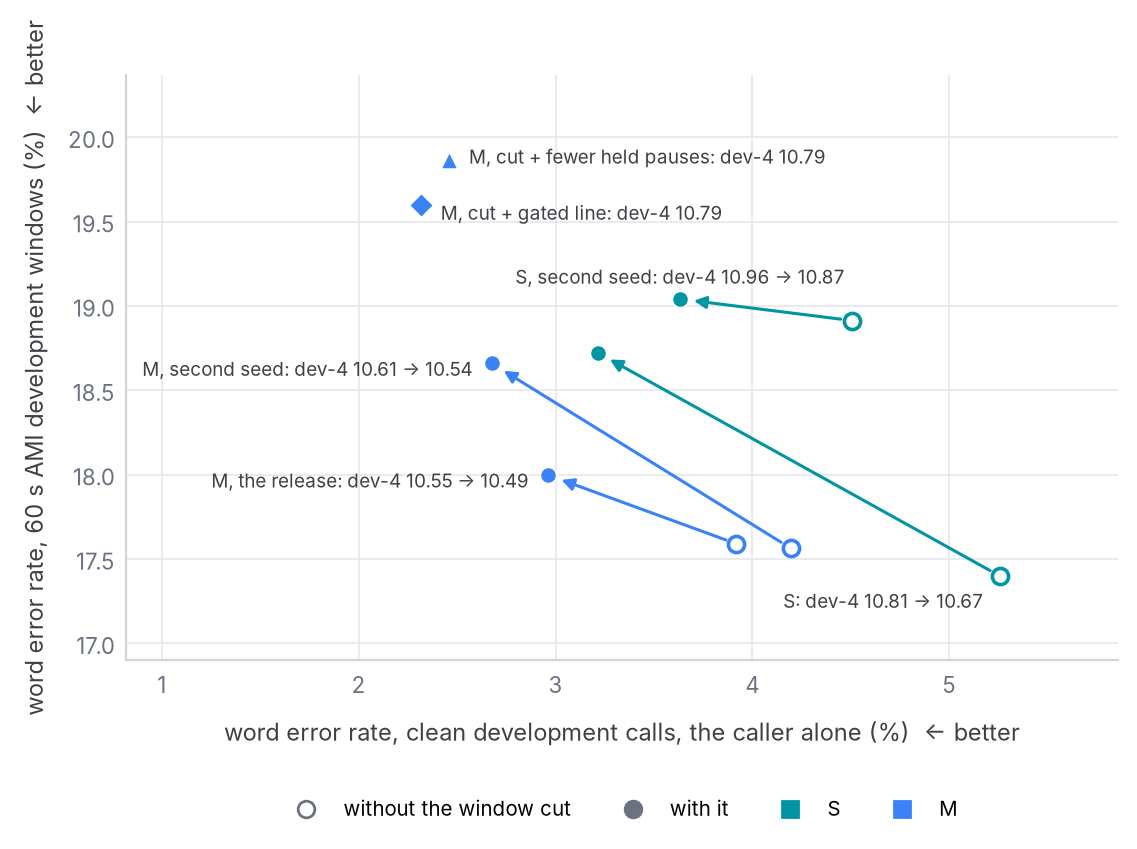}
\caption{The window cut, two seeds at S and at M: word error rate on clean development calls against the 60 s AMI
development windows, an arrow from each run without the cut to the same seed with it, plus two changes on top at M.
Labels: dev-4 without and with the cut.}
\label{fig:windowcut}
\end{figure}

\textbf{The window cut fixes the start.} Three long-call examples in ten are cut from a longer call, so a voice is
present from the first sample. In two seeds at S and M this lowers the error on clean calls and dev-4, at a cost on
the 60 s windows within seed noise (Figure~\ref{fig:windowcut}); it is the release. On the test calls the error falls
from \val{main/pre_selfl/call_test_clean/wer}\% to \val{main/rel_selfl/call_test_clean/wer}\%.

\subsection{Ablations}

\begin{figure}[H]
\centering
\includegraphics[width=0.92\linewidth]{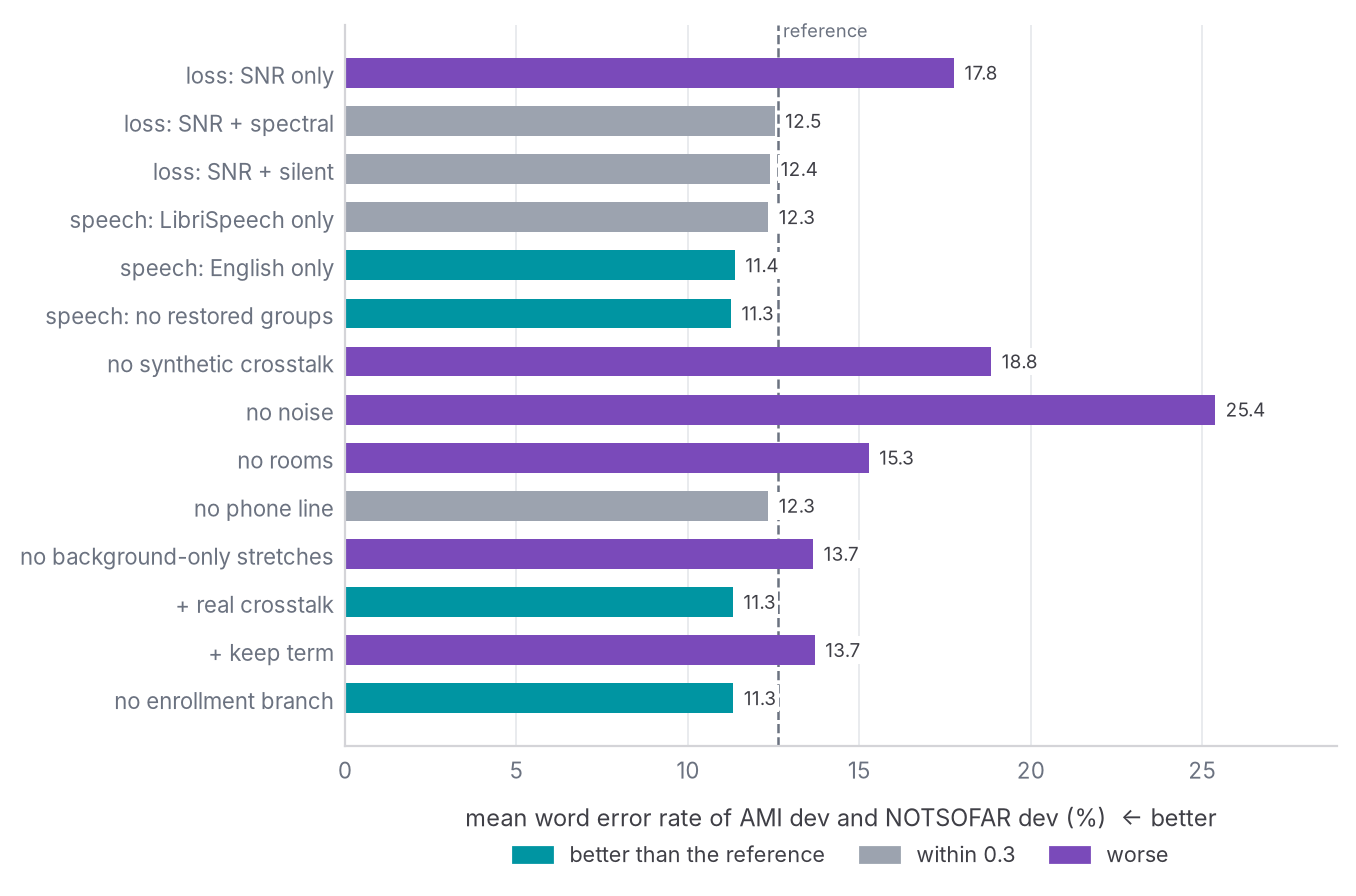}
\caption{Stage 1 of S, 30,000 steps, one change per bar: mean word error rate of AMI dev and NOTSOFAR dev. These short
windows cannot see long calls or a lone caller.}
\label{fig:stage1}
\end{figure}

\textbf{Synthetic data matters most} (Figure~\ref{fig:stage1}). Crosstalk, noise and rooms give the largest gains; the
signal-to-noise loss alone lets crosstalk through. Real crosstalk and dropping the restored speech help, and both
entered the release recipe.

\subsection{Enrollment}
\label{sec:enroll}

\begin{figure}[H]
\centering
\includegraphics[width=0.92\linewidth]{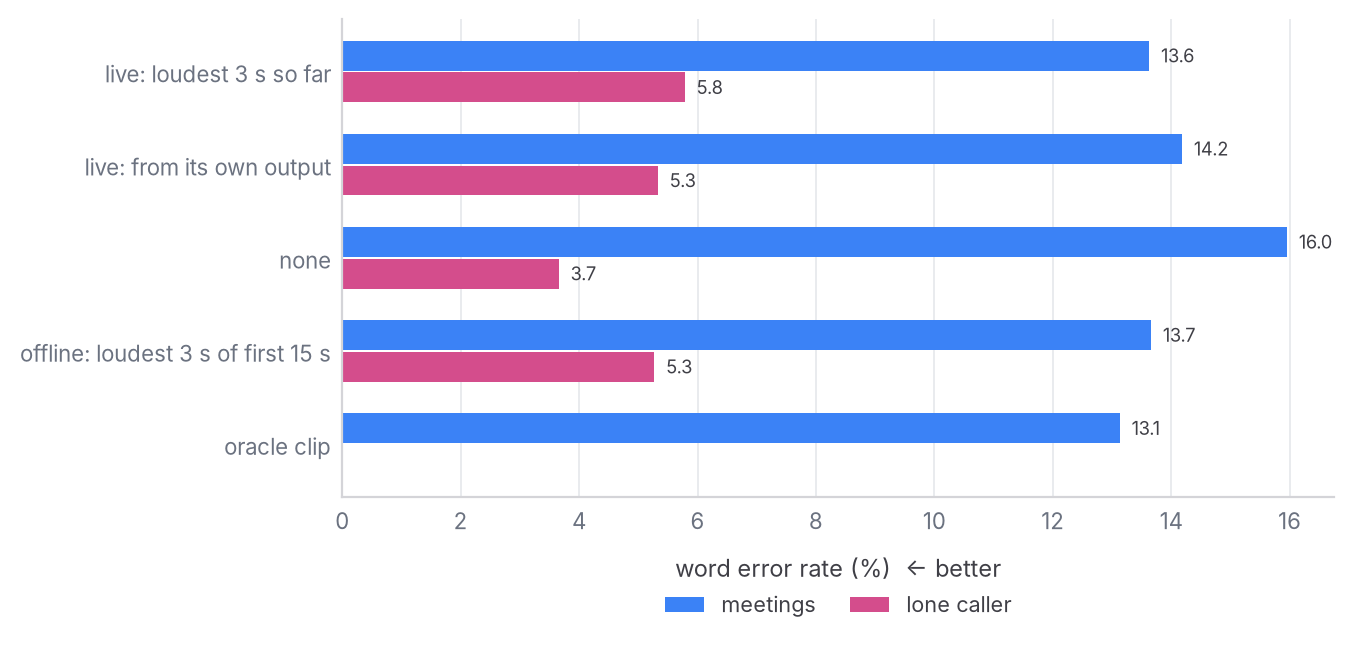}
\caption{Enrollment modes of the S release candidate: meetings (mean of three development sets) and a lone caller
(clean calls). Only the live modes can run in a call.}
\label{fig:enroll}
\end{figure}

\textbf{The loudest 3 s wins.} It beats enrolling from the model's own output; no enrollment is kinder to a lone caller
and worse on long meetings (Figure~\ref{fig:enroll}).

\subsection{Model size}

\begin{figure}[H]
\centering
\includegraphics[width=0.92\linewidth]{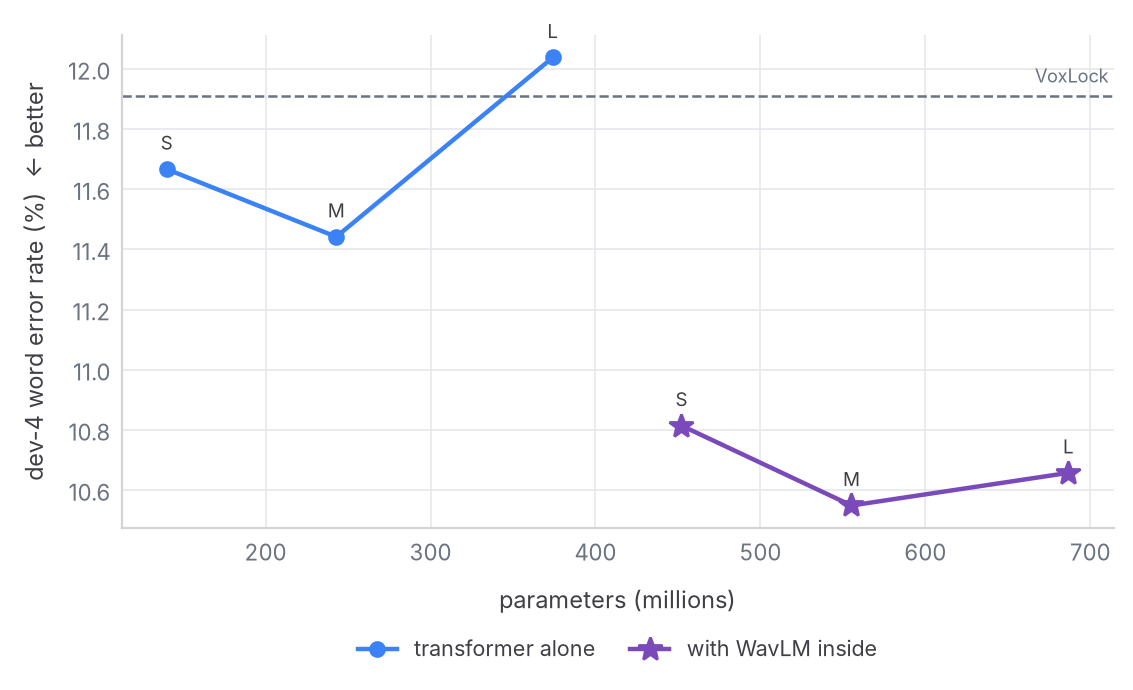}
\caption{Dev-4 (live enrollment) against parameters: S, M and L with the release recipe, alone and with WavLM inside,
before the window cut. Dashed: VoxLock.}
\label{fig:scale}
\end{figure}

\textbf{M is enough.} M beats S, and L does not beat M, with or without WavLM (Figure~\ref{fig:scale}). WavLM is worth
more than width: S with it (\val{r/SB_ftlong_ws_cwlt2_selfl/dev4}\%) beats every size without it.

\subsection{Self-supervised speech models}
\label{sec:ssl}

\begin{figure}[H]
\centering
\includegraphics[width=0.92\linewidth]{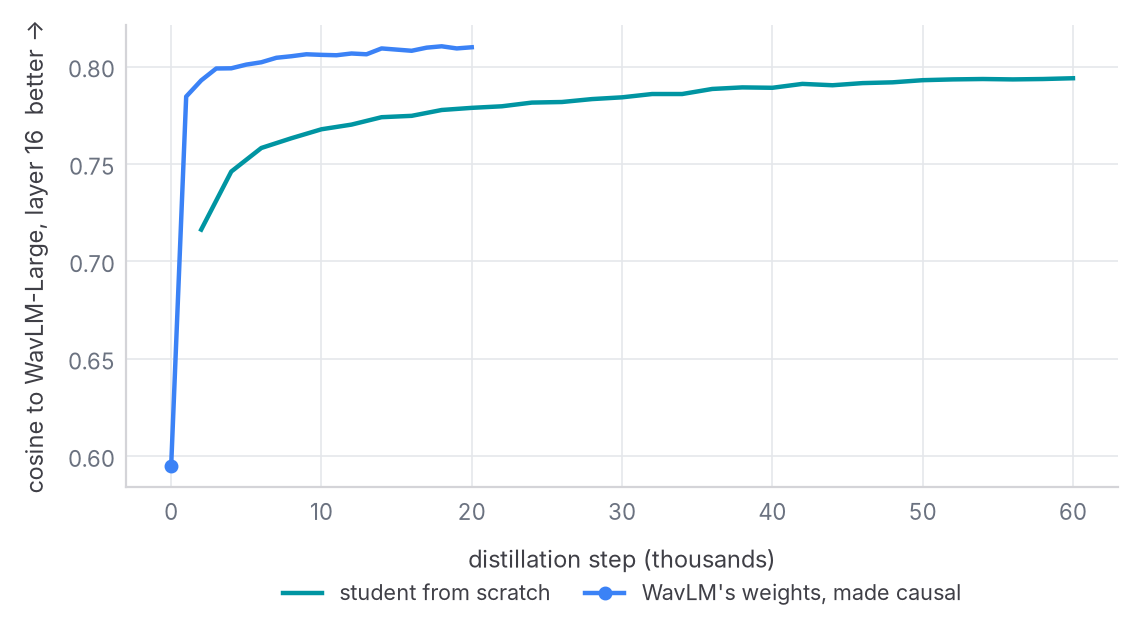}
\caption{Cosine to WavLM-Large's layer 16 on held-out simulated calls during distillation: a student from scratch, and
WavLM's own weights made causal (dot: before training). The causal WavLM overtakes within a thousand steps.}
\label{fig:ssldistill}
\end{figure}

\textbf{WavLM's outputs do not help.} As a teacher, a speaker encoder or a causal student, WavLM gives no gain: a
student learns WavLM's outputs on our calls, not what WavLM learned from 94k hours. WavLM's own weights made causal
overtake such a student within a thousand steps of distillation (Figure~\ref{fig:ssldistill}).

\begin{figure}[H]
\centering
\includegraphics[width=0.92\linewidth]{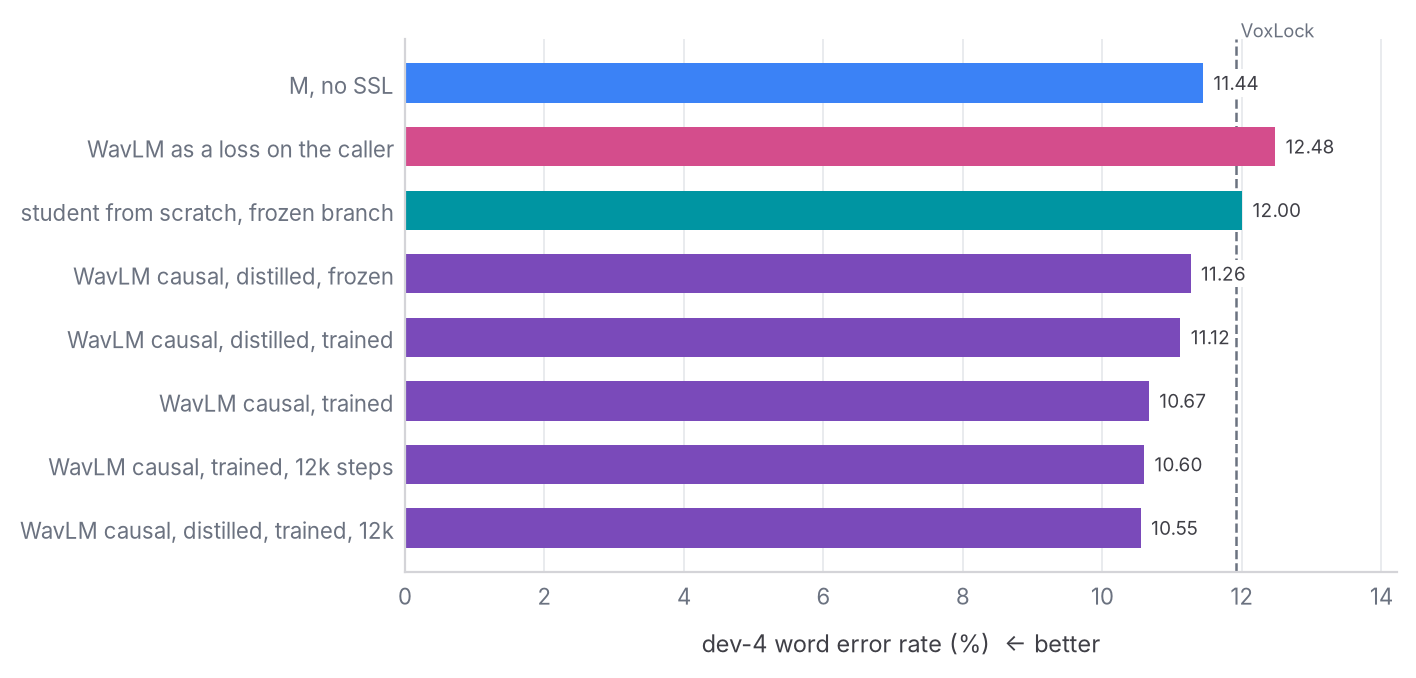}
\caption{Dev-4 of M with each use of WavLM, live enrollment. Dashed: VoxLock. Only WavLM's weights inside the model
help.}
\label{fig:ssl}
\end{figure}

\begin{figure}[H]
\centering
\includegraphics[width=0.92\linewidth]{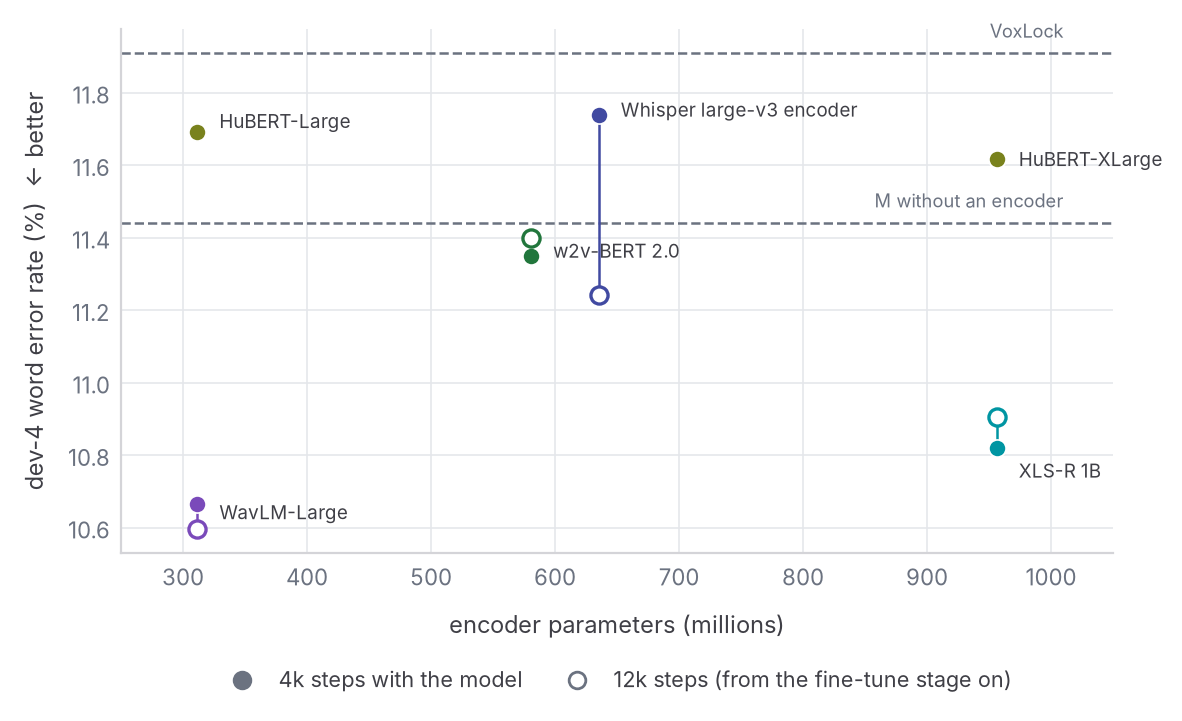}
\caption{Dev-4 of M with each pre-trained encoder made causal and trained with the model, against the encoder's size.
Filled: long-call stage only (4k steps); open: from the real-crosstalk fine-tune on (12k steps).}
\label{fig:sslscale}
\end{figure}

\textbf{WavLM's weights do.} Trained inside the model, WavLM lowers dev-4 from \val{r/MB_ftlong_ws_selfl/dev4}\% to
\val{r/pre_selfl/dev4}\% (Figure~\ref{fig:ssl}), helping meetings and the lone caller at once. Larger pre-training
sets do not explain it (Figure~\ref{fig:sslscale}): w2v-BERT 2.0, Whisper's encoder and HuBERT-Large help less or
hurt, and XLS-R 1B comes closest at three times the cost. WavLM's pre-training task, the main speaker of overlapped
speech, is the nearest to ours.

\subsection{Masking or resynthesis}

\begin{figure}[H]
\centering
\includegraphics[width=0.92\linewidth]{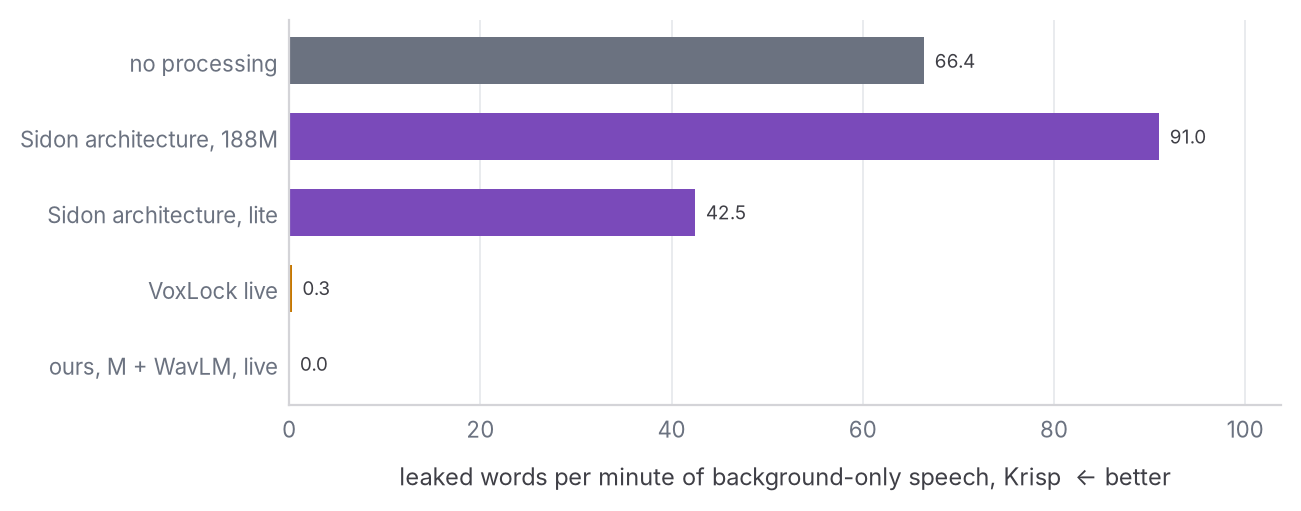}
\caption{Krisp: leaked words per minute of background-only speech. The Sidon-architecture models are trained on the
same calls as ours.}
\label{fig:resynth}
\end{figure}

\textbf{Resynthesis keeps every voice.} The Sidon-architecture models resynthesise self-supervised features, and a
mixture's features carry everyone (Figure~\ref{fig:resynth}).

\subsection{Controlled conditions}

\begin{figure}[H]
\centering
\includegraphics[width=0.92\linewidth]{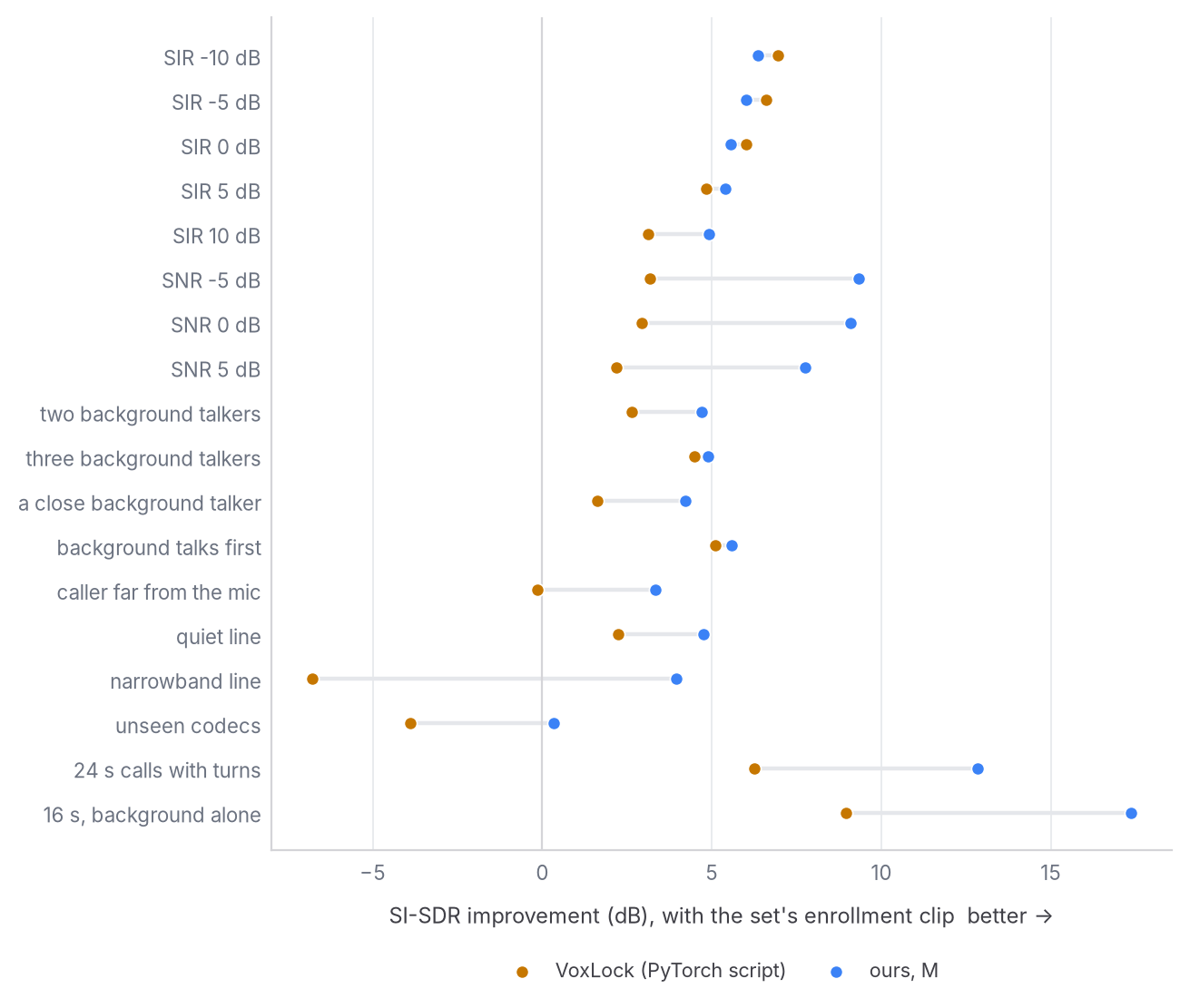}
\caption{Simulated stress sets: SI-SDR improvement of our M and VoxLock's PyTorch script, both given the set's
enrollment clip.}
\label{fig:stress}
\end{figure}

\textbf{Stress sets.} On simulated sets from held-out speakers, rooms and noises, M leads VoxLock under noise, on
narrow or unseen phone lines, with a distant caller and on long calls, and trails slightly when the background talker
is as loud as the caller or louder (Figure~\ref{fig:stress}).

\section{Serving}
\label{sec:serving}

\begin{figure}[H]
\centering
\includegraphics[width=\linewidth]{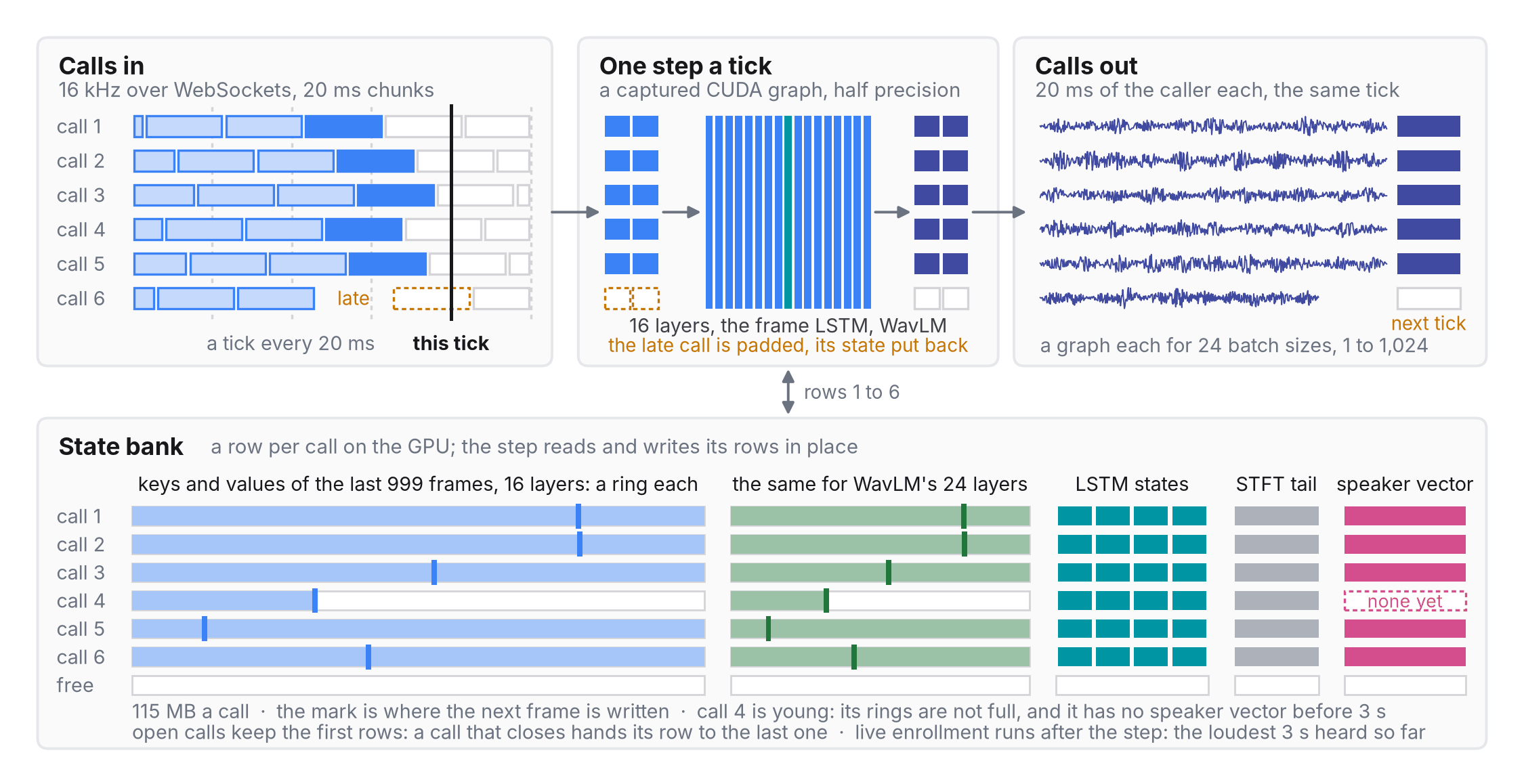}
\caption{The server at one tick, six calls drawn. Chunks arrive on each call's own clock; the tick steps every call
that has one and returns its audio. A late call is padded and its state put back. Open calls keep the first rows of
the state bank, so the step works in place.}
\label{fig:serving}
\end{figure}

\textbf{One step for every call.} Each call keeps a state bank: key-value rings of the last 999 frames per layer, the
LSTM states and the speaker vector (Figure~\ref{fig:serving}). Every 20 ms one CUDA-graph step runs every call with a
chunk in half precision, the LSTMs in single precision. The release keeps \val{model/MW/state-mb} MB a call and
matches the offline model to \val{serve/MW/equiv-snr} dB.

\subsection{Stress tests}

\begin{figure}[H]
\centering
\includegraphics[width=0.92\linewidth]{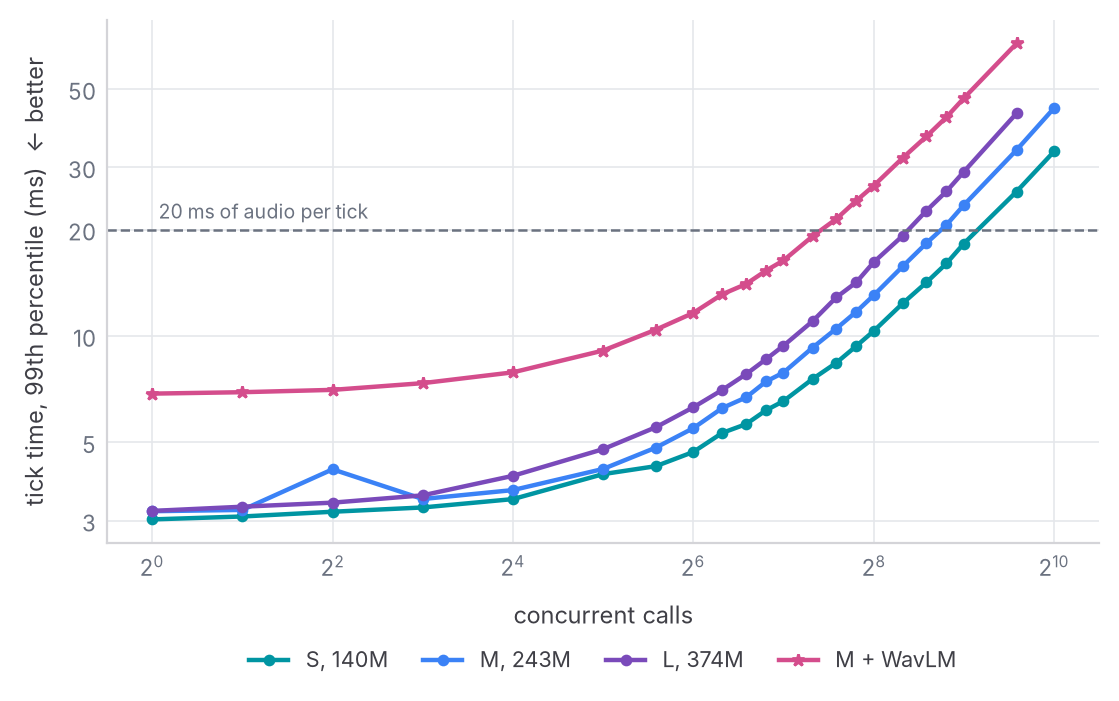}
\caption{Step time (99th percentile) against concurrent calls, fp16 with CUDA graphs, 20 ms chunks, one GPU. Below the
dashed line a step keeps up with real time.}
\label{fig:engine}
\end{figure}

\textbf{The step.} On one GPU with 20 ms chunks a step serves \val{serve/S/cap}, \val{serve/M/cap} and
\val{serve/L/cap} calls in real time with S, M and L (Figure~\ref{fig:engine}). WavLM cuts the release to
\val{serve/MW/cap-c2} calls, \val{serve/MW/cap-c4} with 40 ms chunks.

\begin{figure}[H]
\centering
\IfFileExists{img/fig_ws_latency.png}{\includegraphics[width=0.92\linewidth]{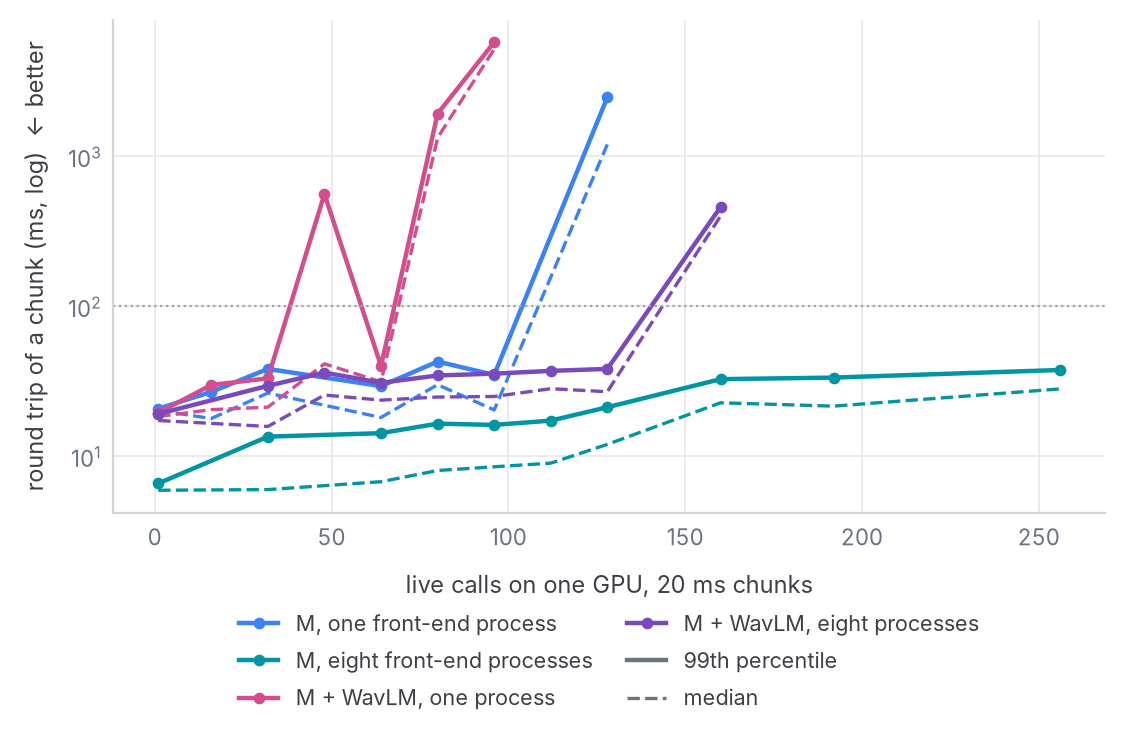}}{\pending{WebSocket load test}}
\caption{End to end over WebSockets on one GPU, live enrollment, 20 ms chunks: a chunk's round trip (99th percentile
solid, median dashed) after each call's first 15 s, for M and the release, with one front-end process or eight. Runs
where calls timed out are omitted.}
\label{fig:wslat}
\end{figure}

\textbf{End to end} over WebSockets (Figure~\ref{fig:wslat}), one Python front end saturates first, at
\val{serve/wsMW/hold} calls. With eight front-end processes the release holds \val{serve/mpMW/hold} calls on 20 ms
chunks (\val{serve/mpMW/hold-sp99} ms at the 99th percentile) and \val{serve/mpMW40/hold} on 40 ms.

\begin{figure}[H]
\centering
\includegraphics[width=0.92\linewidth]{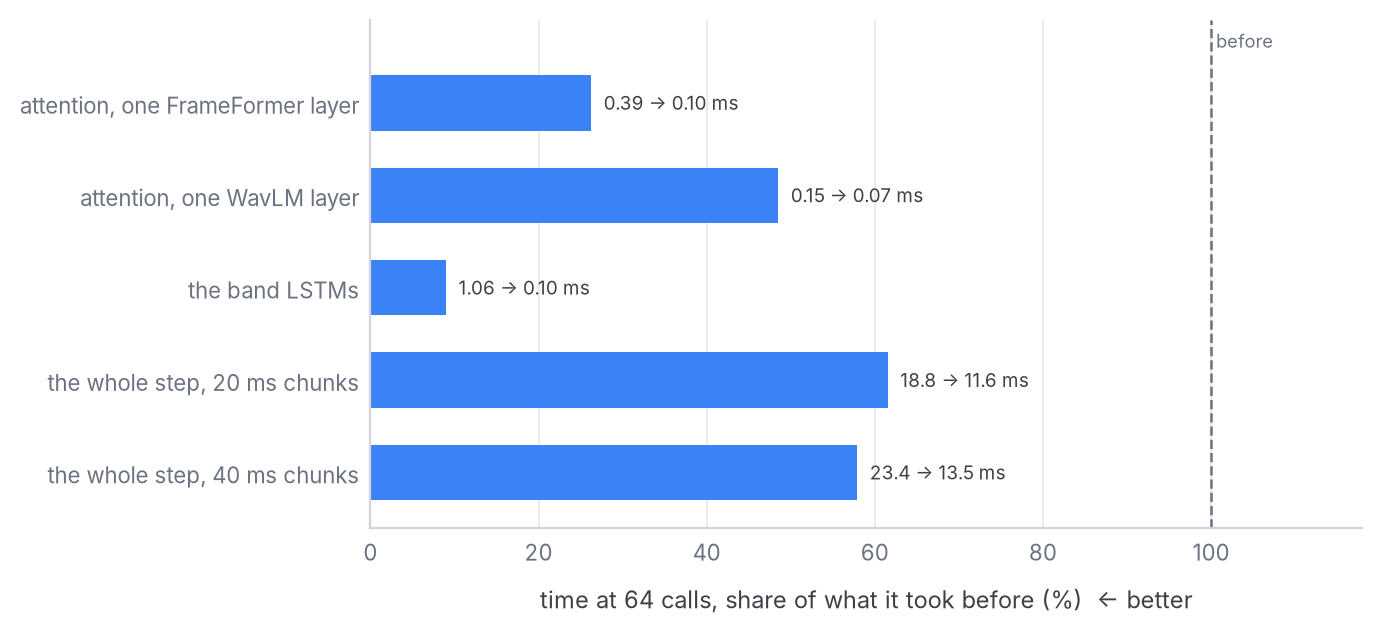}
\caption{The serving kernels at 64 calls of the release: each part's time as a share of what it took before them,
milliseconds on the bars.}
\label{fig:kernels}
\end{figure}

\begin{figure}[H]
\centering
\includegraphics[width=0.92\linewidth]{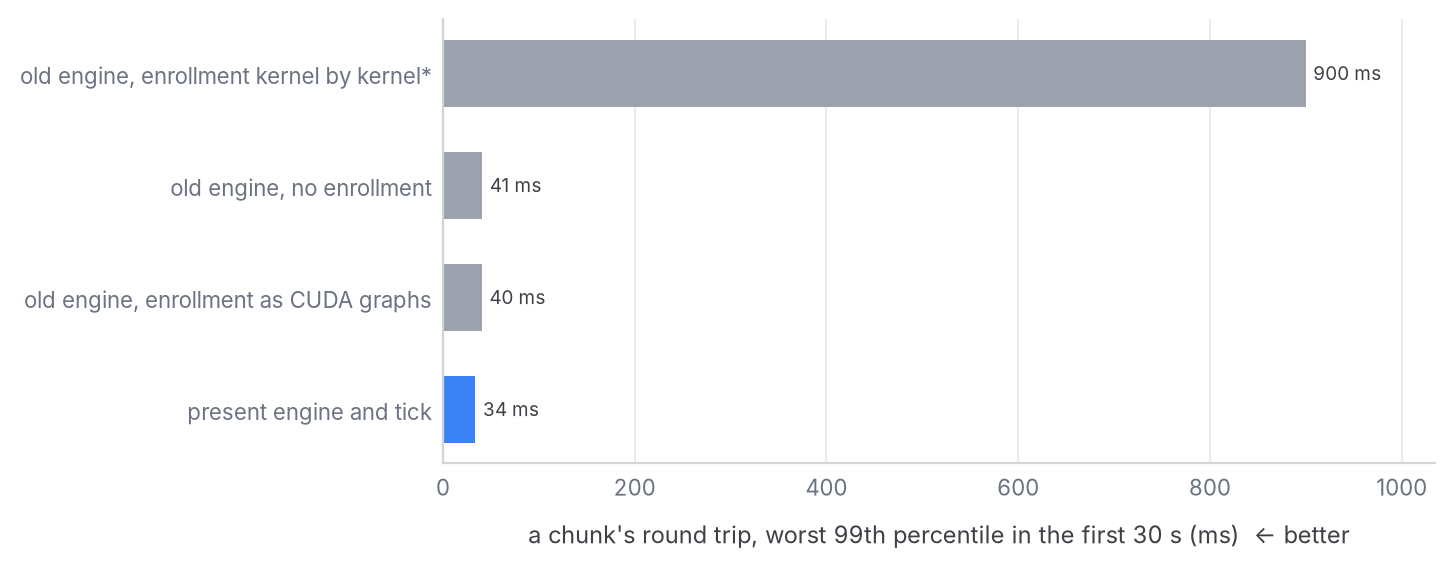}
\caption{A fresh server given 64 calls of the release at once on 20 ms chunks: the worst 99th percentile of a chunk's
round trip in its first 30 s (*: first minute), as the enrollment's cost was taken out of the step.}
\label{fig:coldstart}
\end{figure}

\begin{figure}[H]
\centering
\includegraphics[width=0.92\linewidth]{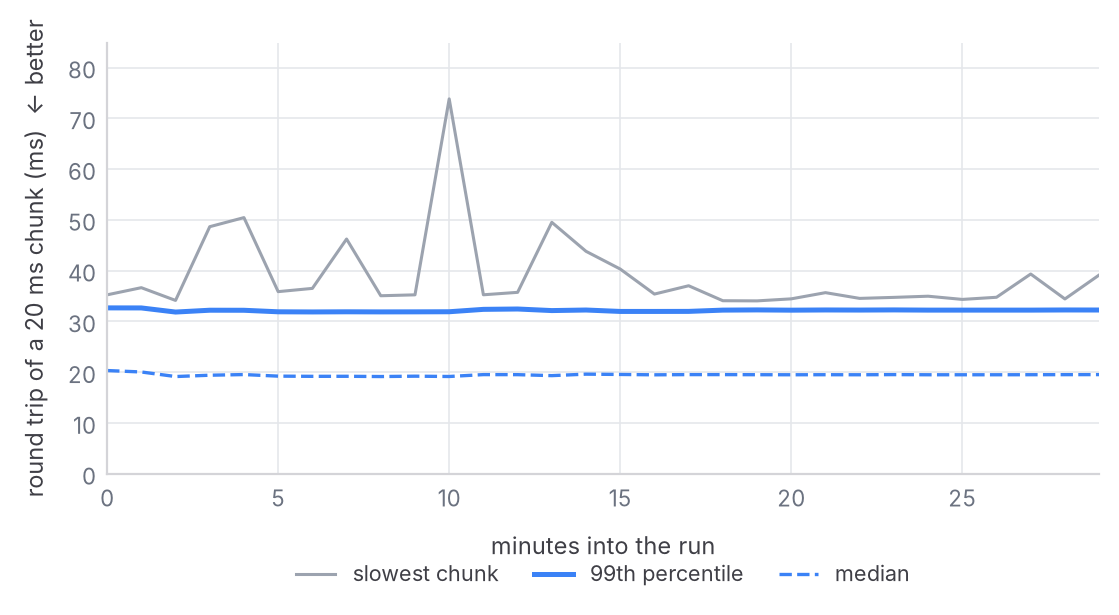}
\caption{Thirty minutes at 64 calls of the release on 20 ms chunks: a chunk's round trip in each minute.}
\label{fig:soak}
\end{figure}

\textbf{Two kernels} speed up the step (Figure~\ref{fig:kernels}): a Triton ring-attention kernel and band LSTMs off
cuDNN take a 64-call step from \val{kern/step/2/old} to \val{kern/step/2/new} ms with the same output. A CUDA graph
for the enrollment encoder removes a \val{serve/soakMWcold/p99-first} ms cold-start lag (Figure~\ref{fig:coldstart}),
and \val{serve/soakMW/minutes} minutes at \val{serve/soakMW/calls} calls lost no call, 99th percentile
\val{serve/soakMW/p99} ms (Figure~\ref{fig:soak}).

\subsection{More than one GPU}

\textbf{Tensor and data parallelism.} Tensor parallelism (\texttt{-{}-tp N}) splits every layer's heads and
feed-forward over GPUs, sharing the per-call part of the step and the call's memory (\val{tp/tp1/c2/state-mb} MB on
one GPU, \val{tp/tp2/c2/state-mb} MB on each of two); the output agrees to \val{tp/check/tp2/agree} dB. Data
parallelism (\texttt{-{}-dp N}) runs a copy per GPU behind one port.

\begin{figure}[H]
\centering
\IfFileExists{img/fig_tp_step.png}{\includegraphics[width=0.92\linewidth]{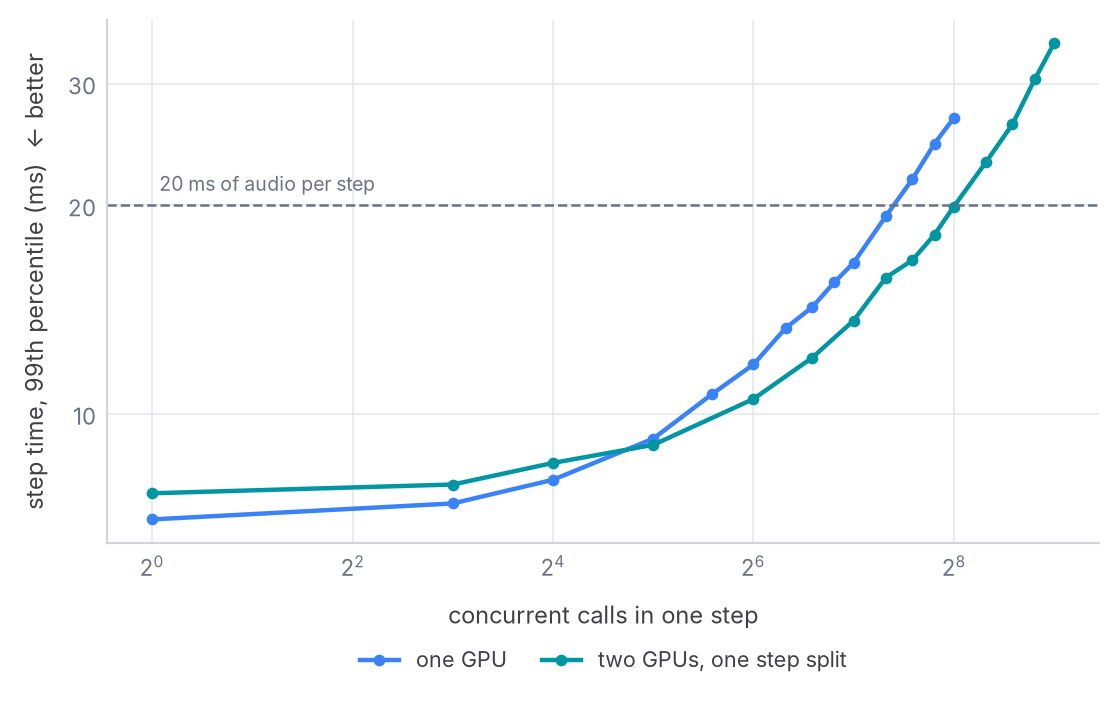}}{\pending{tensor parallelism}}
\caption{Step time (99th percentile) against concurrent calls, the release on one GPU and split over two, 20 ms chunks,
on otherwise idle GPUs. Below the dashed line the server keeps up with real time.}
\label{fig:tpstep}
\end{figure}

\textbf{The step.} Two GPUs carry \val{tp/tp2/c2/cap} calls in real time against \val{tp/tp1/c2/cap} on one
(Figure~\ref{fig:tpstep}), \val{tp/tp2/c2/cap-gain}$\times$, since the step's fixed cost is not shared. Below about
forty calls one GPU is quicker.

\begin{figure}[H]
\centering
\IfFileExists{img/fig_tp_ws.png}{\includegraphics[width=0.92\linewidth]{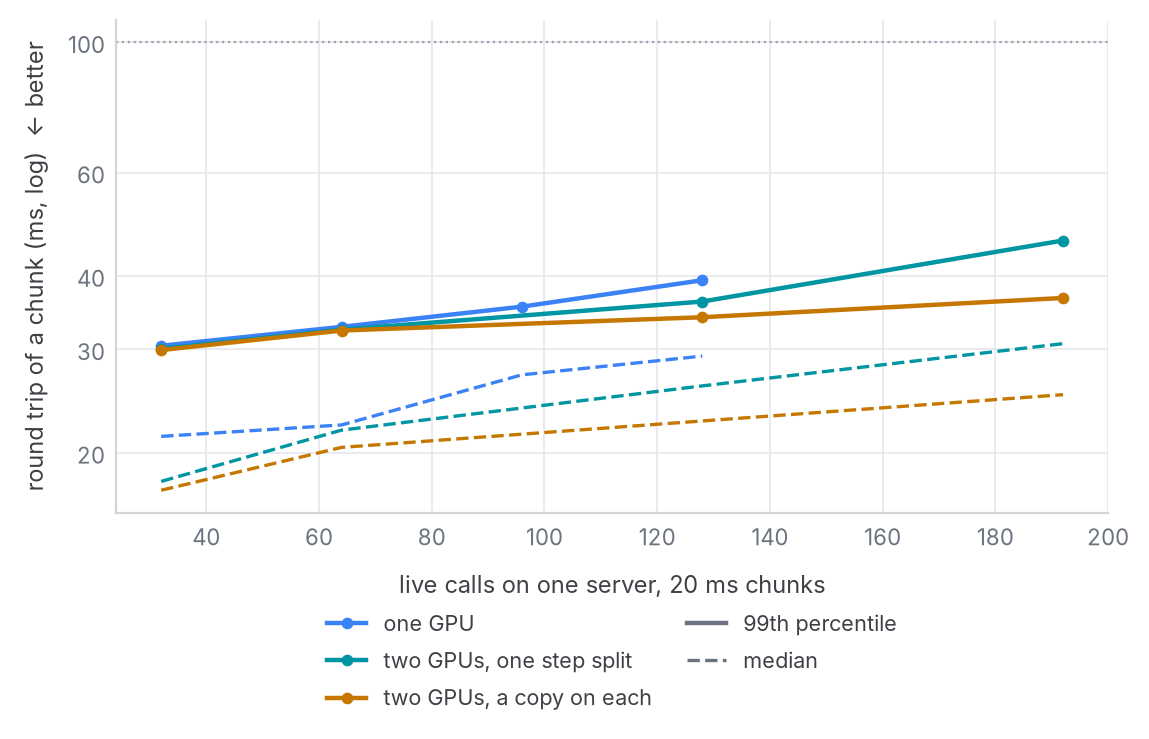}}{\pending{several GPUs end to end}}
\caption{End to end over WebSockets on one and two GPUs, 20 ms chunks, live enrollment: a chunk's round trip (99th
percentile solid, median dashed) after each call's first 15 s. Runs that lost a call, or sent more than one chunk in a
thousand later than 100 ms, are omitted.}
\label{fig:tpws}
\end{figure}

\textbf{End to end,} the front end limits first (Figure~\ref{fig:tpws}): with twelve front-end processes one GPU and
the split step both hold \val{tp/ws/tp1/hold} calls, and a copy per GPU \val{tp/ws/dp2/hold}.

\section{In a voice agent}
\label{sec:agent}

\begin{figure}[H]
\centering
\includegraphics[width=\linewidth]{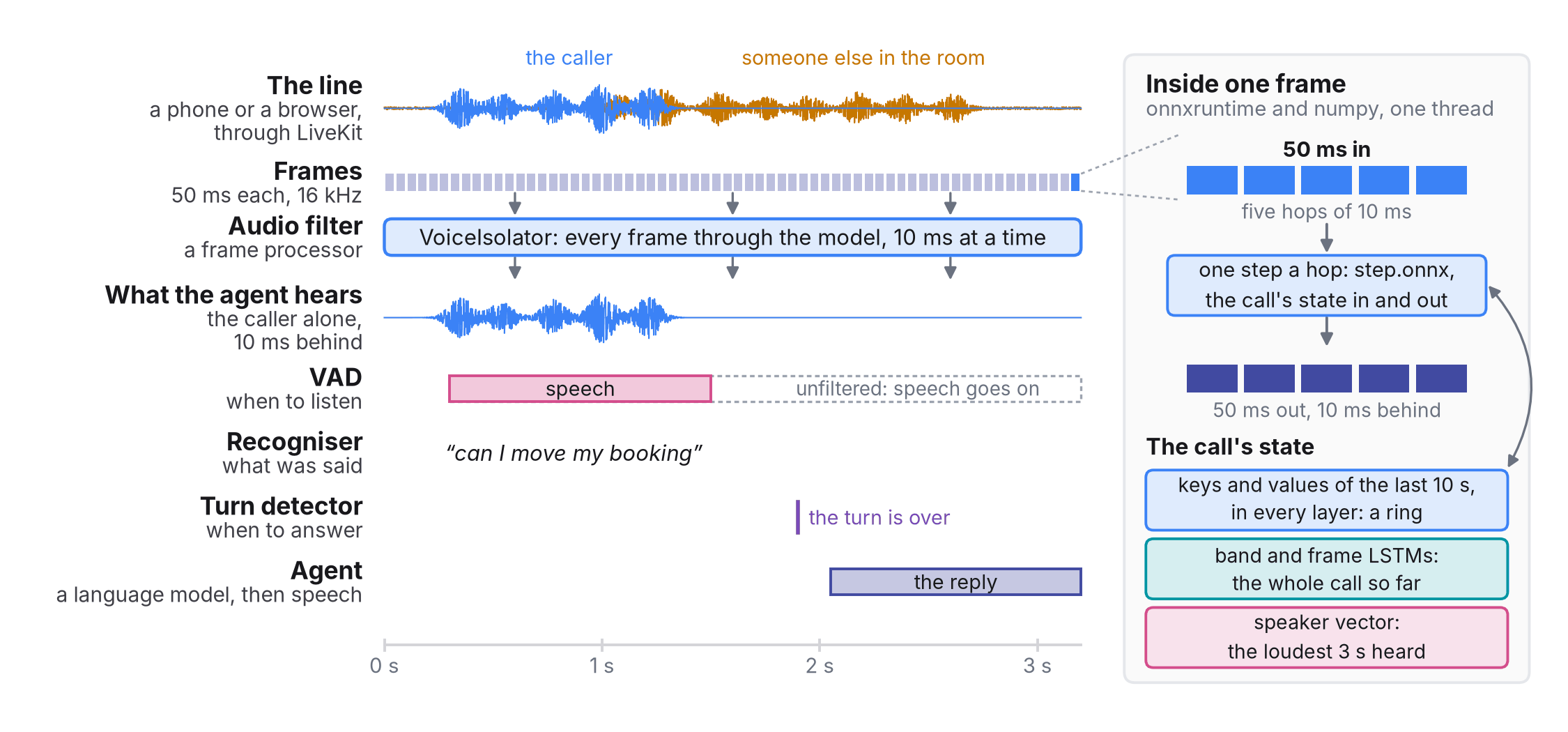}
\caption{The isolator in a LiveKit agent, drawn over three seconds of a call: a frame processor on the caller's audio,
ahead of the VAD, the recogniser and the turn detector. Right: one 50 ms frame, stepped 10 ms at a time. Each call keeps
its own state; one onnxruntime session serves every call in the process.}
\label{fig:livekit}
\end{figure}

\textbf{A LiveKit plugin.} In LiveKit Agents an audio filter hears the caller before the voice activity detector
(VAD), the turn detector and the recogniser; we ship the isolator as one (Figure~\ref{fig:livekit}). One streaming
step runs in ONNX with the call's state passed in and out, without a GPU or PyTorch, and matches the PyTorch model to
\val{agent/onnx/SB_ftlong_ws/snr-live} dB.

\begin{figure}[H]
\centering
\includegraphics[width=\linewidth]{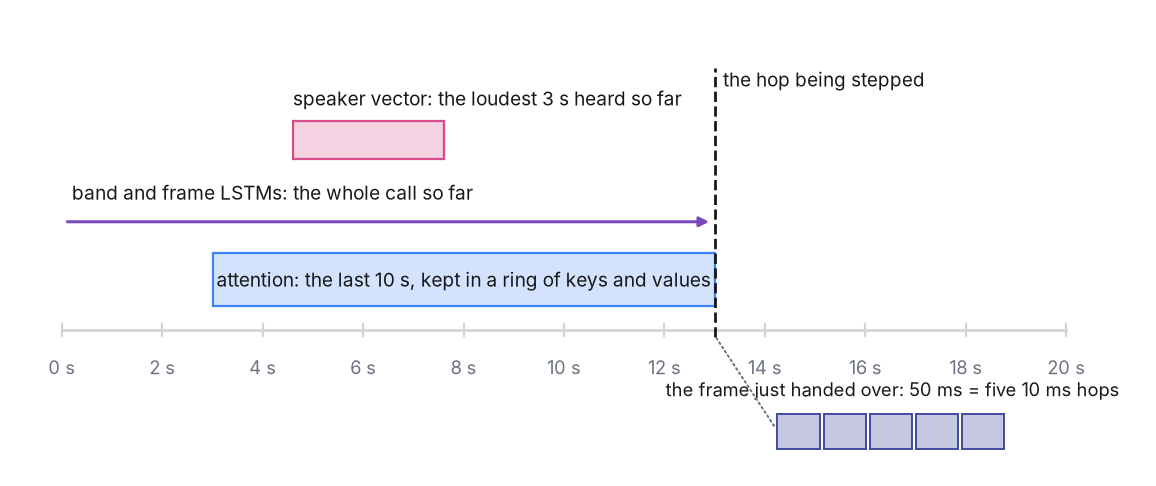}
\caption{What one 10 ms hop sees, to scale, over a 20 s call handed over in 50 ms frames: the attention window, the
LSTMs and the speaker vector belong to the call, not the frame.}
\label{fig:chunks}
\end{figure}

\textbf{Frames do not cut the context.} The speaker vector, the 10 s of attention and the LSTMs belong to the call,
not to LiveKit's 50 ms frame (Figure~\ref{fig:chunks}), so the output equals the offline output, 10 ms later.

\begin{figure}[H]
\centering
\includegraphics[width=0.92\linewidth]{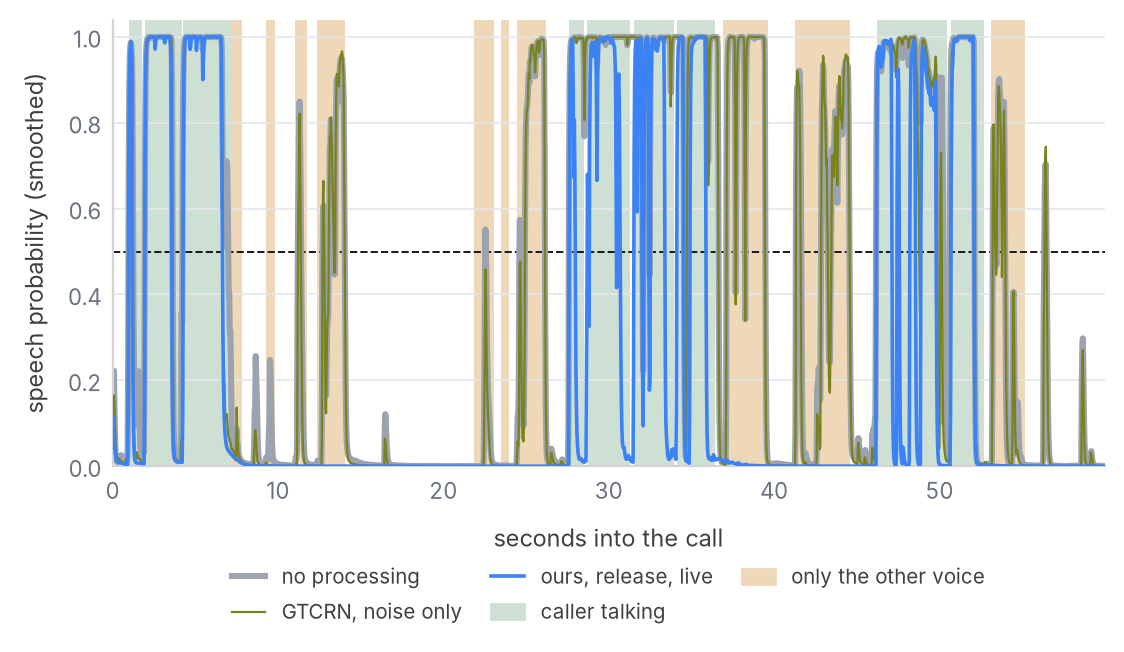}
\caption{One call with another voice: LiveKit's smoothed Silero probability on the raw call, behind GTCRN and behind
the release. Green: the caller talks; orange: only the other voice. Above the dashed line a window counts as speech.}
\label{fig:vadtrace}
\end{figure}

\begin{figure}[H]
\centering
\includegraphics[width=0.92\linewidth]{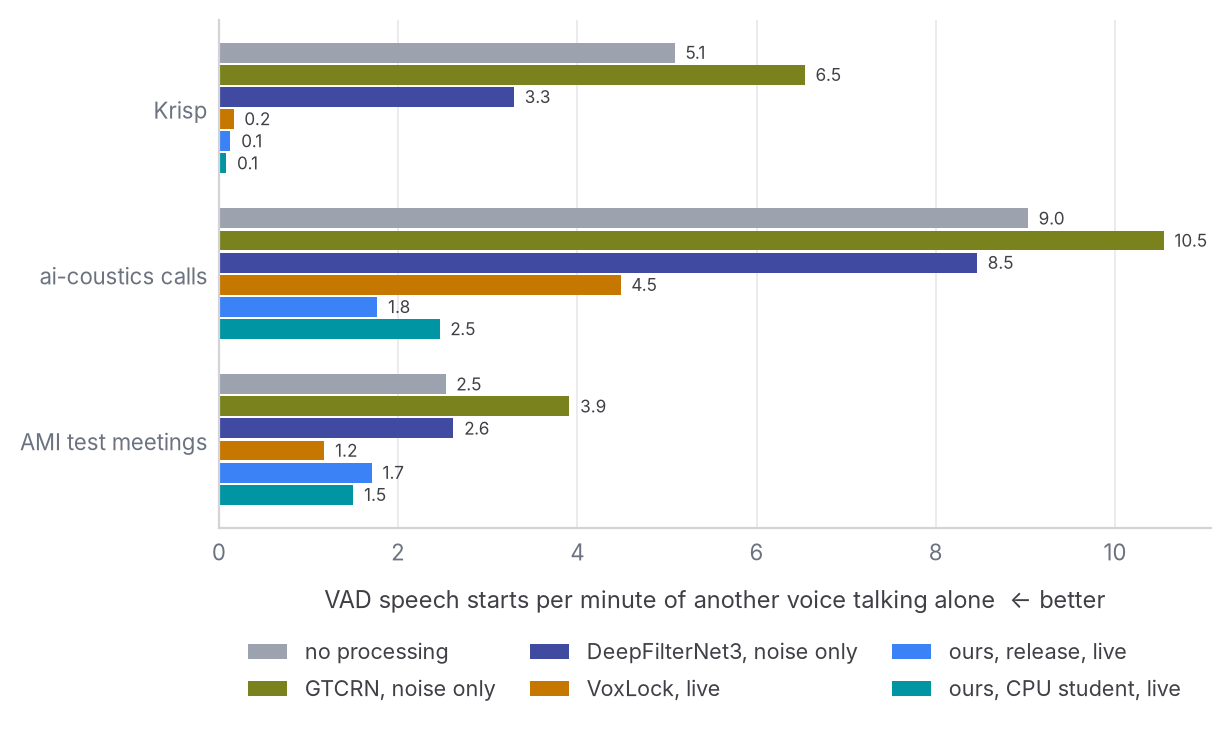}
\caption{LiveKit's VAD behind each system: speech starts inside the labelled stretches where only another voice talks,
per minute of them, on three sets of real recordings.}
\label{fig:vad}
\end{figure}

\textbf{The VAD hears fewer other voices.} Where only another voice talks on Krisp, LiveKit's Silero VAD starts speech
\val{vad/krisp/raw/bg_starts_per_min} times a minute on the raw call and \val{vad/krisp/rel/bg_starts_per_min} behind
the release (Figures~\ref{fig:vadtrace} and~\ref{fig:vad}). Behind GTCRN, a noise suppressor, it rises to
\val{vad/krisp/gtcrn/bg_starts_per_min}: a voice cleaned of noise sounds more like speech.

\textbf{The caller's turns.} The release cuts the share of the caller's pauses heard as speech from
\val{vad/call_test_mix/raw/false}\% to \val{vad/call_test_mix/rel/false}\%, but the VAD misses
\val{vad/call_test_clean/rel/missed}\% of a lone caller's speech (VoxLock
\val{vad/call_test_clean/voxlock_live/missed}\%), mostly in the first seconds (Appendix~\ref{app:tables}).

\section{On one CPU core}
\label{sec:cpu}

\begin{figure}[H]
\centering
\includegraphics[width=\linewidth]{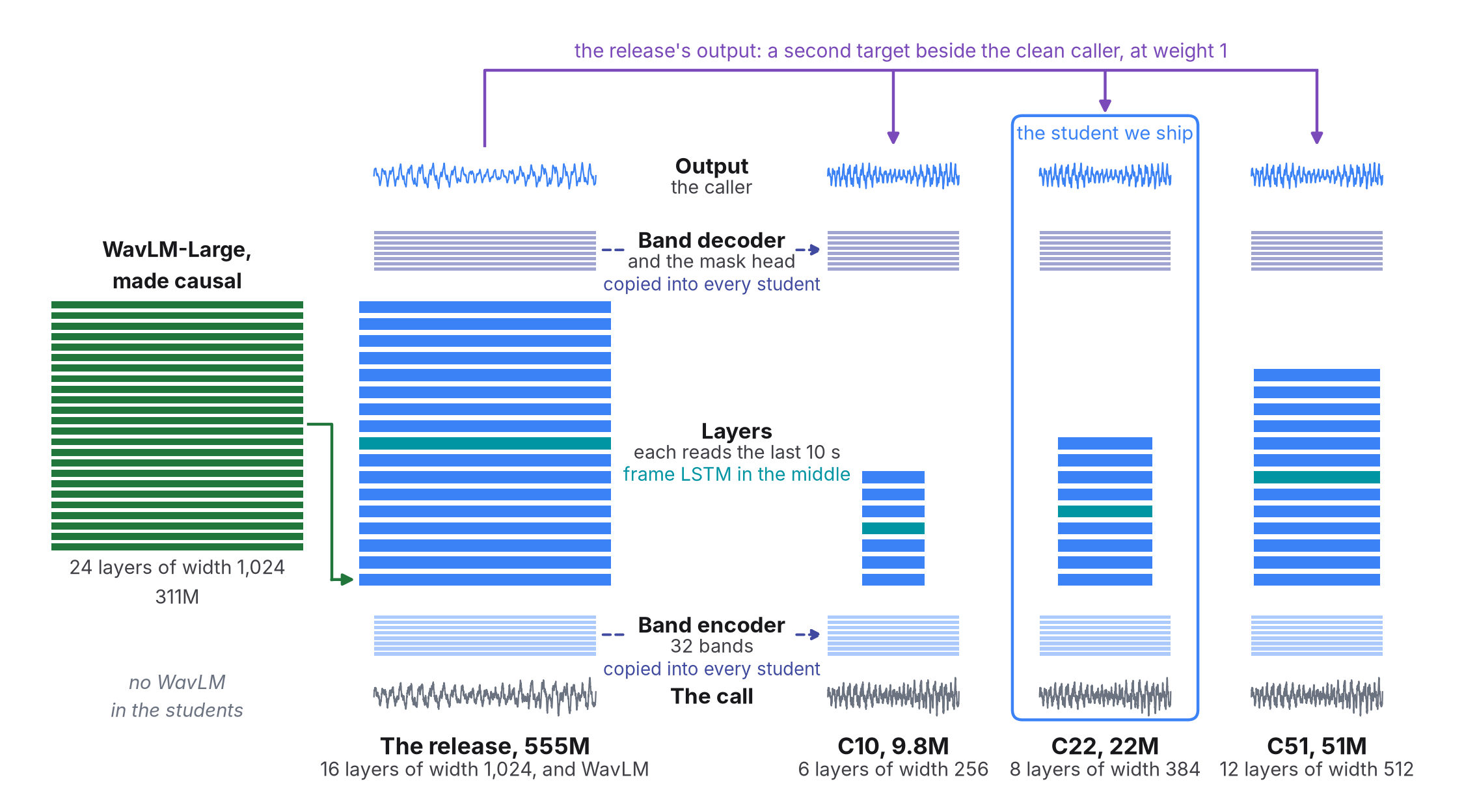}
\caption{The CPU students beside the release, each layer a bar as long as the model is wide. The band encoder and
decoder, the mask head and the enrollment encoder keep their shapes and start from the release's weights; the
transformer shrinks but keeps the 10 s window and the frame LSTM; WavLM is dropped. The release's output is a second
target beside the clean caller.}
\label{fig:cpuarch}
\end{figure}

\textbf{Students for the CPU.} VoxLock's 8-bit ONNX model needs \val{cpu/cmp/voxlock/20/core}\% of a CPU thread per
call; even S without WavLM needs \val{agent/onnx/SB_ftlong_ws/rtf} s of a thread per second of audio. We distil the
release into three FrameFormers for one core, C10, C22 and C51 (Figure~\ref{fig:cpuarch}).

\begin{figure}[H]
\centering
\includegraphics[width=0.92\linewidth]{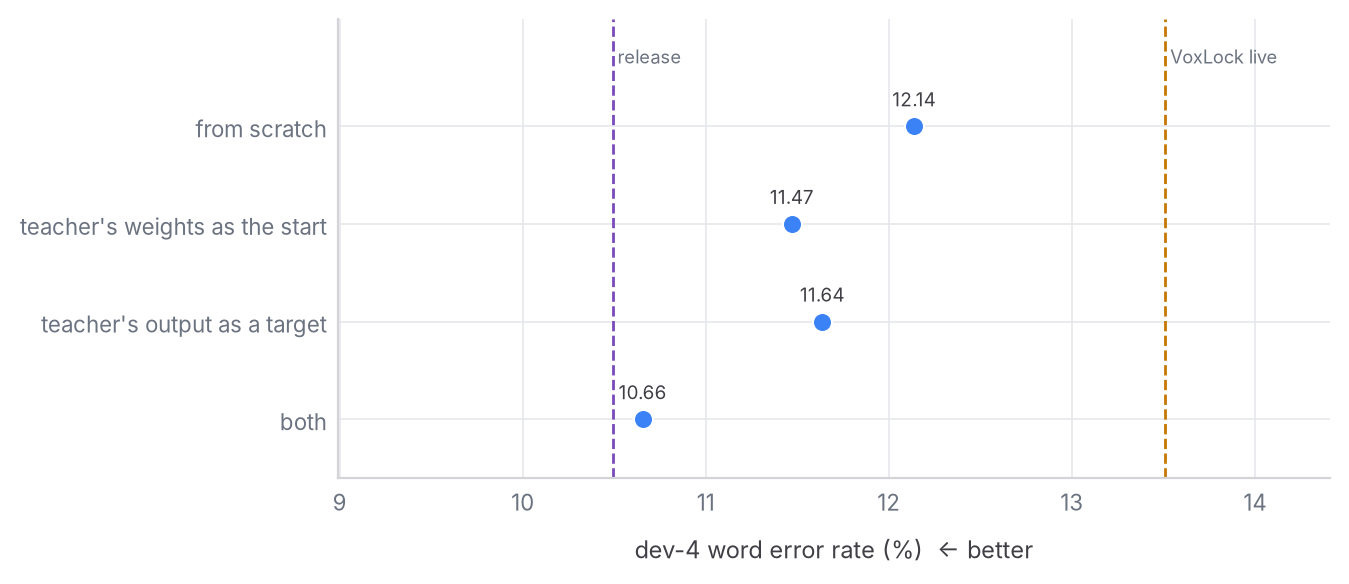}
\caption{Distilling the release into C22, release recipe throughout: dev-4 from scratch, from the release's weights,
with the release's output as a second target, and both. Dashed: VoxLock live and the release.}
\label{fig:cpukd}
\end{figure}

\textbf{The teacher is worth more than size} (Figure~\ref{fig:cpukd}). C22 from scratch reaches
\val{r/C22_ftlong_sc_selfl/dev4}\% on dev-4; taught by the release, \val{r/C22_ftlong_kd_selfl/dev4}\%, near the
release's \val{r/rel_selfl/dev4}\%. Most of the gain is the lone caller: the student takes from the release's output
what WavLM gave the release.

\begin{figure}[H]
\centering
\includegraphics[width=0.92\linewidth]{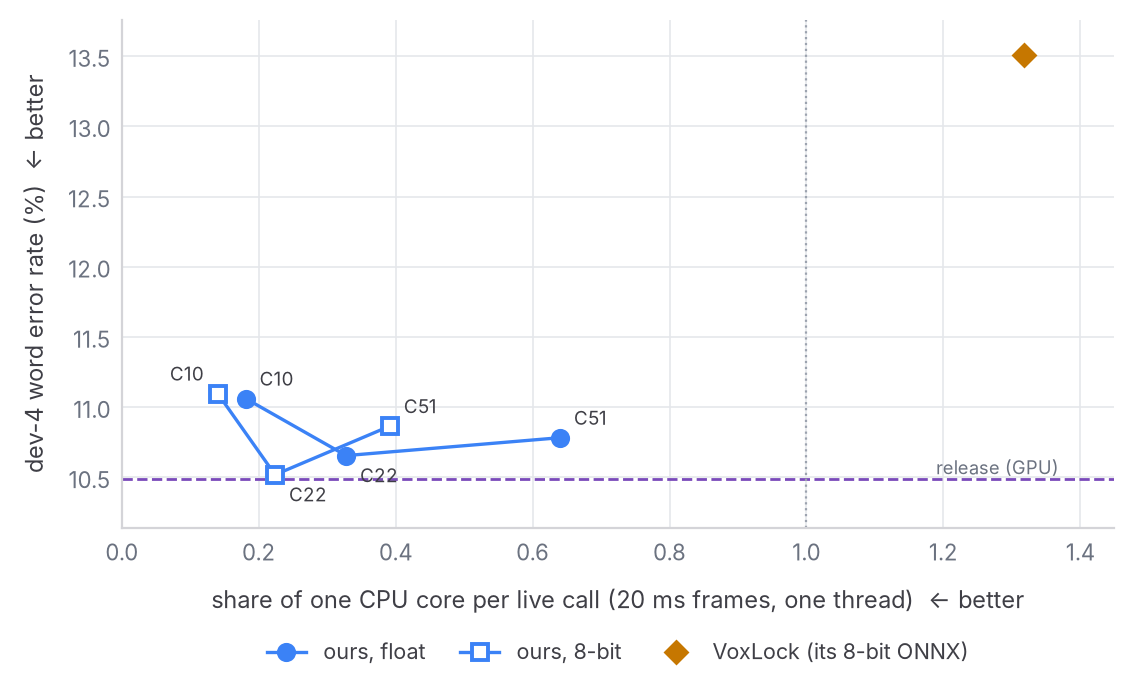}
\caption{Dev-4 against the share of one CPU thread a live call takes through the plugin in 20 ms frames, each
distilled size in float and 8 bits, and VoxLock's 8-bit package. Dashed: the release, on a GPU.}
\label{fig:cpucost}
\end{figure}

\textbf{Choosing the size.} A rule fixed before scoring takes the lowest dev-4 unless a larger size wins by more than
0.2 points. C10, C22 and C51 reach \val{r/C10_ftlong_kd_selfl/dev4}\%, \val{r/C22_ftlong_kd_selfl/dev4}\% and
\val{r/C51_ftlong_kd_selfl/dev4}\% (Figure~\ref{fig:cpucost}), so the student is \val{cpu/student}.

\begin{figure}[H]
\centering
\includegraphics[width=0.92\linewidth]{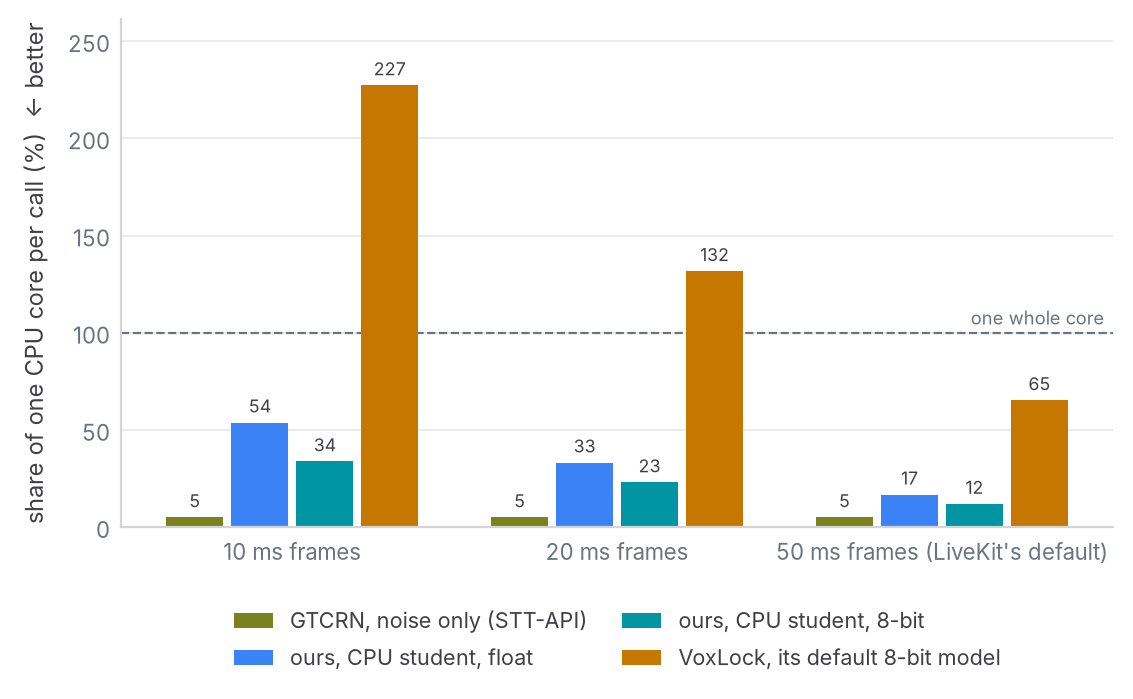}
\caption{Share of one CPU thread per live call by LiveKit's frame size: GTCRN (noise only), the CPU student in float
and 8 bits, and VoxLock's default 8-bit model, timed in one run. Above the dashed line a call needs more than a
thread.}
\label{fig:cpucmp}
\end{figure}

\textbf{CPU use} (Figure~\ref{fig:cpucmp}). In 50 ms frames the student takes \val{cpu/cmp/ours/50/core}\% of a core
per call, GTCRN \val{cpu/cmp/gtcrn/50/core}\% and VoxLock \val{cpu/cmp/voxlock/50/core}\%. In 8 bits it takes
\val{cpu/cmp/ours-int8/50/core}\% at no measurable cost on dev-4 (Appendix~\ref{app:tables}).

\begin{figure}[H]
\centering
\includegraphics[width=0.92\linewidth]{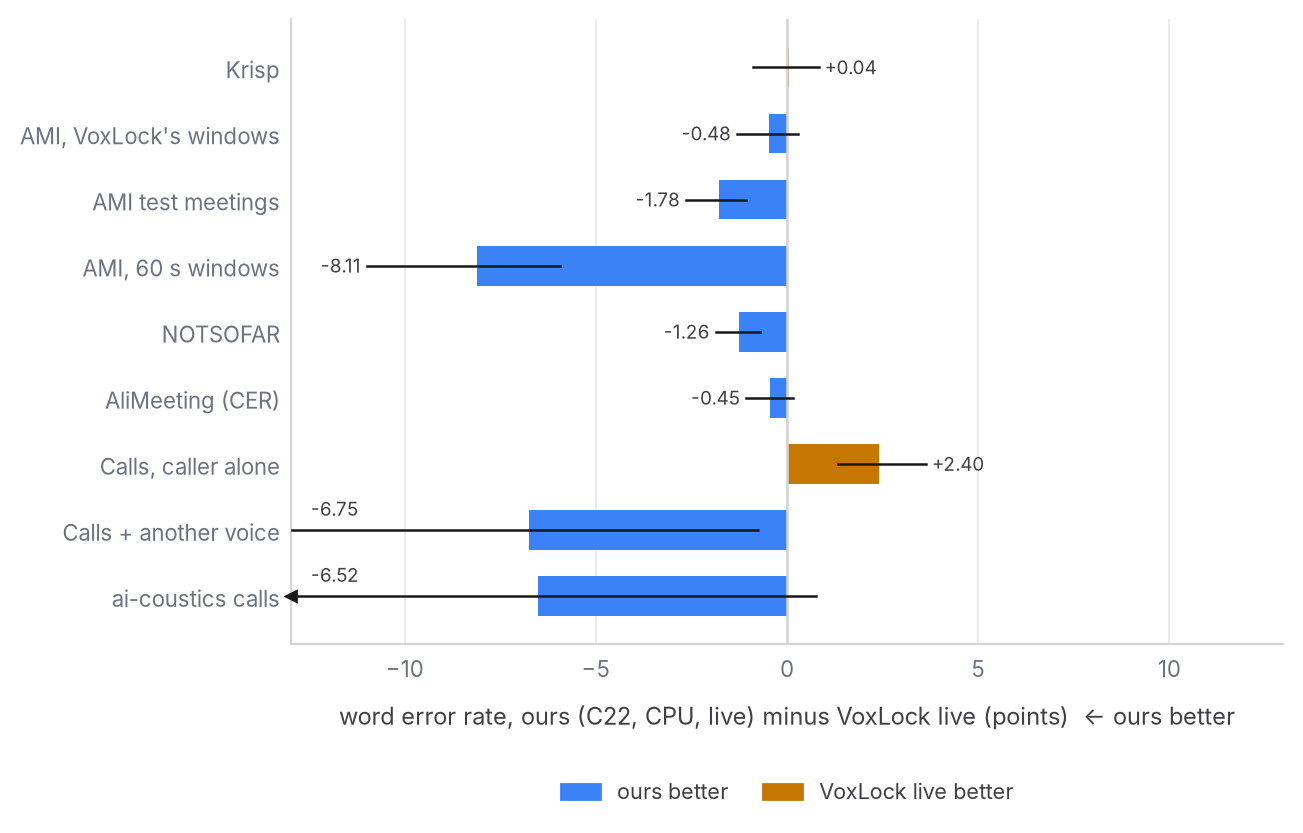}
\caption{Every test set: word error rate of the CPU student, live on one thread, minus VoxLock live, with 95\%
intervals over callers. Left of zero: the student makes fewer errors.}
\label{fig:cputests}
\end{figure}

\textbf{On the test sets} the student leads VoxLock live beyond the interval on four sets, beats the release on the 60
s windows, is level on Krisp and trails on the lone caller (Figure~\ref{fig:cputests}).

\section{In a running agent}
\label{sec:lkbench}

\textbf{Setup.} \val{lk/final/none/calls} callers stream the test calls with another voice into real LiveKit agents at
once, each running Silero VAD on the isolator's output; the reference is each caller's own channel with no filter.
Appendix~\ref{app:tables} has the agent's cost and the VAD's counts.

\textbf{The isolator keeps up.} Through the GPU server the caller arrives \val{lk/final/remote/delay_ms_p50} ms late
at \val{lk/final/remote/calls} calls (\val{lk/final64/remote/delay_ms_p50} ms at \val{lk/final64/remote/calls});
replies later than 40 ms go out unfiltered, \val{lk/final/remote/unfiltered_pct}\% of the audio. On the CPU the
student is \val{lk/final/local/delay_ms_p50} ms behind.

\begin{figure}[H]
\centering
\includegraphics[width=0.92\linewidth]{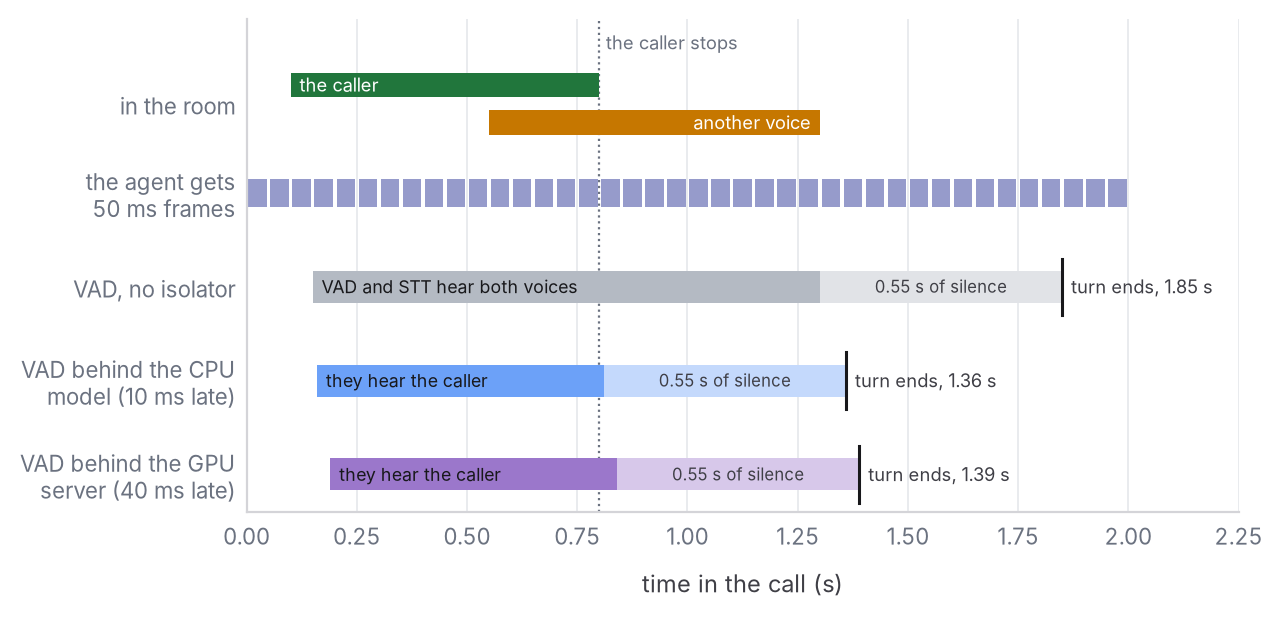}
\caption{One call through a running agent. The caller stops at 0.8 s while another voice goes on; LiveKit's VAD ends a
turn after 0.55 s of silence. On the raw frames it waits for the other voice too; behind the isolator it hears only
the caller, late by the isolator's delay (\val{lk/final/local/delay_ms_p50} ms on the CPU,
\val{lk/final/remote/delay_ms_p50} ms through the GPU server). A schematic with LiveKit's rule and the measured
delays.}
\label{fig:lkstream}
\end{figure}

\textbf{The VAD} behaves as on recordings (Figure~\ref{fig:lkstream}): behind the GPU and CPU models it hears speech
in \val{lk/final/remote/pauses_heard_pct}\% and \val{lk/final/local/pauses_heard_pct}\% of the caller's pauses,
against \val{lk/final/none/pauses_heard_pct}\% with no filter.

\textbf{Three fixes} came from real agents: the GPU server now steps again only within its tick's budget (64 calls
opening at once stay within \val{serve/start/final/p99-30s} ms); the plugin caps how long it waits for a late server
and lets later audio through unfiltered; and it enrolls from a clip of the caller when one is given, since a stream
that opens with silence enrolls on it.

\begin{figure}[H]
\centering
\includegraphics[width=0.92\linewidth]{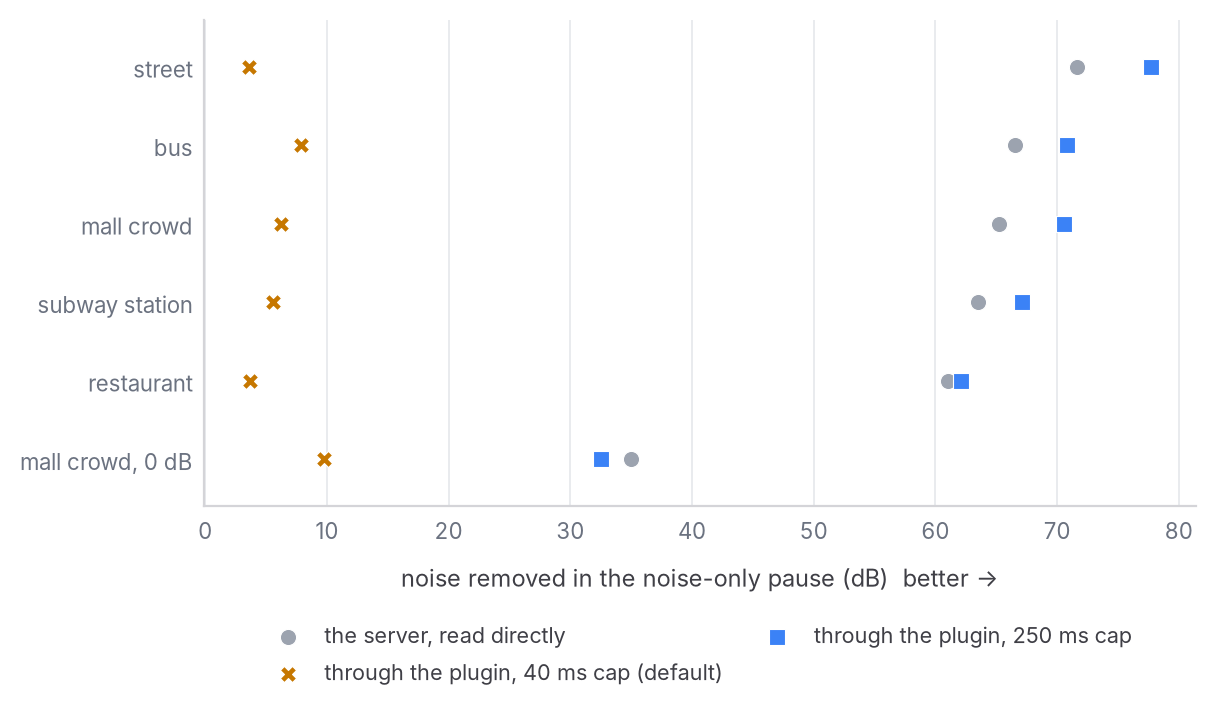}
\caption{The GPU card's calls with a noise-only pause, one call at a time: noise removed in that pause by the server
read directly and through the plugin with a 40 ms (default) and a 250 ms cap. Under the 40 ms cap late audio goes out
unfiltered, noise and all.}
\label{fig:lklone}
\end{figure}

\textbf{A lone call needs a longer cap} (Figure~\ref{fig:lklone}). With one call the server steps on a fixed 40 ms
clock, so most replies miss the plugin's 40 ms cap: through a real agent \val{lklone/livekit/unfiltered-pct}\% of the
audio went out unfiltered and \val{lklone/livekit/others} of other people's words reached the recogniser, against
\val{lklone/direct/others} read from the server directly. With a 250 ms cap the output matched the direct read, at
\val{lklone/livekit_lag/delay-p50} ms of delay.

\section{Interruptions}
\label{sec:interrupt}

\begin{figure}[H]
\centering
\includegraphics[width=\linewidth]{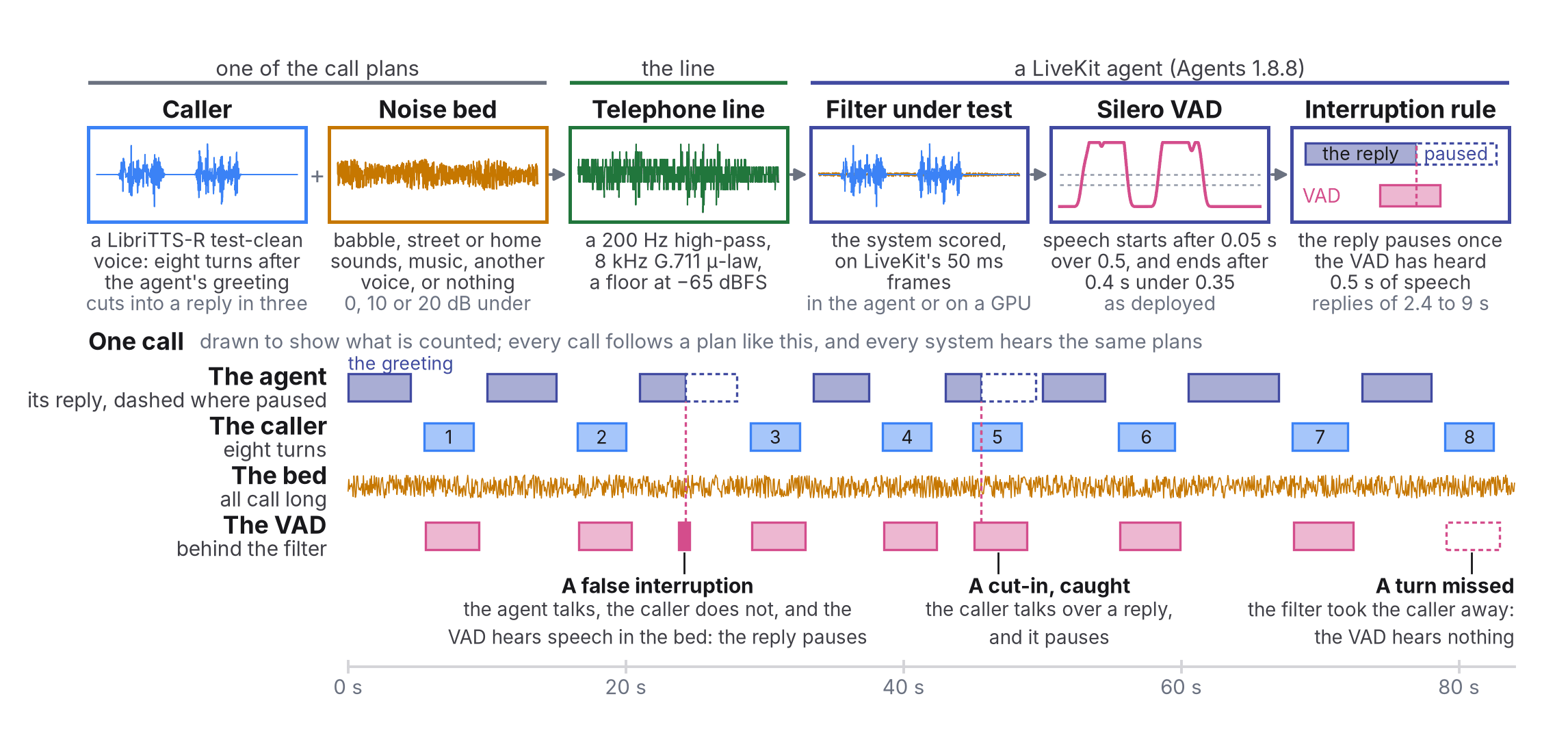}
\caption{The interruption bench. Top: what a call is made of. Callers are LibriTTS-R test-clean speakers
\citep{koizumi2023librittsr}. Beds: babble, street or home sounds from CochlScene's test split
\citep{jeong2022cochlscene} (home with ESC-50 sounds \citep{piczak2015esc} on top), music from MUSAN
\citep{snyder2015musan}, or another voice, FLEURS sentences \citep{conneau2023fleurs} played through a room like a
television. The VAD runs at a deployed agent's settings. Bottom: one call, drawn, with the three things the bench
counts.}
\label{fig:vintbench}
\end{figure}

\textbf{The bench.} An agent that talks can fail one more way: noise stops it mid-sentence. We count how often in
LiveKit agents that hold whole conversations with simulated phone callers (Figure~\ref{fig:vintbench}). Five noise beds
at three levels, plus a clean line, make sixteen conditions; there are \val{vint/plans} call plans, three per
condition, and every system hears the same plans.

\textbf{Systems,} each as it ships for LiveKit: no filter; GTCRN \citep{rong2024gtcrn}, STT-API's noise suppressor, in
the agent; VoxLock's live mode in the agent, full precision on three threads; NVIDIA's Real-time RE-USE
\citep{fu2026rethinking,fu2026one}, a universal speech enhancer with no notion of the caller, on a GPU server (exit
layer 8; research licence only); Inworld Voice Isolator v1 \citep{inworld2026voiceisolator}, its realtime model on a
GPU (offline only); our CPU student in the agent, 8 bits; our release on a GPU server; and the release fine-tuned for
agents (below). Both GPU servers step late calls again within their tick (Section~\ref{sec:lkbench}).

\begin{figure}[H]
\centering
\includegraphics[width=0.92\linewidth]{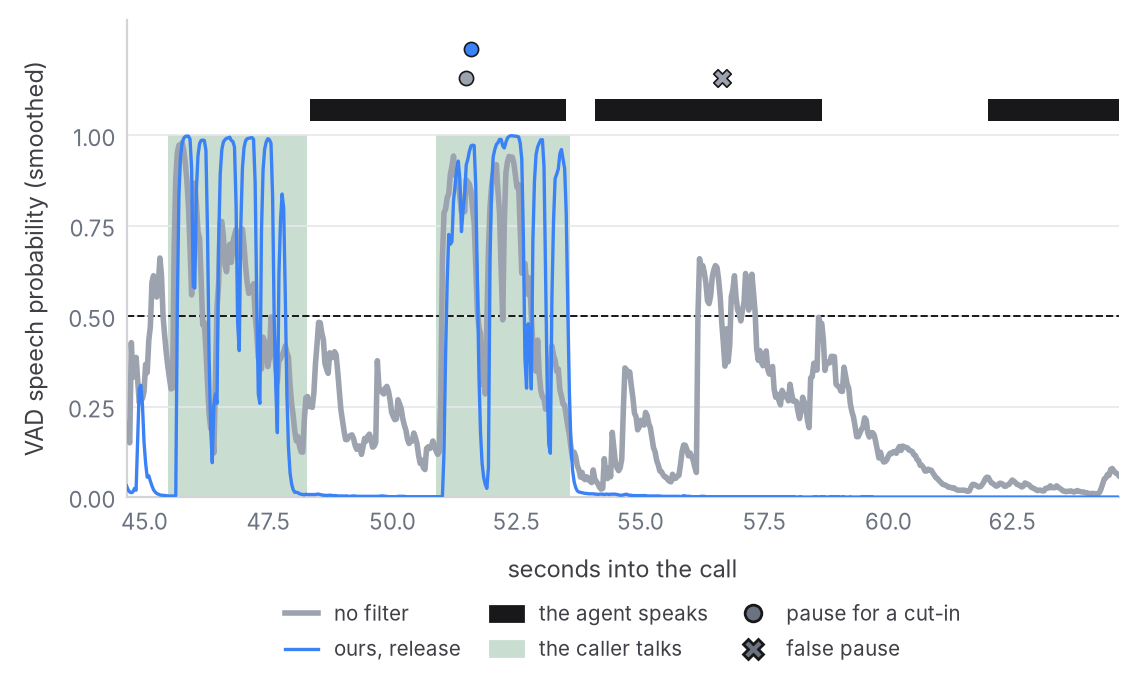}
\caption{\val{vint/trace/after/len} s of a held-out call over babble \val{vint/trace/after/snr} dB under the caller.
Curves: the VAD's smoothed speech probability; dashed: its threshold. Above the agent's speech, each filter's pauses.
Both filters pause for the caller's cut-in; with no filter, babble alone paused the next reply. The call is the
session's first that shows all three.}
\label{fig:vinttrace}
\end{figure}

\textbf{What we count.} LiveKit pauses the agent's reply once its VAD has heard half a second of speech, hangover
included, so about a tenth of a second of noise is enough. A pause is false when the caller did not talk in the 2 s
before it (Figure~\ref{fig:vinttrace}). We report false pauses per minute of agent speech.

\begin{figure}[H]
\centering
\includegraphics[width=0.92\linewidth]{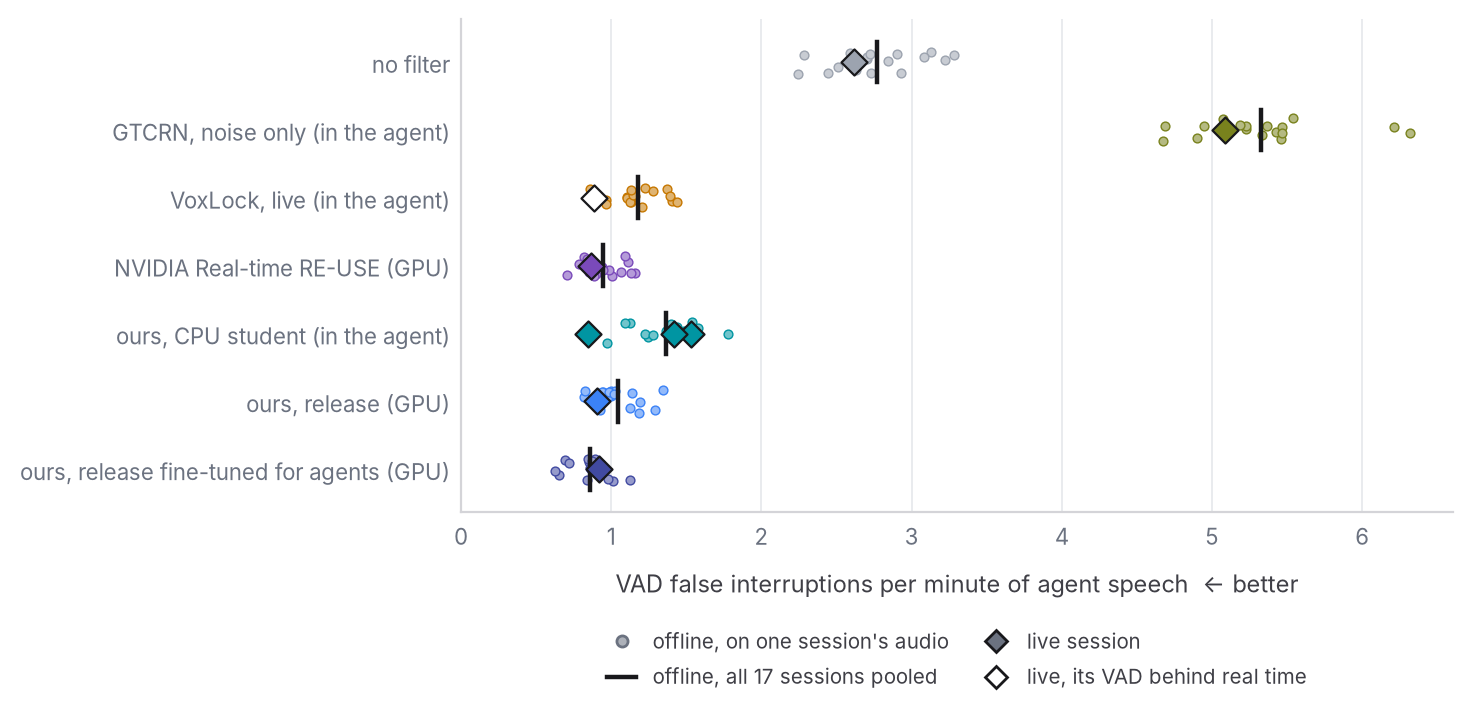}
\caption{False interruptions per minute of agent speech: each system offline on each held-out session's audio (dots)
and on all \val{vint/sessions} together (black lines), and live (diamonds; the CPU student ran live three times).
VoxLock's live diamond is hollow: its VAD fell behind real time, so that count does not compare.}
\label{fig:vintsessions}
\end{figure}

\textbf{Every session's audio.} Live runs of one system vary more than systems differ (Figure~\ref{fig:vintsessions}),
so every system also runs offline on the audio each held-out session's agent received (\val{vint/sessions} sessions,
\val{vint/calls} calls), scored by the bench's replay of LiveKit's VAD and interruption rule.

\begin{figure}[H]
\centering
\includegraphics[width=0.92\linewidth]{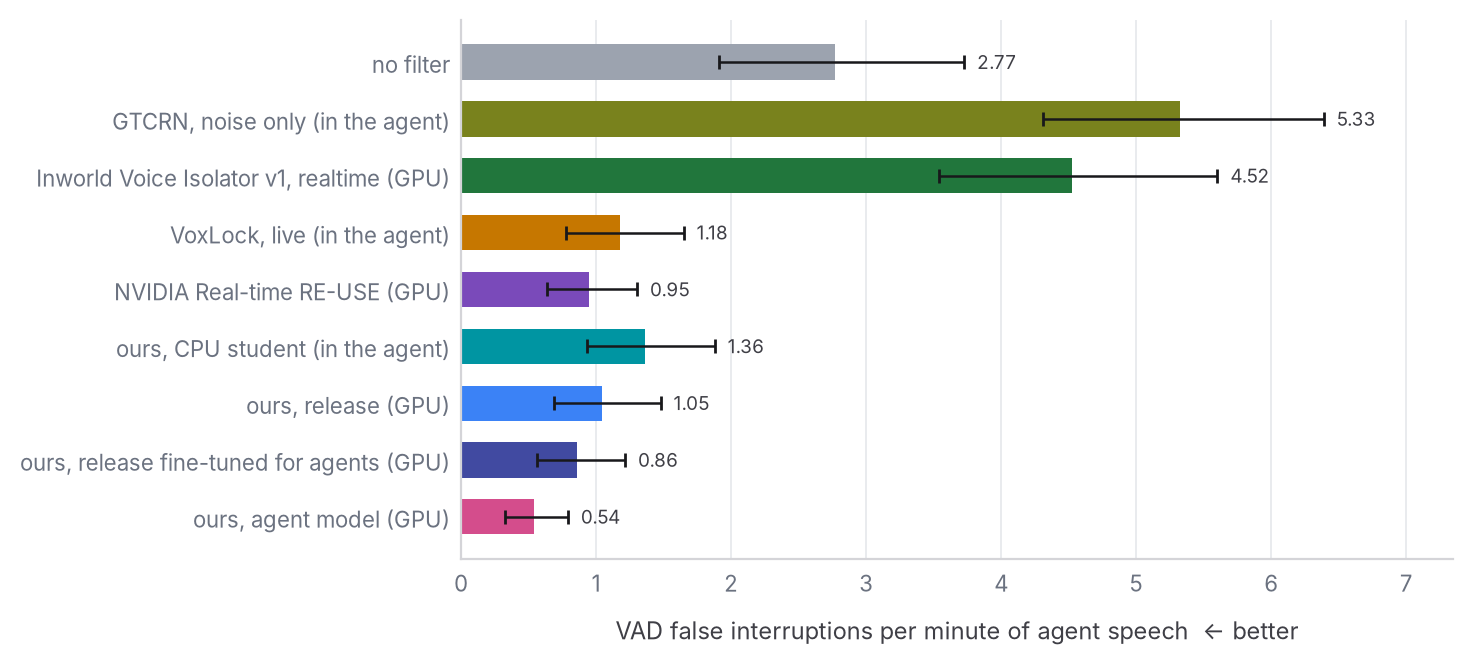}
\caption{False interruptions per minute of agent speech (the VAD paused the reply although the caller had not talked
in the 2 s before), every system on the audio every held-out session's agent received (\val{vint/calls} calls). Bars:
90\% intervals over resampled call plans. The last bar is the agent model (Section~\ref{sec:gate}).}
\label{fig:vint}
\end{figure}

\textbf{False interruptions} (Figure~\ref{fig:vint}, Table~\ref{tab:vint}). With no filter, noise paused the agent
\val{vint/none/fi} times a minute of its speech. Noise removal raises that: GTCRN to \val{vint/gtcrn/fi}, Inworld to
\val{vint/iw/inworld/all}. Every isolator lowers it: the release to \val{vint/rel/fi}, RE-USE to \val{vint/reuse/fi},
VoxLock to \val{vint/voxlock/fi} and our CPU student to \val{vint/cpu/fi}; the agent model of Section~\ref{sec:gate}
reaches \val{vint/gate/agent/all}.

\begin{table}[H]
\centering
\caption{The interruption bench, every system on the audio every held-out session's agent received. False starts:
VAD speech starts while the caller is silent, per minute of that silence. Cut-ins: the caller's interruptions the
agent paused for within 2.5 s. Turns missed: caller turns the VAD never heard; held: turns whose end it did not see
before the caller spoke again. Caller heard: share of the caller's speech the VAD covered. Whisper: large-v3 on the
caller's turns after the filter (two sessions). CPU: share of one core per call in the agent, over all of the filter's
threads; GPU systems compute on their server.}
\label{tab:vint}
\begin{adjustbox}{max width=\linewidth}
\begin{tabular}{lrrrrrrrr}
\toprule
 & false interruptions & false starts & cut-ins & turns & turns & caller & Whisper & CPU \% \\
System & per agent-min (90\% int.) & per silent min & caught \% & missed \% & held \% & heard \% & WER \% & per call \\
\midrule
No filter & 2.77 (1.91 to 3.73) & 3.04 & 83.7 & 0.4 & 9.3 & 94.4 & 7.4 & 0 \\
GTCRN, noise only, in the agent & 5.33 (4.31 to 6.39) & 5.37 & 81.8 & 0.1 & 8.6 & 95.4 & 11.0 & 6 \\
VoxLock, live, in the agent & 1.18 (0.78 to 1.65) & 1.34 & 82.5 & 6.8 & 12.7 & 82.8 & 17.0 & 220 \\
NVIDIA Real-time RE-USE, GPU server & 0.95 (0.63 to 1.30) & 1.01 & 87.3 & 2.6 & 5.8 & 87.2 & 16.9 & GPU \\
Ours, CPU student, in the agent & 1.36 (0.93 to 1.88) & 1.58 & 90.9 & 0.3 & 3.4 & 93.4 & 11.8 & 18 \\
Ours, release, GPU server & 1.05 (0.69 to 1.48) & 1.23 & 89.9 & 0.7 & 3.5 & 90.9 & 13.4 & GPU \\
Ours, release fine-tuned for agents & 0.86 (0.56 to 1.21) & 1.03 & 89.7 & 0.6 & 3.3 & 91.1 & 13.0 & GPU \\
\bottomrule
\end{tabular}
\end{adjustbox}

\end{table}

\begin{figure}[H]
\centering
\includegraphics[width=0.92\linewidth]{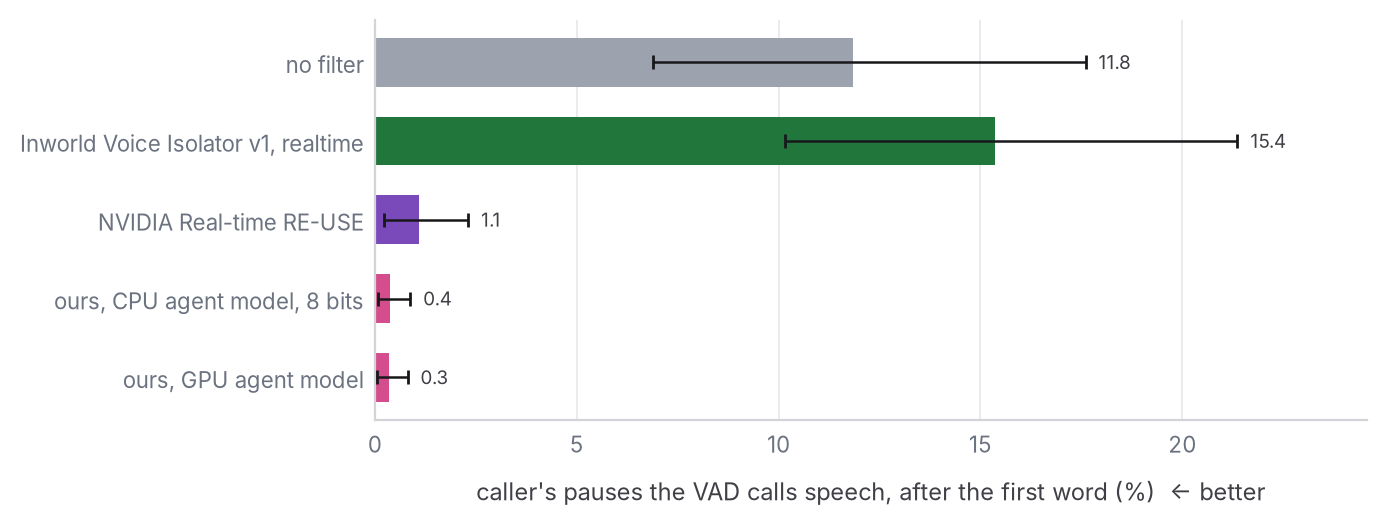}
\caption{The caller's pauses while the agent speaks, after the caller's first word: share of the VAD's 32 ms windows
called speech, every held-out session's audio, with 90\% intervals over resampled call plans.}
\label{fig:iwvad}
\end{figure}

\textbf{Noise removal makes the VAD's job harder} (Figure~\ref{fig:iwvad}). Inworld turns the caller's pauses down by
a median of only \val{vint/iw/silence/inworld/after/down} dB, and the VAD calls
\val{vint/iw/silence/inworld/after/vad}\% of them speech, against \val{vint/iw/silence/none/after/vad}\% on the raw
line. The filters that do well send the pauses out as digital silence.

\textbf{The caller.} Behind VoxLock the VAD missed \val{vint/voxlock/missed}\% of the caller's turns, against
\val{vint/rel/missed}\% behind the release, because VoxLock can lock onto music while the caller waits for the
greeting. Whisper's error on the caller's turns is \val{vint/none/wer}\% with no filter, \val{vint/rel/wer}\% after
the release and \val{vint/reuse/wer}\% after RE-USE.

\begin{figure}[H]
\centering
\includegraphics[width=0.92\linewidth]{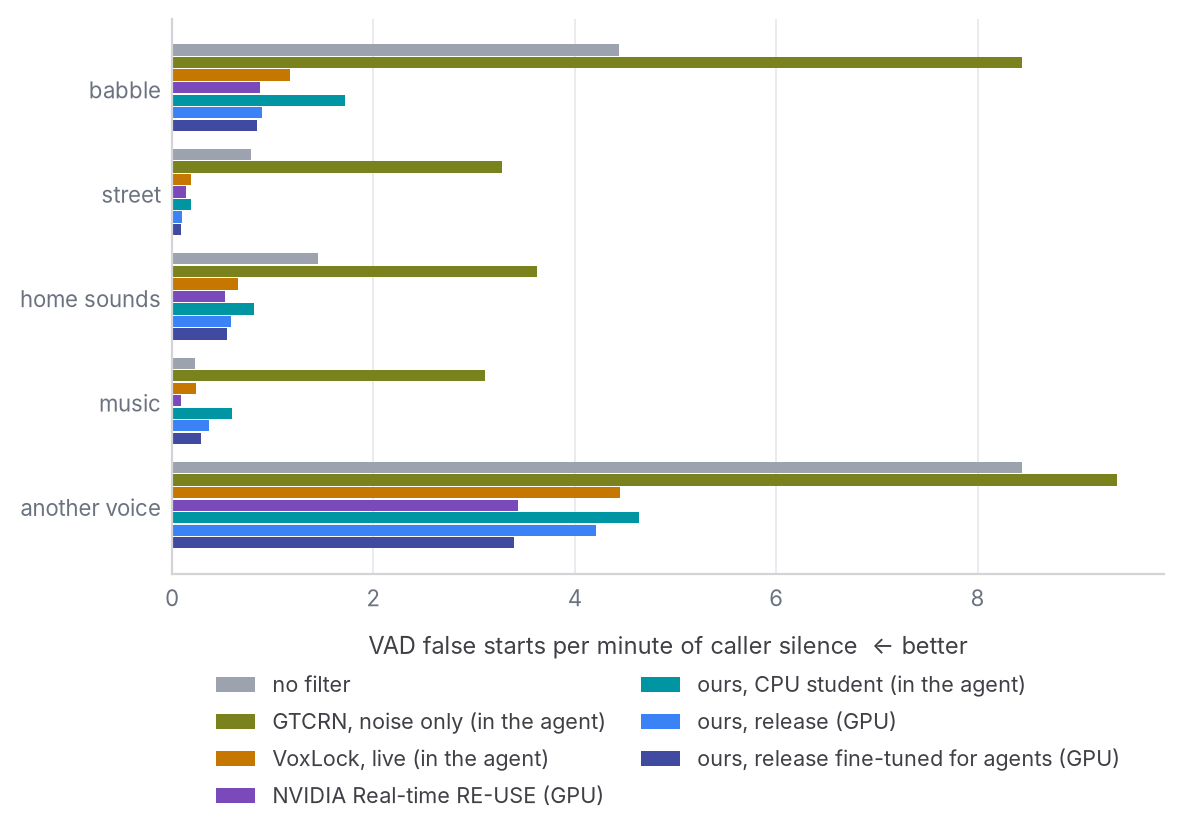}
\caption{VAD false starts per minute of the caller's silence, by the noise under the call (0, 10 and 20 dB under the
caller together), every system on every held-out session's audio.}
\label{fig:vintkinds}
\end{figure}

\textbf{By kind of noise} (Figure~\ref{fig:vintkinds}), another voice is the hardest case for every isolator, since its
sentences are speech. RE-USE leads on music: it removes music, while our models keep sung music that the VAD hears as
speech, slightly more often than with no filter.

\begin{figure}[H]
\centering
\includegraphics[width=0.92\linewidth]{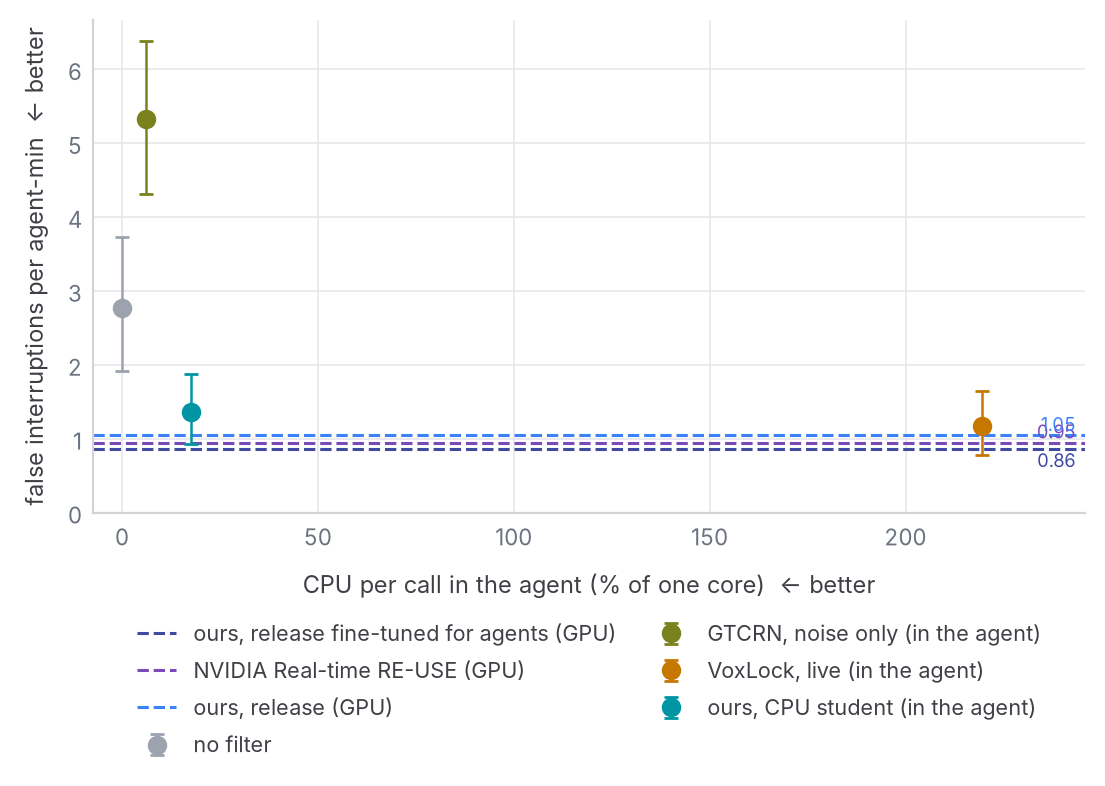}
\caption{False interruptions against each filter's CPU cost per call in the agent. Points: filters in the agent (no
filter at zero). Dashed: GPU systems, whose agent only sends and receives audio.}
\label{fig:vintcpu}
\end{figure}

\textbf{CPU} (Figure~\ref{fig:vintcpu}). In the agent GTCRN costs \val{vint/gtcrn/core-pct}\% of a core per call, our
CPU student \val{vint/cpu/core-pct}\% and VoxLock \val{vint/voxlock/core-pct}\%, more than a core.

\begin{figure}[H]
\centering
\includegraphics[width=0.92\linewidth]{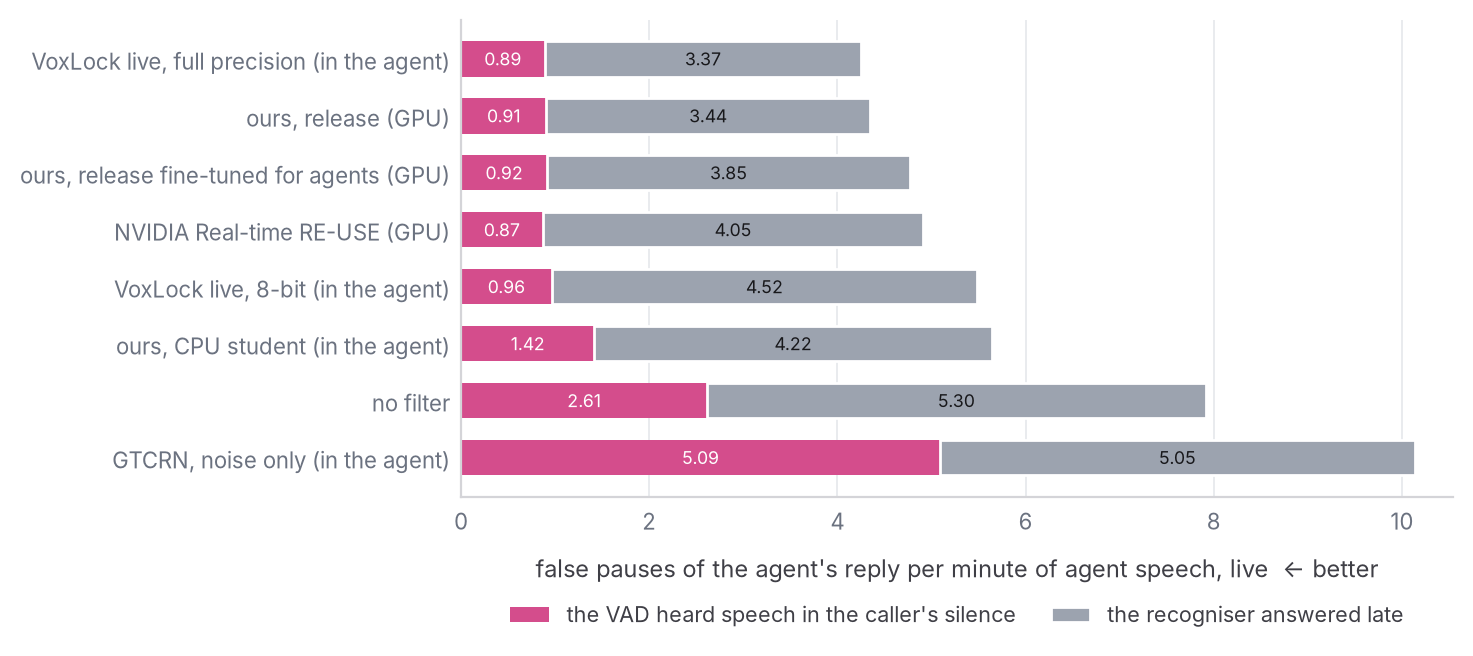}
\caption{What falsely paused the agent in each system's live session, per minute of agent speech: the VAD hearing
speech while the caller was silent, or the recogniser's late answer for the caller's last words. Sorted by the sum,
fewest at the top. VoxLock at full precision fell behind real time (Figure~\ref{fig:vintrealtime}), so its count does
not compare.}
\label{fig:vintcauses}
\end{figure}

\begin{figure}[H]
\centering
\includegraphics[width=0.92\linewidth]{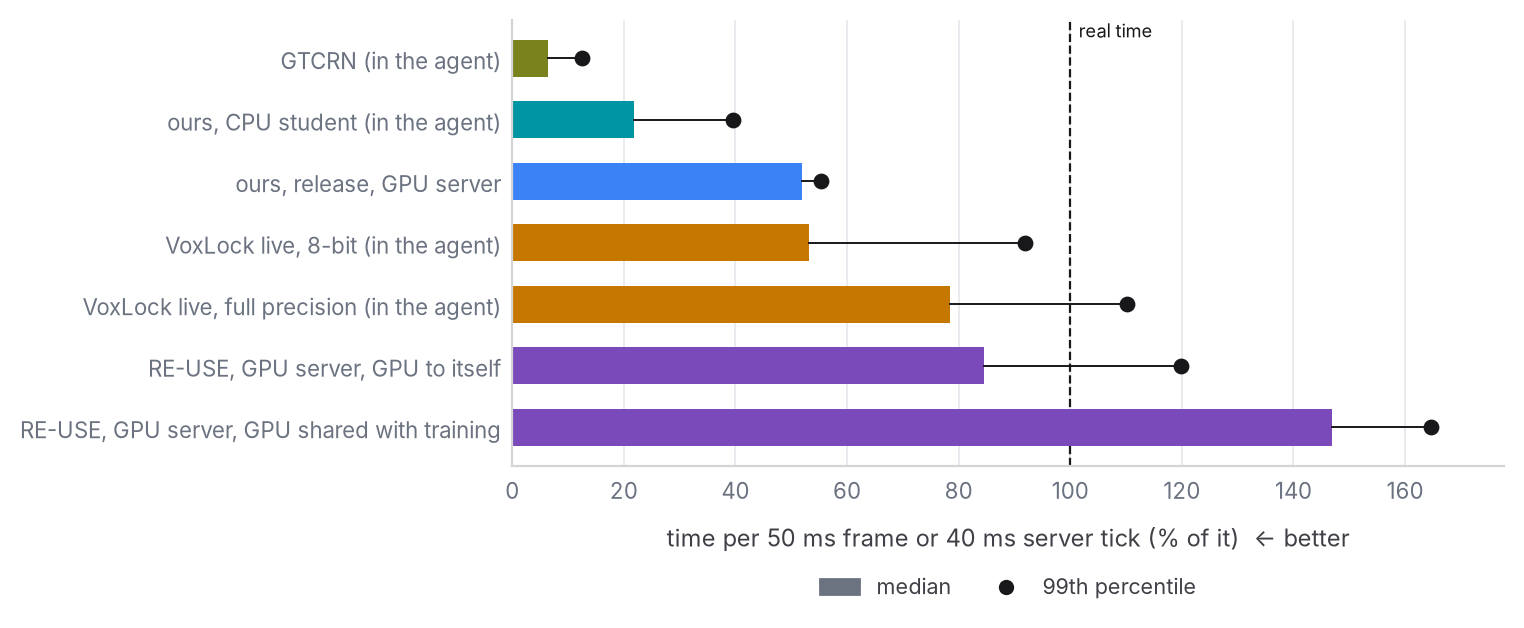}
\caption{Whether each filter kept up live: its time per 50 ms frame in the agent, or a GPU server's per 40 ms tick, as
a share of that audio, sorted by the median. Past the dashed line it falls behind. RE-USE with its GPU shared: its
server's log at \val{vint/reuse_shared/server/calls} calls, outside the live sessions.}
\label{fig:vintrealtime}
\end{figure}

\textbf{Live,} counts fall within the spread between sessions (Figure~\ref{fig:vintsessions}). The recogniser paused
the live agent more often than noise, behind every filter, when its answer came after the turn had ended
(Figure~\ref{fig:vintcauses}). VoxLock fell behind real time, and RE-USE's live row is a session with its GPU to
itself (Figure~\ref{fig:vintrealtime}; Appendix~\ref{app:tables}).

\begin{figure}[H]
\centering
\includegraphics[width=\linewidth]{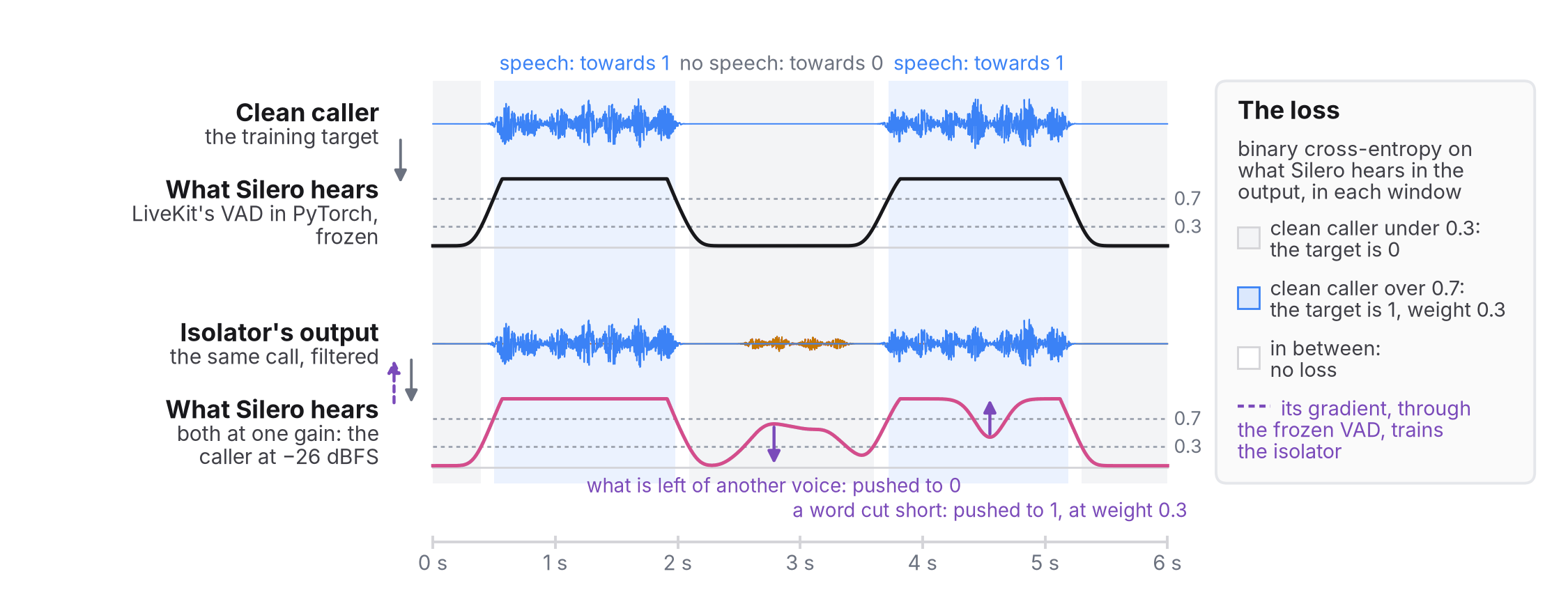}
\caption{The fine-tune's VAD loss over six seconds of a training call, drawn. What the frozen VAD hears in the clean
caller sets the target for what it hears in the output. Both signals are first scaled so the caller's speech sits at
-26 dBFS, so a quiet call is judged as a normal one.}
\label{fig:vintvadloss}
\end{figure}

\begin{figure}[H]
\centering
\includegraphics[width=0.92\linewidth]{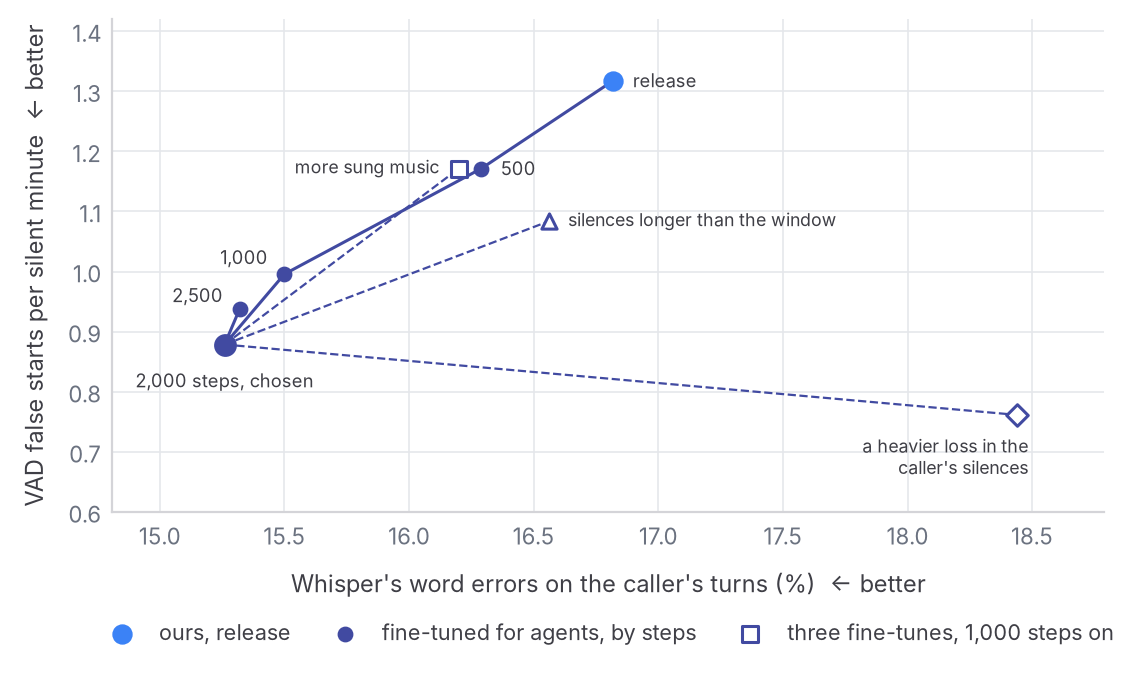}
\caption{The bench's development calls, where the fine-tune was chosen: VAD false starts per minute of the caller's
silence against Whisper's errors on the caller's turns, behind the release, its fine-tune at each saved step, and three
fine-tunes trained 1,000 steps on from the chosen one (dashed; see What is left).}
\label{fig:vintftdev}
\end{figure}

\textbf{Fine-tuned for agents.} We fine-tune the release for 2,000 steps on 24 s calls with turn-taking, telephone
lines, GTZAN music \citep{tzanetakis2002gtzan} and enrollment from the noisy line, adding a loss through LiveKit's
frozen VAD (Figure~\ref{fig:vintvadloss}): the VAD's speech probability on the output is pushed towards what it hears
in the clean caller. Chosen on the bench's development calls, the fine-tune cuts VAD false starts from
\val{vint/dev/rel/fs} to \val{vint/dev/ft/fs} per minute of the caller's silence (Figure~\ref{fig:vintftdev}).

\begin{figure}[H]
\centering
\includegraphics[width=0.92\linewidth]{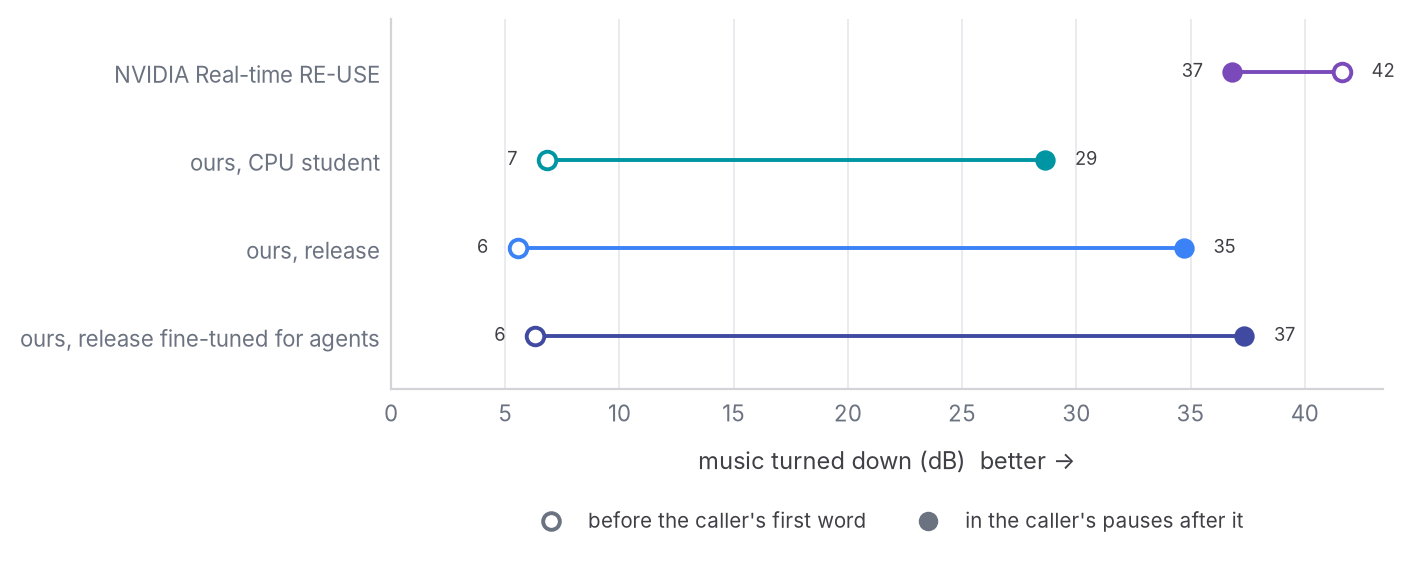}
\caption{Music 20 dB under the caller, bench development material, no agent: how far each filter turns it down before
the caller's first word (hollow) and in the caller's pauses after it (filled), over the
\val{vint/probe/lead/release/music20/n} calls where the caller first says nothing for 6 to 24 s.}
\label{fig:vintmusic}
\end{figure}

\begin{figure}[H]
\centering
\includegraphics[width=0.92\linewidth]{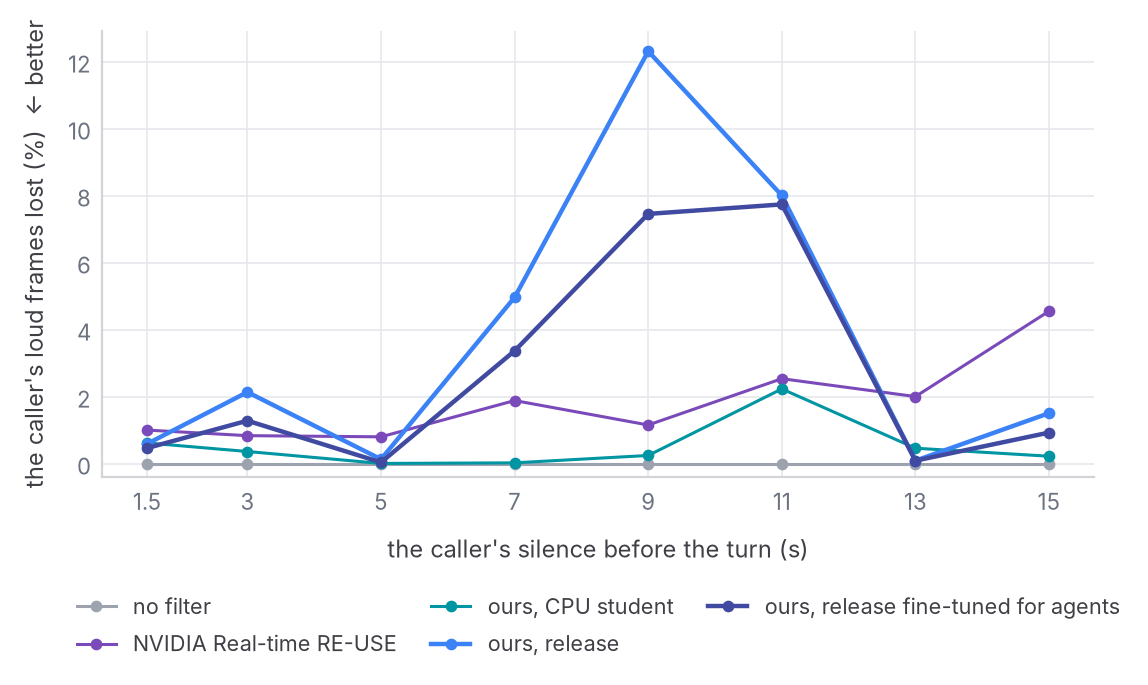}
\caption{A returning caller, bench development material, no agent, clean lines: share of the caller's loud frames lost
in a turn, by the silence before it.}
\label{fig:vintreturn}
\end{figure}

\textbf{What is left.} Two probes isolate it: music before the caller's first word, which the release turns down by
only \val{vint/probe/lead/release/music20/down} dB against RE-USE's \val{vint/probe/lead/reuse/music20/down}
(Figure~\ref{fig:vintmusic}), and a caller returning after 7 s of silence, who loses
\val{vint/probe/ret/release/long/lost}\% of their loud frames (Figure~\ref{fig:vintreturn}). Three more fine-tunes
fixed neither without a cost elsewhere (Figure~\ref{fig:vintftdev}).

\begin{figure}[H]
\centering
\includegraphics[width=0.92\linewidth]{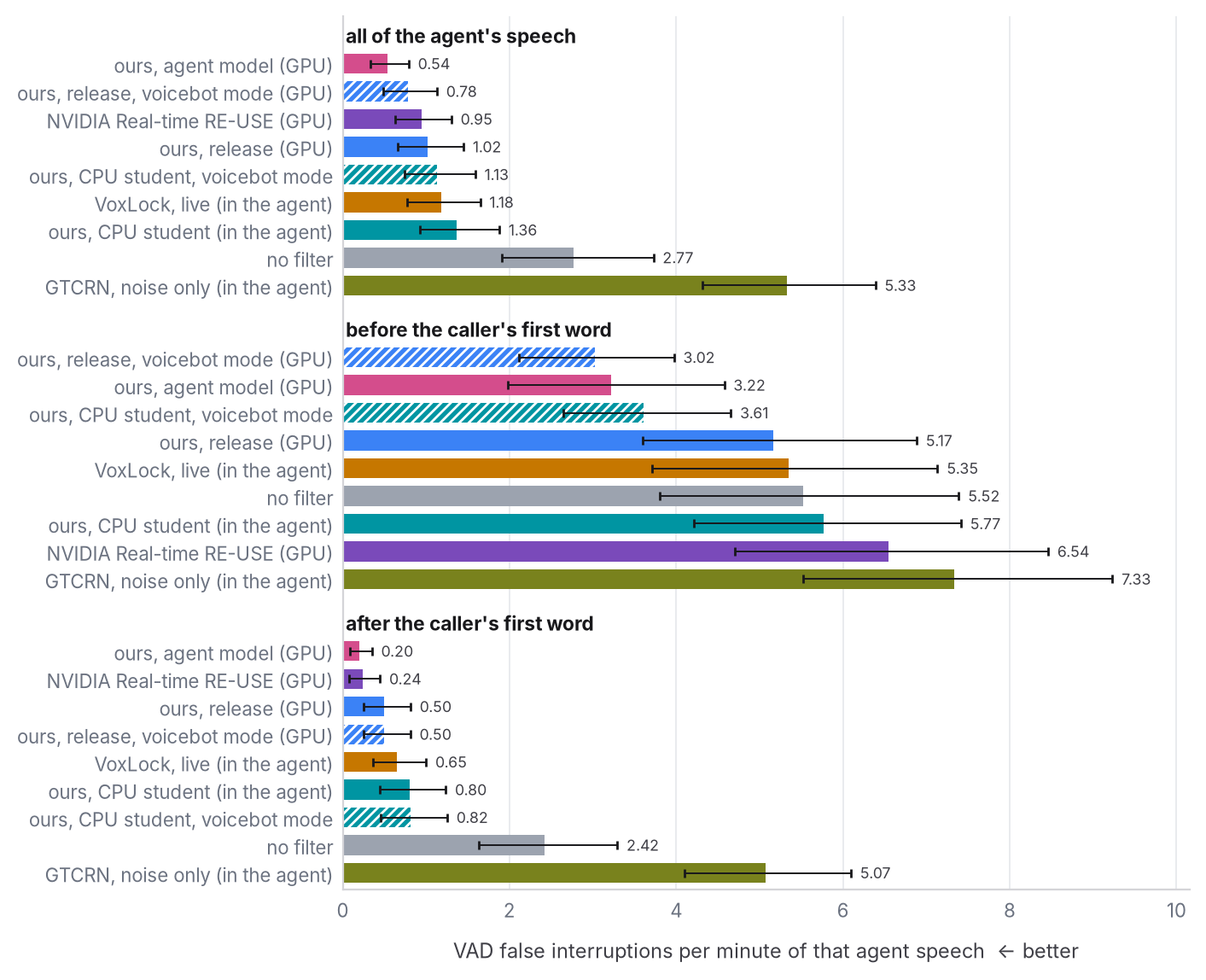}
\caption{False interruptions per minute of the agent's speech: all of it, before the caller's first word (the greeting,
\val{vint/greet/lead-pct}\% of its speech) and after it, every system on every held-out session's audio. Ours as
shipped, in the default and voicebot (hatched) modes, and the agent model with its presence gate
(Section~\ref{sec:gate}). Groups sorted, fewest at the top; 90\% intervals over resampled call plans.}
\label{fig:vintgreet}
\end{figure}

\textbf{The greeting holds most false interruptions.} It is \val{vint/greet/lead-pct}\% of the agent's speech but
\val{vint/greet/iso/share-min} to \val{vint/greet/iso/share-max}\% of the false interruptions behind every isolator
(Figure~\ref{fig:vintgreet}). Before the caller speaks, the loudest 3 s is the television, the music or the other
person, and at a call's start the model lets a voice through regardless.

\begin{figure}[H]
\centering
\includegraphics[width=0.92\linewidth]{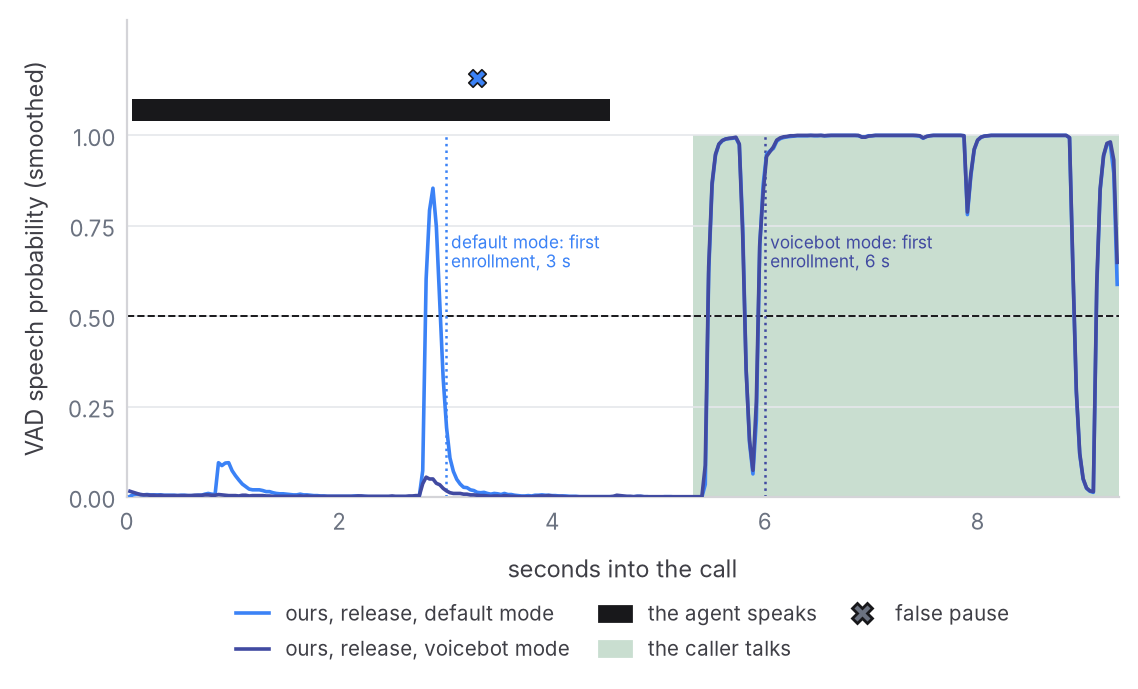}
\caption{The start of a held-out call over babble \val{vint/trace/greet/snr} dB under the caller, behind our release.
By default babble passed before the first enrollment (\val{vint/trace/greet/rel/switched} s) and the VAD paused the
greeting; in the voicebot mode nothing passed until the caller spoke, and enrollment came at
\val{vint/trace/greet/rel_vb/switched} s. The call is the session's first with a false pause in the greeting by
default and none in the voicebot mode.}
\label{fig:vintvbtrace}
\end{figure}

\textbf{A voicebot mode.} An agent that speaks first sets \texttt{voicebot=True} on the plugin, or
\texttt{?voicebot=1} or \texttt{-{}-voicebot} on the servers. Until the caller is enrolled the model keeps a decoy, 4
s of a training speaker in no call, so it keeps no voice instead of every voice; the first enrollment waits until the
loudest 3 s is 6 dB above the call's first 3 s, or 10 s (Figure~\ref{fig:vintvbtrace}). Held out, the release falls
from \val{vint/ship/rel/fi} to \val{vint/ship/rel_vb/fi} false interruptions a minute, below RE-USE's
\val{vint/ship/reuse/fi}, and the 8-bit CPU student from \val{vint/ship/cpu8/fi} to \val{vint/ship/cpu8_vb/fi}; turns
missed and Whisper do not move. (The servers enroll a step later than our offline runs, hence \val{vint/ship/rel/fi}
rather than \val{vint/rel/fi}.)

\begin{figure}[H]
\centering
\includegraphics[width=0.92\linewidth]{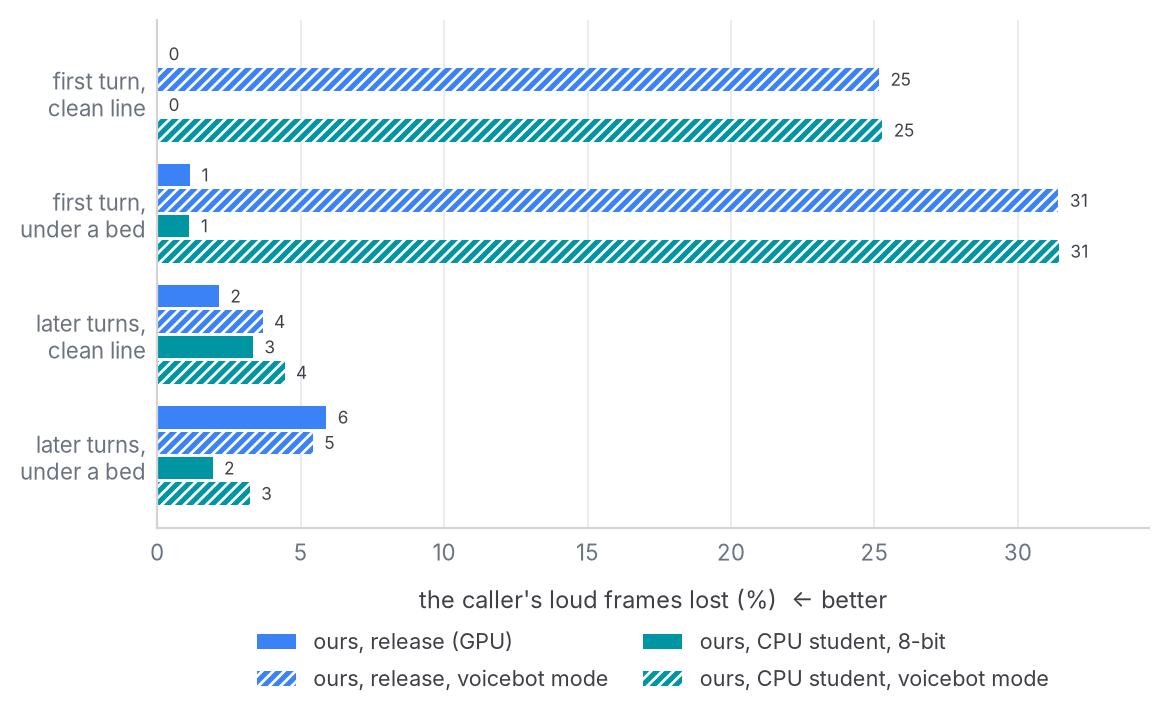}
\caption{What the voicebot mode (hatched) costs a caller who talks first, bench development material, no agent: loud
frames lost in a first turn starting 0.3 to 1 s in, and in later turns, on clean lines
(\val{vint/ship/probe/clean/n} calls) and under beds (\val{vint/ship/probe/beds/n}).}
\label{fig:vintvbcost}
\end{figure}

\textbf{Its cost} falls on a caller who talks first: one who speaks 0.3 to 1 s into the call loses
\val{vint/ship/probe/clean/release_vb/first}\% of their loud frames in that turn, against
\val{vint/ship/probe/clean/release/first}\% by default (Figure~\ref{fig:vintvbcost}). The mode is opt-in.

\begin{figure}[H]
\centering
\includegraphics[width=0.92\linewidth]{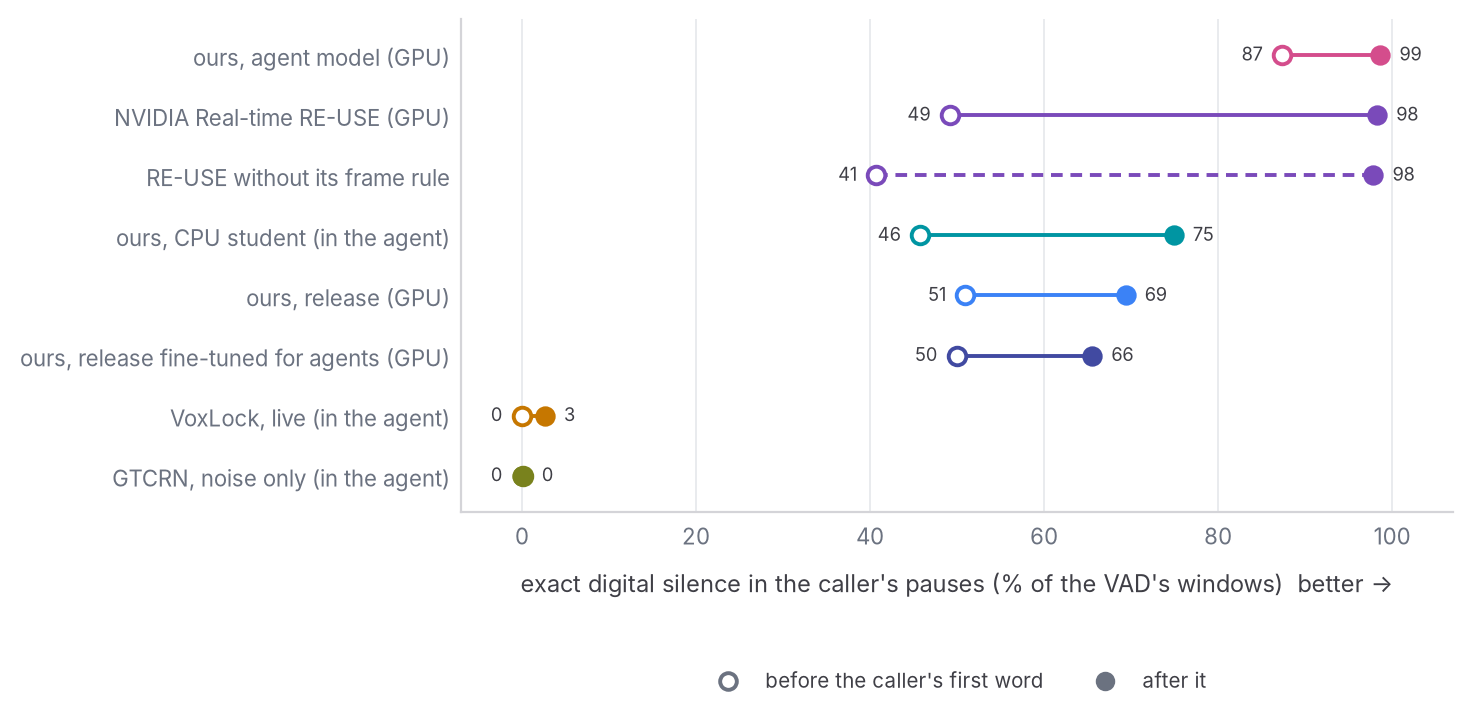}
\caption{Share of the VAD's 32 ms windows in the caller's pauses that are exact digital silence, before the caller's
first word (hollow) and after it (filled), on the \val{vint/silence/bed-calls} held-out calls with a noise bed (no
filter omitted: a line under a bed is never silent). Dashed: RE-USE without its released frame rule (zero a frame when
more than half its bins are zero). The agent model is Section~\ref{sec:gate}'s. Sorted by the share after the first
word.}
\label{fig:vintsilence}
\end{figure}

\textbf{After the first word RE-USE leads:} \val{vint/greet/reuse/after} false interruptions a minute against
\val{vint/greet/rel_ship/after} behind the release (Figure~\ref{fig:vintgreet}). Behind RE-USE
\val{vint/silence/reuse/after/silence}\% of the VAD's windows in the caller's pauses are digital silence, behind the
release \val{vint/silence/rel/after/silence}\% (Figure~\ref{fig:vintsilence}); in digital silence the VAD has nothing
to hear.

\textbf{Why silence.} RE-USE predicts magnitudes and clips negatives to zero, so a noise bin leaves as nothing; our
masks turn down by degrees. Without its released frame rule, RE-USE's after-word rate barely moves
(\val{vint/norule/reuse_norule/after} against \val{vint/norule/reuse/after}): the clip inside the network does the
work.

\begin{figure}[H]
\centering
\includegraphics[width=0.92\linewidth]{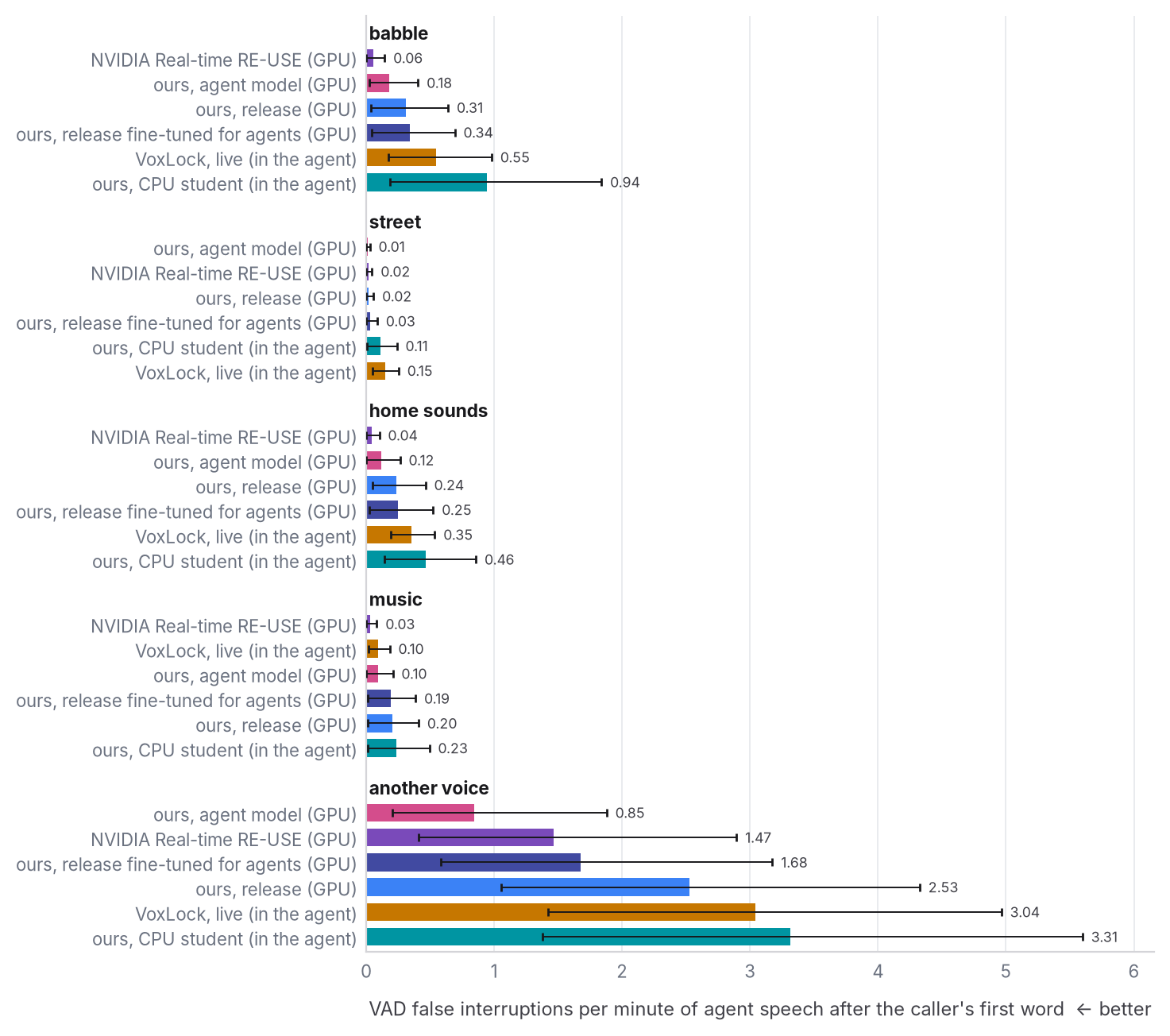}
\caption{False interruptions per minute of the agent's speech after the caller's first word, by the noise under the
call, every isolator on every held-out session's audio. Groups sorted, fewest at the top. No filter and GTCRN are
omitted: with another voice they reach \val{vint/silence/none/kind/talker/after} and
\val{vint/silence/gtcrn/kind/talker/after}. No system paused the agent on a clean line after the first word. The agent
model is Section~\ref{sec:gate}'s. Bars: 90\% intervals over resampled call plans.}
\label{fig:vintafter}
\end{figure}

\textbf{By the noise} (Figure~\ref{fig:vintafter}), another voice is most of what remains for every isolator:
\val{vint/silence/reuse/kind/talker/after} false interruptions a minute behind RE-USE and
\val{vint/silence/rel/kind/talker/after} behind the release, out of \val{vint/silence/none/kind/talker/after} with no
filter. The release had turned those passages down by a median of only \val{vint/silence/rel/trig/down-p50} dB.

\section{Ahead of RE-USE}
\label{sec:gate}

\textbf{The gap.} RE-USE sends digital silence into the caller's pauses, and our models pass a voice as loud as the
caller whole (Section~\ref{sec:interrupt}). A presence gate, trained with the isolator and set on more development
calls, closes the gap. The result is the agent model, released beside the release.

\begin{figure}[H]
\centering
\includegraphics[width=\linewidth]{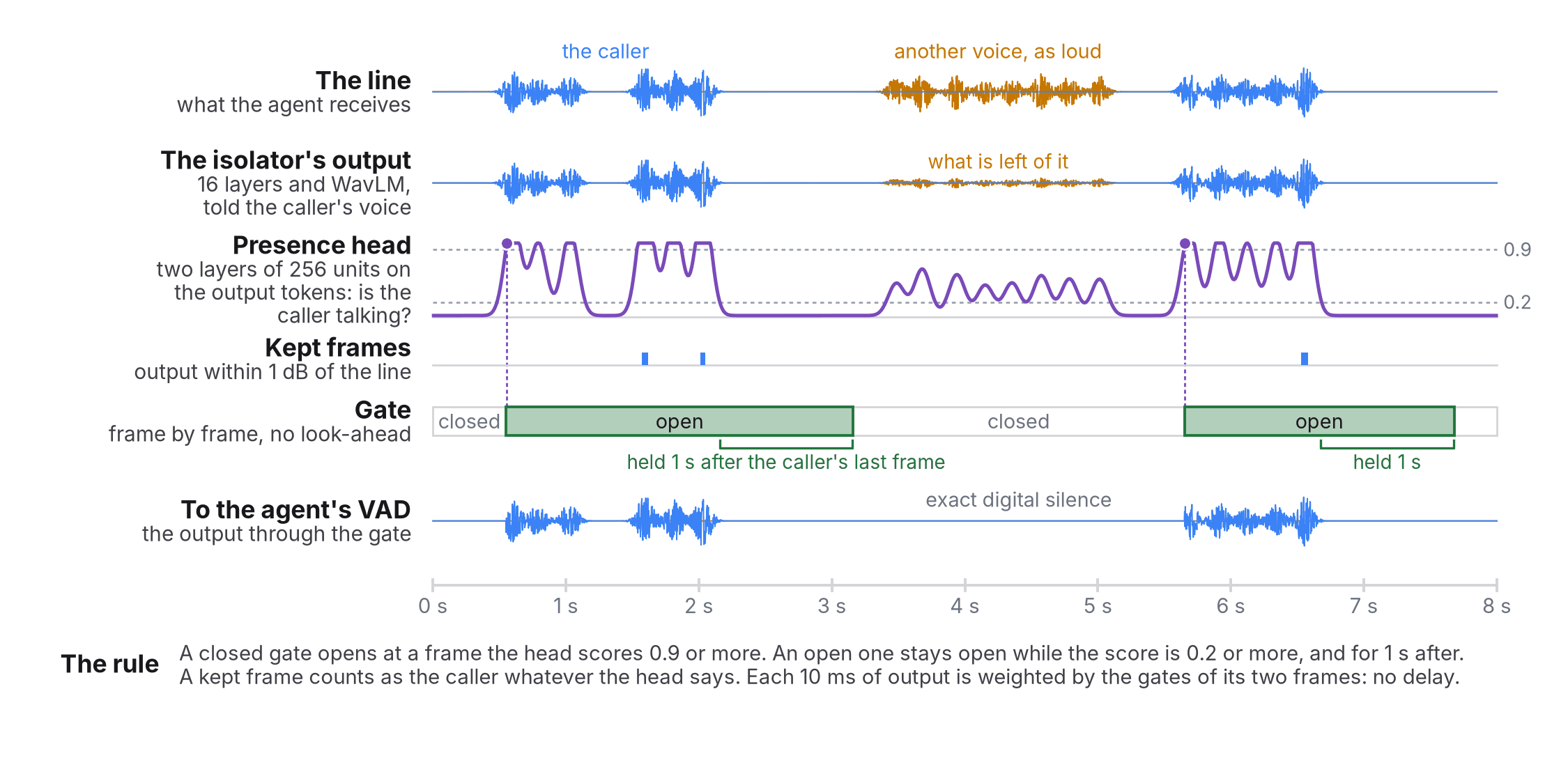}
\caption{The presence gate over eight seconds of a call. The head's score is drawn; the gate is that score through the
agent model's rule: open from \val{vint/agent/theta}, held while the score is \val{vint/agent/theta-off} or more and for
\val{vint/agent/hold-s} s after. Frames kept within \val{vint/agent/keep-db} dB of the line count as the caller; closed
runs leave as exact digital silence.}
\label{fig:gate}
\end{figure}

\textbf{A presence gate} (Figure~\ref{fig:gate}). A small head on the isolator's output tokens scores each 20 ms frame
for the caller. The gate opens at \val{vint/agent/theta}, stays open while the score is \val{vint/agent/theta-off} or
more and for \val{vint/agent/hold-s} s after, and passes any frame the isolator kept within \val{vint/agent/keep-db}
dB of the line. Closed runs leave as digital silence, with no added delay.

\begin{figure}[H]
\centering
\includegraphics[width=0.88\linewidth]{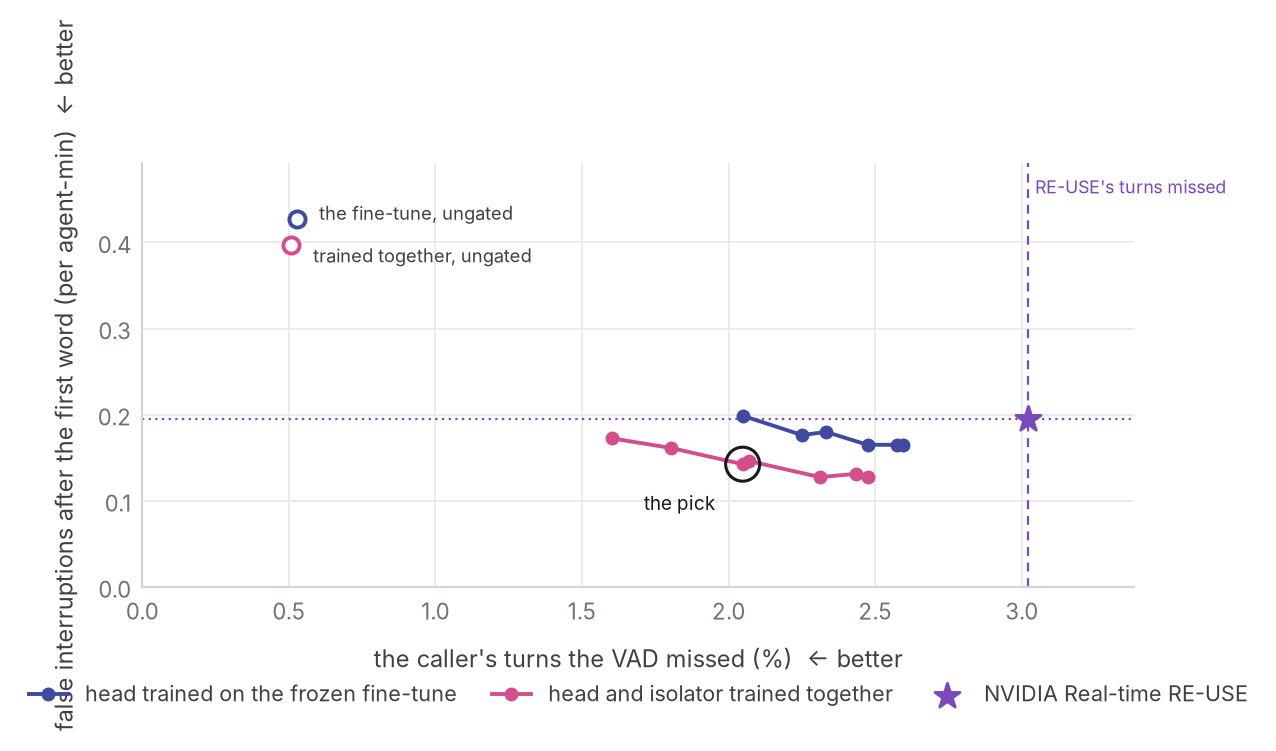}
\caption{Development calls (\val{vint/gate/dev/select/calls}): false interruptions after the first word against the
caller's turns missed, each model over its gate settings (hollow: ungated). Star: RE-USE, dashed line its turns
missed. Ringed: the rule's pick. The joint model's curve lies under the frozen head's.}
\label{fig:gatedev}
\end{figure}

\textbf{Trained together} (Figure~\ref{fig:gatedev}). A head on the frozen fine-tune lowered the after-word rate but
made the VAD miss \val{vint/gate/dev/select/vbg2kg_t90_h100_o10_k1/missed}\% of the caller's turns instead of
\val{vint/gate/dev/select/vbg2k/missed}\%. Training isolator and head together for \val{vint/agent/steps} more steps
lowers the whole curve.

\textbf{Chosen on more calls.} On \val{vint/gate/dev/select/calls} development calls, a rule fixed first takes the
lowest after-word rate among settings that keep the caller as well as RE-USE. It picks the joint model at the setting
above: \val{vint/gate/dev/select/vbj15g_t90_h100_o20_k1/after} false interruptions a minute after the first word,
against RE-USE's \val{vint/gate/dev/select/reuse8/after}.

\begin{figure}[H]
\centering
\includegraphics[width=0.88\linewidth]{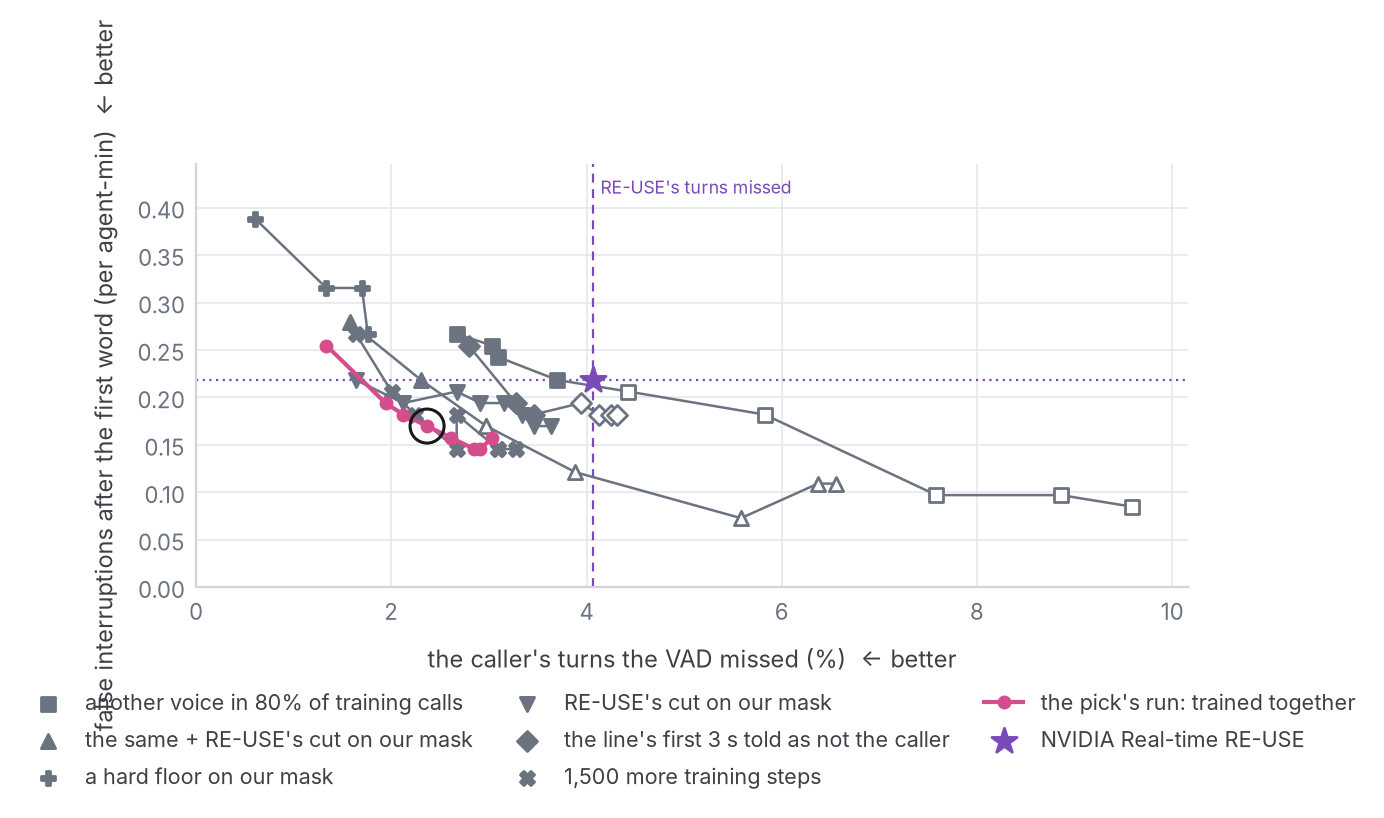}
\caption{The other recipes on \val{vint/gate/dev/screen/calls} development calls, each over its gate settings, against
the chosen run (pick ringed) and RE-USE (star). Hollow: the agent stops for fewer of the caller's cut-ins than RE-USE
less 2 points.}
\label{fig:gateother}
\end{figure}

\textbf{What did not help} (Figure~\ref{fig:gateother}). Training with another voice in \val{vint/agent/tv-p}\% of
calls lowered the rate by losing the caller. RE-USE's clip on our mask, a mask floor, telling the model the line's
first 3 s and more steps did not lower the curve. A gate aware of the agent's speech delayed the slowest cut-ins.

\begin{figure}[H]
\centering
\includegraphics[width=0.92\linewidth]{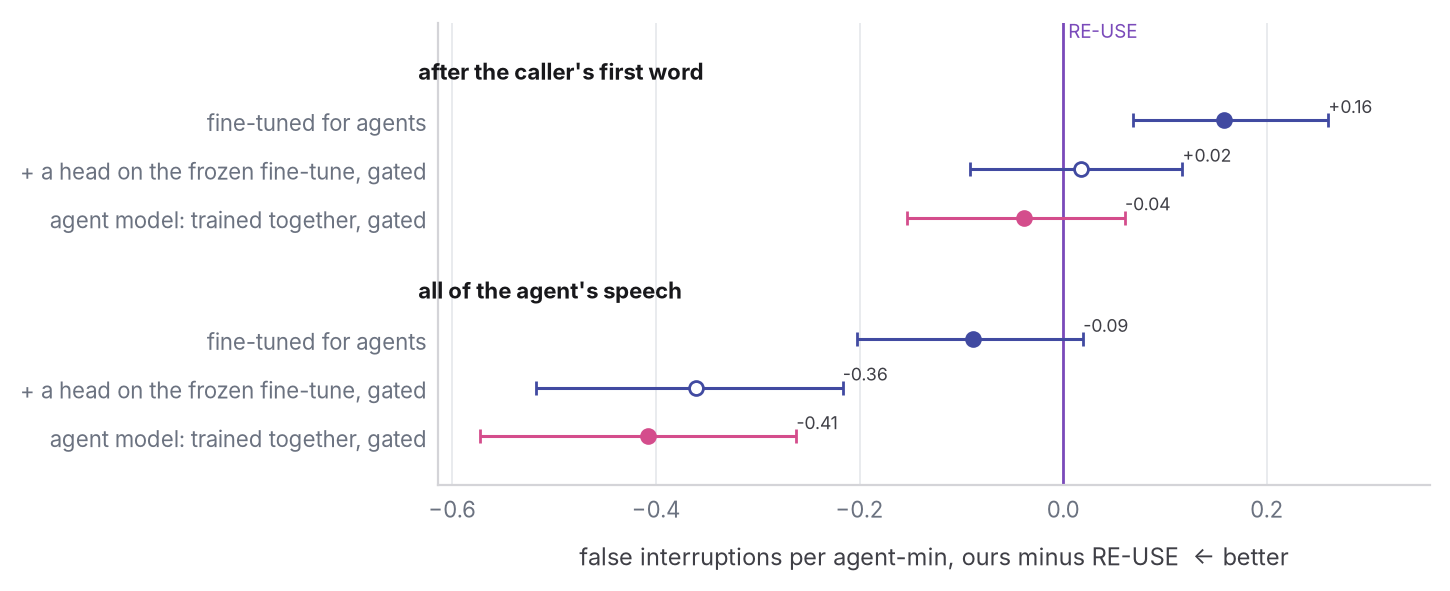}
\caption{Held out (\val{vint/gate/calls} calls, \val{vint/gate/sessions} sessions' audio): false interruptions per
minute of the agent's speech, each of ours minus RE-USE, after the caller's first word and over all of it, with 90\%
intervals over resampled call plans. Left of zero: fewer than RE-USE.}
\label{fig:gatetest}
\end{figure}

\textbf{Held out, ahead of RE-USE} (Figure~\ref{fig:gatetest}). On \val{vint/gate/calls} held-out calls, scored once,
the agent model paused the agent \val{vint/gate/agent/all} times a minute of its speech and RE-USE
\val{vint/gate/reuse/all}, a difference of \val{vint/gate/diff/agent-reuse/all} (90\% interval
\val{vint/gate/diff/agent-reuse/all-lo} to \val{vint/gate/diff/agent-reuse/all-hi}). After the first word it is lower,
\val{vint/gate/agent/after} against \val{vint/gate/reuse/after}, but the interval includes zero.

\begin{figure}[H]
\centering
\includegraphics[width=0.92\linewidth]{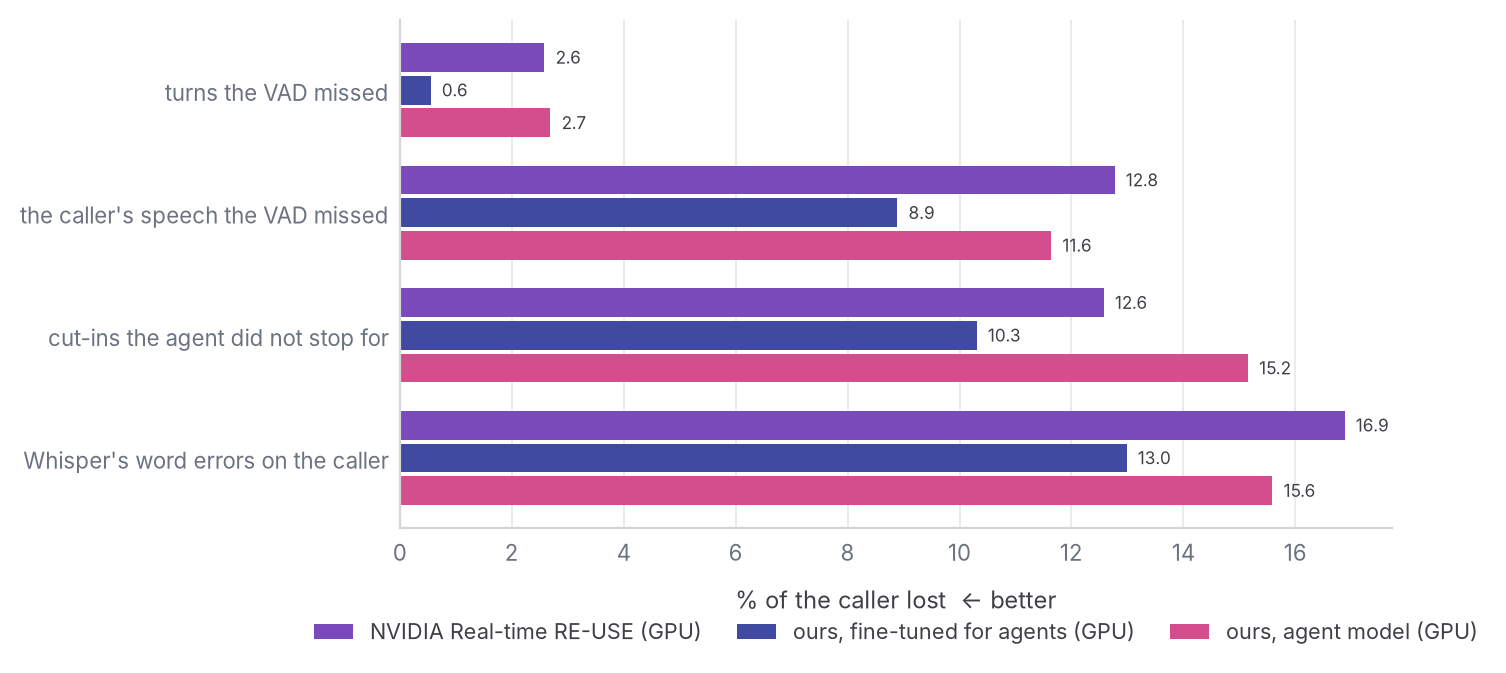}
\caption{What each system costs the caller, held out, every measure as a share lost: turns the VAD missed, speech it
missed, cut-ins the agent did not stop for, and Whisper's word errors on the caller's turns.}
\label{fig:gatecaller}
\end{figure}

\textbf{The caller pays about what RE-USE costs} (Figure~\ref{fig:gatecaller}): \val{vint/gate/silence/agent/missed}\%
of turns missed (RE-USE \val{vint/gate/silence/reuse/missed}\%) and \val{vint/gate/silence/agent/wer}\% Whisper errors
(\val{vint/gate/silence/reuse/wer}\%). RE-USE stays ahead only on cut-ins caught, \val{vint/gate/test/reuse/cut}\%
against \val{vint/gate/test/agent/cut}\%.

\textbf{By the noise} (Figure~\ref{fig:vintafter}), the agent model wins with another voice on the line,
\val{vint/gate/silence/agent/kind/talker/after} against RE-USE's \val{vint/gate/silence/reuse/kind/talker/after} after
the first word, and trails on babble, home sounds and music, which our development calls barely hold.

\begin{figure}[H]
\centering
\includegraphics[width=0.92\linewidth]{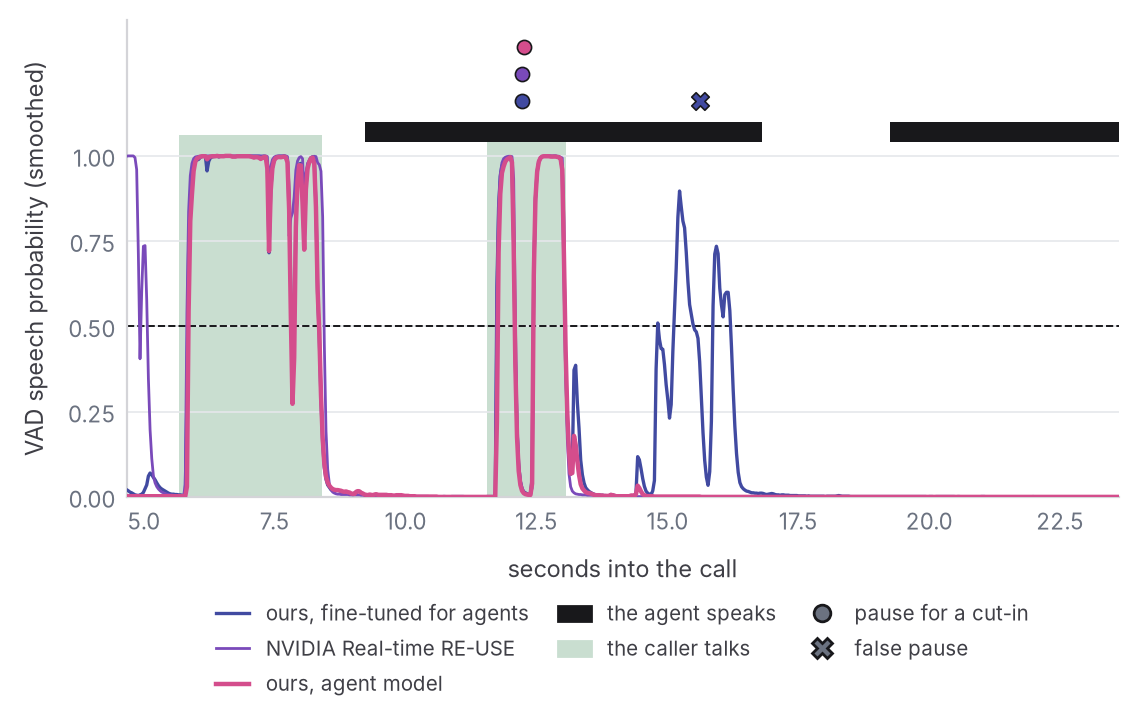}
\caption{One held-out call behind the fine-tune, RE-USE and the agent model: the VAD's smoothed speech probability, the
caller's speech (green), the agent's speech (dark strip) and each system's pauses of the reply (circle: for a cut-in;
cross: false).}
\label{fig:gatetrace}
\end{figure}

\textbf{One call} (Figure~\ref{fig:gatetrace}). In a held-out call picked by rule, the fine-tune lets the VAD pause
the agent after the caller's first word; the agent model does not, and still stops for the caller's cut-in.

\textbf{Serving.} The gate runs inside the GPU server's batched step and matched the offline decision in every 10 ms
of \val{vint/gate/engine/calls} development calls. The agent model is in the release's repository (folder
\texttt{agent}), and the servers read its gate from its configuration.

\textbf{On a CPU.} The recipe carries to the CPU student of Section~\ref{sec:cpu}: \val{vint/cpuagent/steps} more steps with a presence head
and the GPU agent model as teacher, the gate applied in numpy by the plugin, held a shorter \val{vint/cpuagent/hold-s}
s. It was released before Whisper scored it on development calls, where it ended just outside the rule's guard
(\val{vint/cpugate/dev/c22j30g_t90_h30_o20_k1/wer}\% against RE-USE's \val{vint/cpugate/dev/reuse8/wer}\%).

\begin{figure}[H]
\centering
\includegraphics[width=0.92\linewidth]{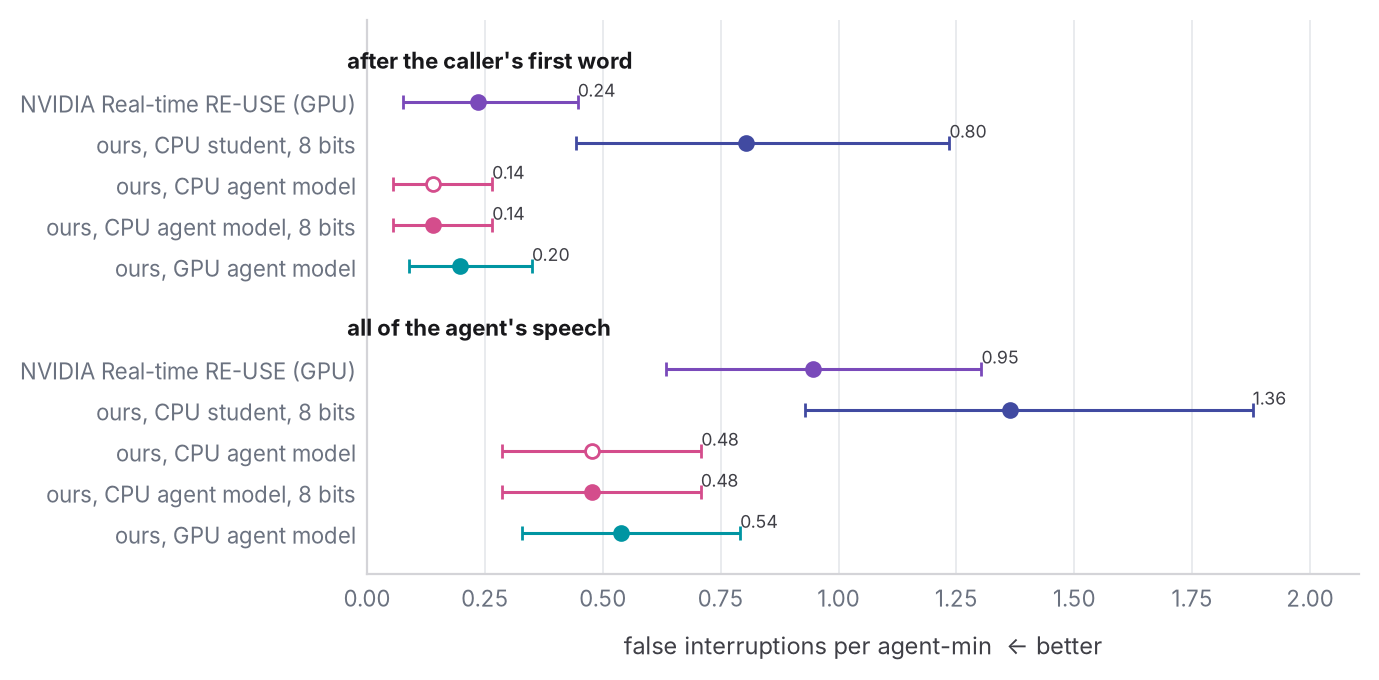}
\caption{Held out (\val{vint/cpugate/calls} calls), through the LiveKit plugin: false interruptions per minute of the
agent's speech after the caller's first word and over all of it, with 90\% intervals over resampled call plans. The CPU
agent model in float32 (hollow) and 8 bits, the 8-bit CPU student, RE-USE, and the GPU agent model as served.}
\label{fig:gatecpu}
\end{figure}

\textbf{The CPU agent model, held out} (Figure~\ref{fig:gatecpu}), in 8 bits through the plugin, paused the agent
\val{vint/cpugate/cpuag8/all} times a minute against RE-USE's \val{vint/cpugate/reuse/all} and the ungated student's
\val{vint/cpugate/cpu8/all}, missing \val{vint/cpugate/silence/cpuag8/missed}\% of turns (RE-USE
\val{vint/cpugate/silence/reuse/missed}\%) at \val{vint/cpugate/core/cpuag8}\% of a core. These are the first released
weights; the CPU release's \texttt{agent} folder now holds their phone-line fine-tune (Section~\ref{sec:phone}).

\section{The clip and the greeting}
\label{sec:clip}

\textbf{Serving only.} This section changes how the agent models are served, not their weights. A rule fixed first
chooses on \val{vint/r4/dev/calls} development calls: the fewest false interruptions among settings that keep the
caller as well as RE-USE, replacing the served setting only if \val{vint/r4/replace-all} lower.

\begin{figure}[H]
\centering
\includegraphics[width=0.92\linewidth]{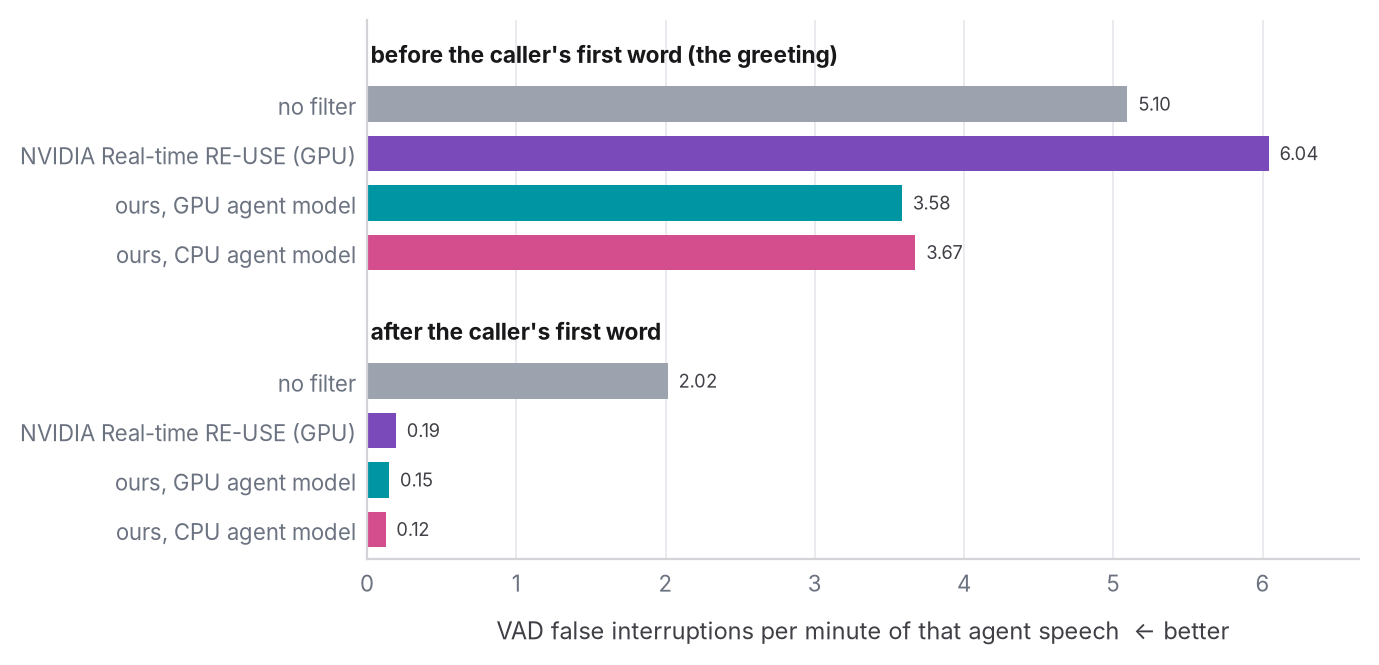}
\caption{The \val{vint/r4/dev/calls} development calls: false interruptions per minute of the agent's speech before and
after the caller's first word, behind no filter, RE-USE and the two agent models as first served.}
\label{fig:r4where}
\end{figure}

\textbf{The rest is in the greeting} (Figure~\ref{fig:r4where}). It is \val{vint/r4/where/agent/greet-minutes}\% of
the agent's speech and \val{vint/r4/where/agent/greet-share}\% of the GPU agent model's false interruptions on
development calls, since before the caller speaks the live rule enrolls the loudest thing on the line.

\begin{figure}[H]
\centering
\includegraphics[width=0.92\linewidth]{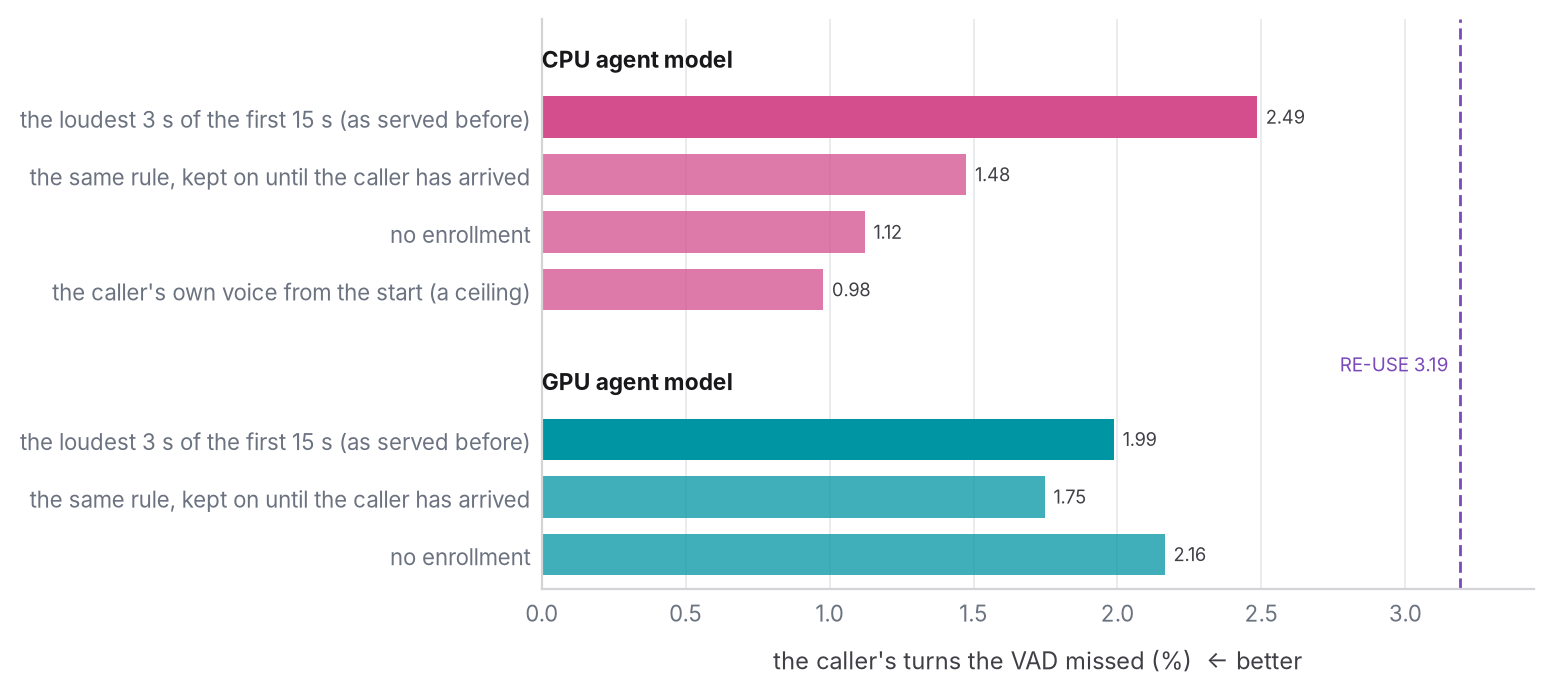}
\caption{Development calls: the caller's turns the VAD missed behind each agent model at its first-served gate, by what
the model is enrolled on. Dashed: RE-USE.}
\label{fig:r4clip}
\end{figure}

\textbf{The clip} (Figure~\ref{fig:r4clip}). The live rule stops at 15 s, and in \val{vint/r4/where/low-calls} of
\val{vint/r4/where/calls} calls the caller is in under a quarter of the final clip. Keeping the rule on until the
caller arrives cuts the turns missed behind the CPU agent model from \val{vint/r4/dev/cpu/shipped/missed}\% to
\val{vint/r4/dev/cpu/longer/missed}\%; no enrollment at all reaches \val{vint/r4/dev/cpu/none_old_gate/missed}\%. The
GPU agent model keeps its default.

\begin{figure}[H]
\centering
\includegraphics[width=0.92\linewidth]{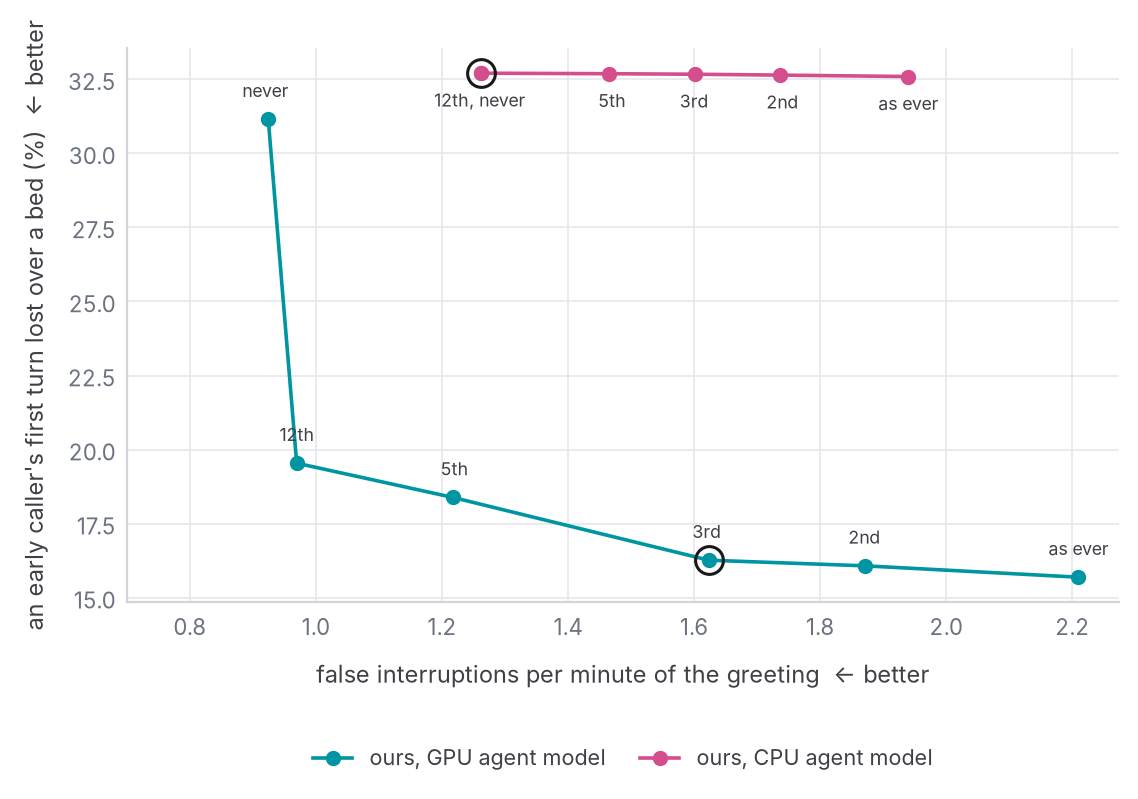}
\caption{The voicebot mode's wait, per agent model: a kept frame counts for the gate from the $n$-th in a row while the
decoy holds (as ever: every one; never: none). Across: false interruptions per minute of the greeting, development
calls. Up: share of the first turn lost by a caller who talks in the first second over a noise bed. Ringed: the setting
served.}
\label{fig:r4wait}
\end{figure}

\textbf{The wait} (Figure~\ref{fig:r4wait}). The voicebot mode carries over: \val{vint/r4/dev/gpu/vb_shipped/all}
false interruptions a minute behind the GPU agent model against \val{vint/r4/dev/gpu/shipped/all} by default. While
the decoy holds, a frame kept near the line says little about the caller, so such frames open the gate only from the
$n$-th in a row: never for the CPU model, from the third for the GPU one, the strictest settings that cost an early
caller at most a point.

\begin{figure}[H]
\centering
\includegraphics[width=0.92\linewidth]{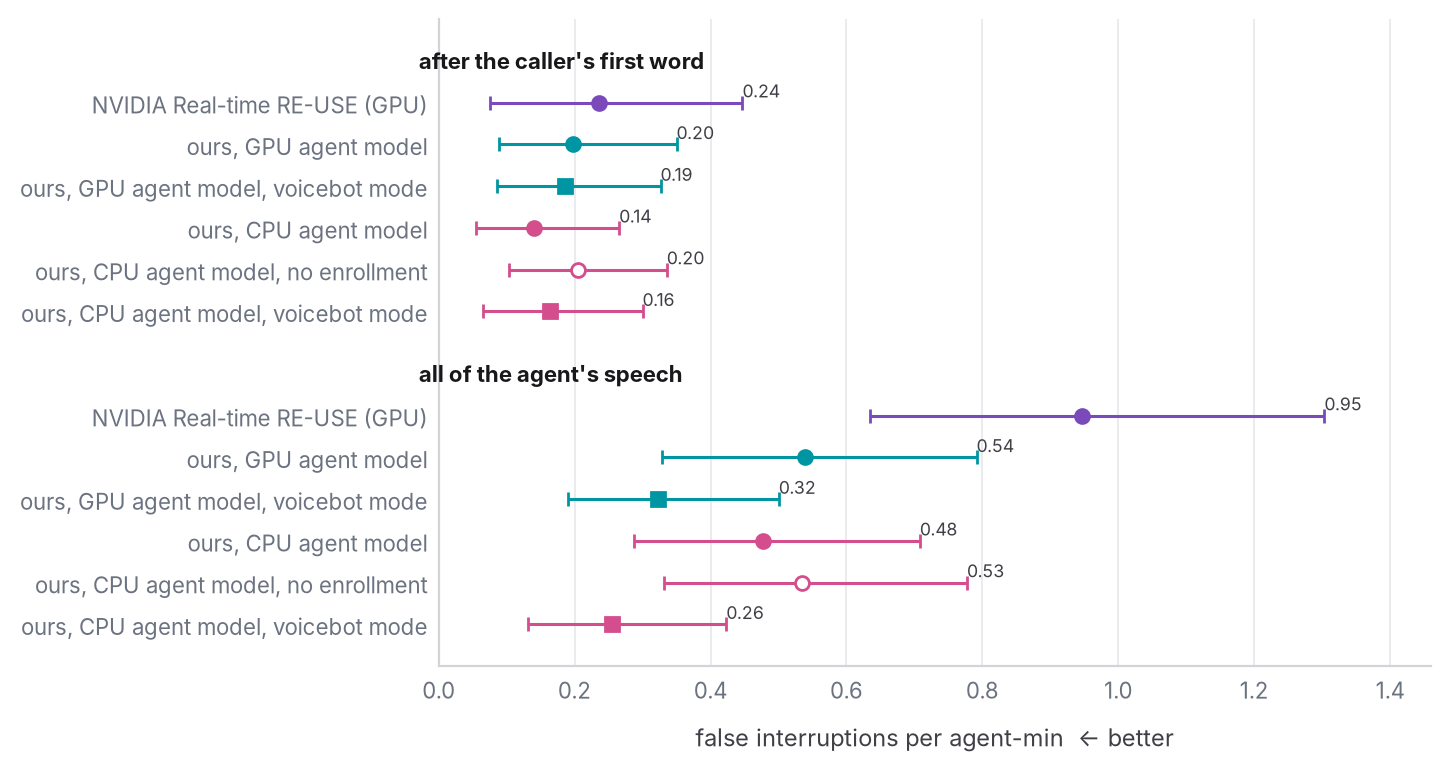}
\caption{Held out (\val{vint/r4/test/calls} calls): false interruptions per minute of the agent's speech after the
caller's first word and over all of it, with 90\% intervals over resampled call plans. Squares: the voicebot mode.
Hollow: the CPU agent model with no enrollment, the default mode's development pick, scored and not made the default.}
\label{fig:r4test}
\end{figure}

\textbf{Held out} (Figure~\ref{fig:r4test}), the voicebot mode paused the agent \val{vint/r4/test/cpuBv8/all} and
\val{vint/r4/test/gpuv/all} times a minute behind the CPU and GPU agent models, against \val{vint/r4/test/cpuag8/all}
and \val{vint/r4/test/agent/all} by default and \val{vint/r4/test/reuse/all} behind RE-USE. Nearly all of the gain is
the greeting.

\begin{figure}[H]
\centering
\includegraphics[width=0.92\linewidth]{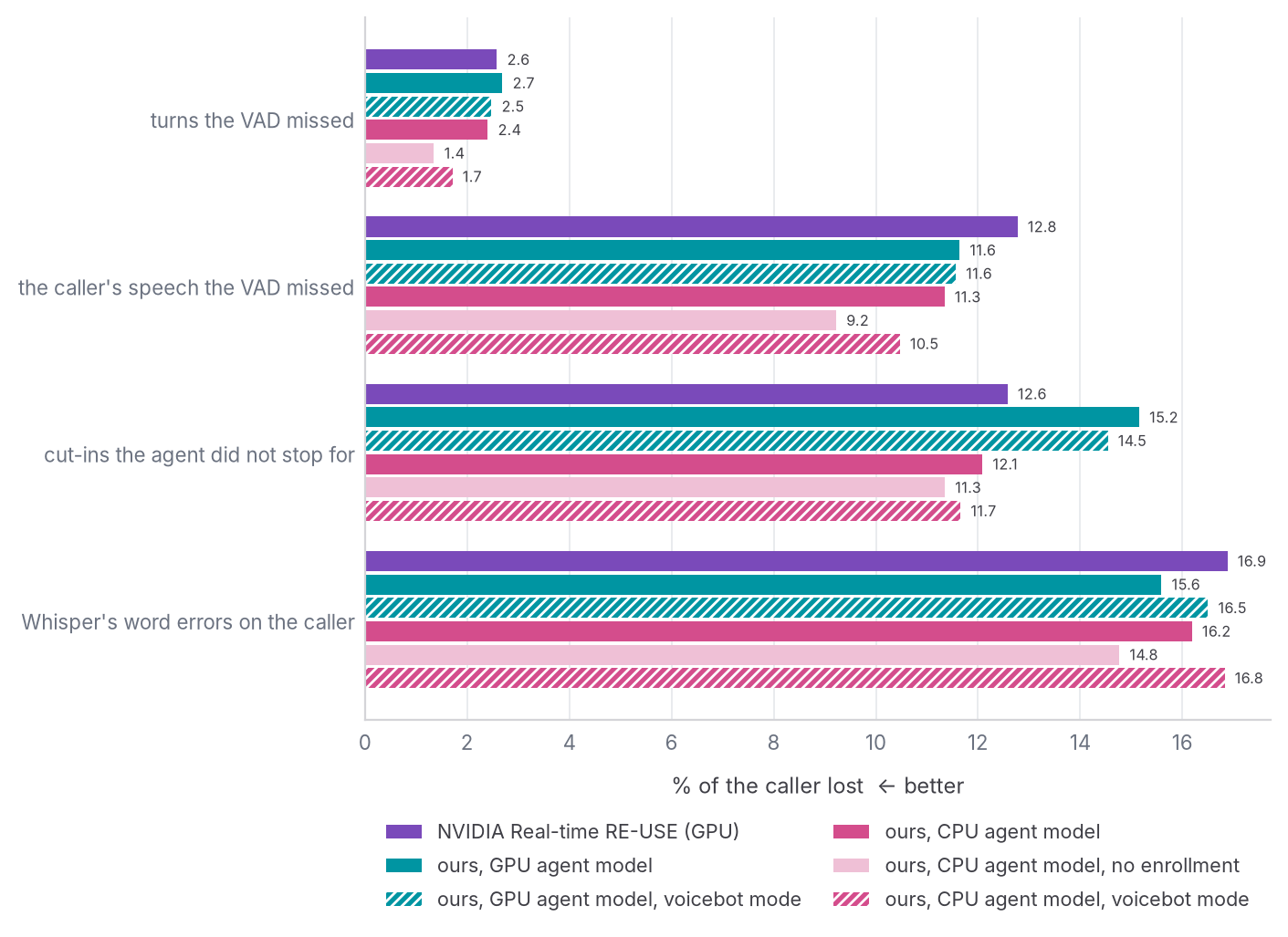}
\caption{What each system costs the caller, held out, every measure as a share lost, for the systems of
Figure~\ref{fig:r4test}. Hatched: the voicebot mode. Pale: the CPU agent model with no enrollment.}
\label{fig:r4caller}
\end{figure}

\textbf{The caller, held out} (Figure~\ref{fig:r4caller}). In the mode the VAD misses fewer turns
(\val{vint/r4/test/cpuBv8/missed}\% and \val{vint/r4/test/gpuv/missed}\%, RE-USE \val{vint/r4/test/reuse/missed}\%),
and Whisper errs slightly more (\val{vint/r4/test/cpuBv8/wer}\% and \val{vint/r4/test/gpuv/wer}\%, RE-USE
\val{vint/r4/test/reuse/wer}\%), since the first turn begins under the decoy.

\textbf{No enrollment, held out.} The CPU model's default-mode development pick, no enrollment, kept the caller best
(\val{vint/r4/test/cpuB8/missed}\% of turns missed) but paused the agent \val{vint/r4/test/cpuB8/all} times a minute
against the released default's \val{vint/r4/test/cpuag8/all}. We kept the default and show both.

\textbf{Its cost.} Behind the CPU agent model a caller who talks in the first second loses
\val{vint/r4/probe/cpu/voicebot/clean-first}\% of the first turn on a clean line in the mode, against
\val{vint/r4/probe/cpu/default/clean-first}\% by default, so it stays a mode for agents that speak first.

\textbf{Serving.} Each agent model's configuration names what the mode changes, so it is still one argument. The
plugin and the GPU engine matched the PyTorch model's gate in every hop of \val{vint/r4/check/plugin/calls}
development calls.

\section{Smaller models}
\label{sec:small}

\textbf{The CPU students on a GPU.} The release serves \val{serve/MW/cap-c2} calls per GPU in real time. The CPU
students share its architecture without WavLM, so the same server runs them: we publish the 22M agent model beside the
555M one and train 10M CPU models.

\begin{figure}[H]
\centering
\includegraphics[width=0.92\linewidth]{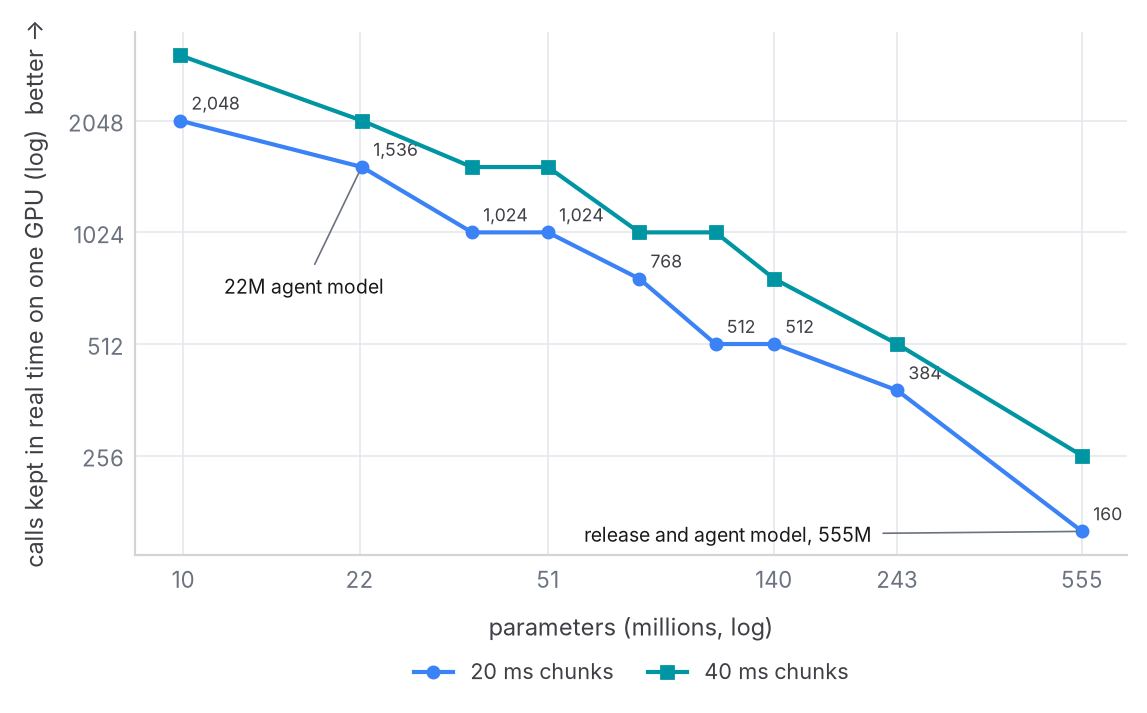}
\caption{Calls one GPU's step keeps in real time against parameters, float16 with CUDA graphs and the presence gate in
the step. 10M to 243M: the released architecture at other widths and depths, without WavLM; 555M: the release, with
it.}
\label{fig:smallengine}
\end{figure}

\textbf{Ten times the calls per GPU} (Figure~\ref{fig:smallengine}). At 20 ms chunks one GPU keeps
\val{small/C22/cap-c2} calls of the 22M model and \val{small/C10/cap-c2} of the 10M one in real time, against
\val{serve/MW/cap-c2} of the release. In float16 the served 22M model gates the same hops as in float32.

\begin{figure}[H]
\centering
\includegraphics[width=0.92\linewidth]{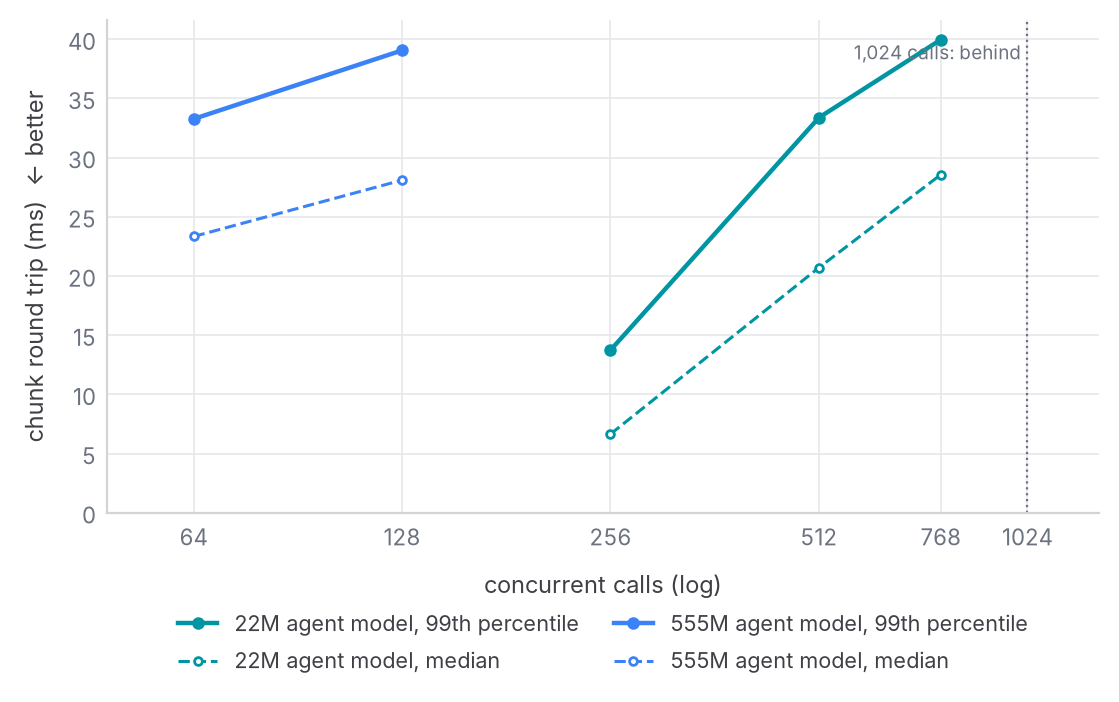}
\caption{End to end over WebSockets on one GPU, 20 ms chunks, live enrollment and the gate: a chunk's round trip after
each call's first 15 s, the 22M agent model against the 555M one, measured the same day.}
\label{fig:smallws}
\end{figure}

\textbf{End to end} (Figure~\ref{fig:smallws}), the 22M agent model holds \val{small/ws/hold} calls with every chunk
back within 100 ms (\val{small/ws/hold-sp99} ms at the 99th percentile), limited by the front end, not the GPU.

\begin{figure}[H]
\centering
\includegraphics[width=0.92\linewidth]{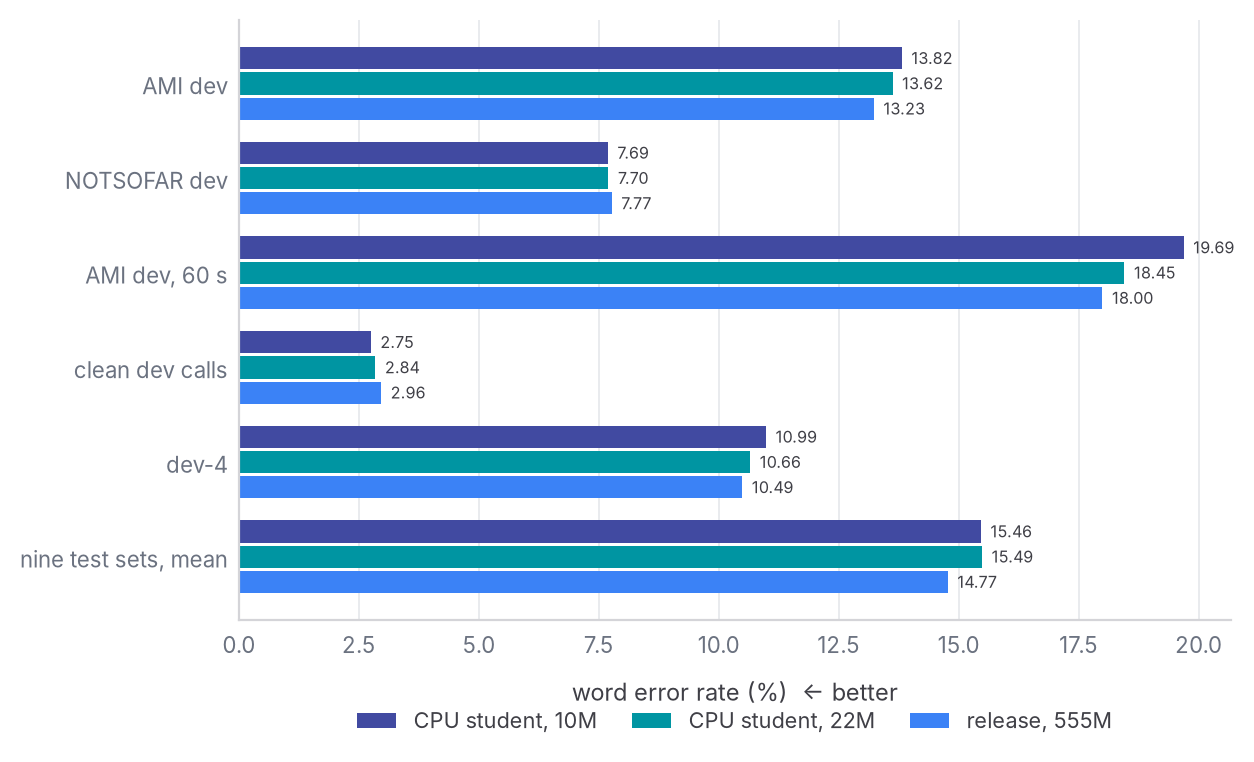}
\caption{The 10M and 22M CPU students against the release, live: each development set, dev-4 and the mean of the nine
test sets. The smaller models lose only on the long windows.}
\label{fig:smallsets}
\end{figure}

\textbf{What it costs} (Figure~\ref{fig:smallsets}). The 22M student makes \val{r/C22_ftlong_kd_selfl/dev4}\% on dev-4
against the release's \val{r/rel_selfl/dev4}\%, and \val{nine/C22_ftlong_kd_selfl/wer}\% over the nine test sets
against \val{nine/rel_selfl/wer}\%; M without WavLM or teacher makes \val{r/MB_ftlong_ws_selfl/dev4}\%. The 10M
student reaches \val{r/C10n_ftlong_kd_selfl/dev4}\%, losing only on the 60 s windows, and is level over the test sets.

\begin{figure}[H]
\centering
\includegraphics[width=0.92\linewidth]{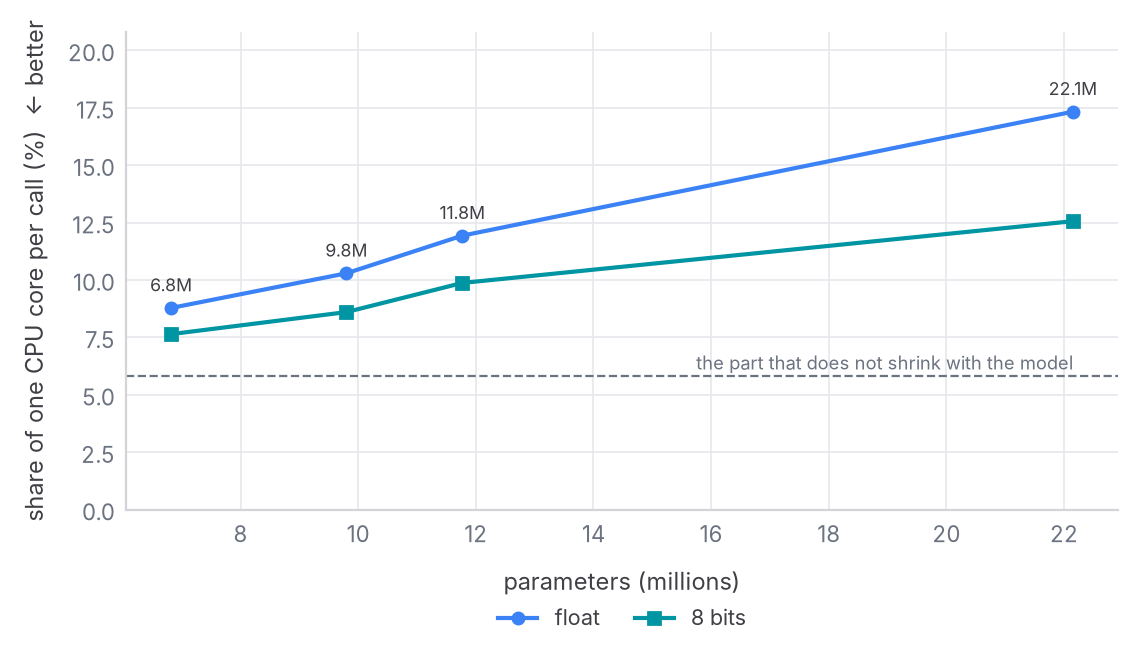}
\caption{Share of one CPU core per call through the LiveKit plugin against parameters, one thread, 50 ms frames, the
released architecture at four sizes.}
\label{fig:smallcpu}
\end{figure}

\textbf{On one CPU core} (Figure~\ref{fig:smallcpu}), the 10M student takes \val{vint/c10/core/c108}\% of a core per
call in 8 bits, against \val{vint/c10/core/cpu8}\% at 22M. About \val{small/c10/fixed-pct}\% is fixed cost, so going
smaller buys little.

\begin{figure}[H]
\centering
\includegraphics[width=0.92\linewidth]{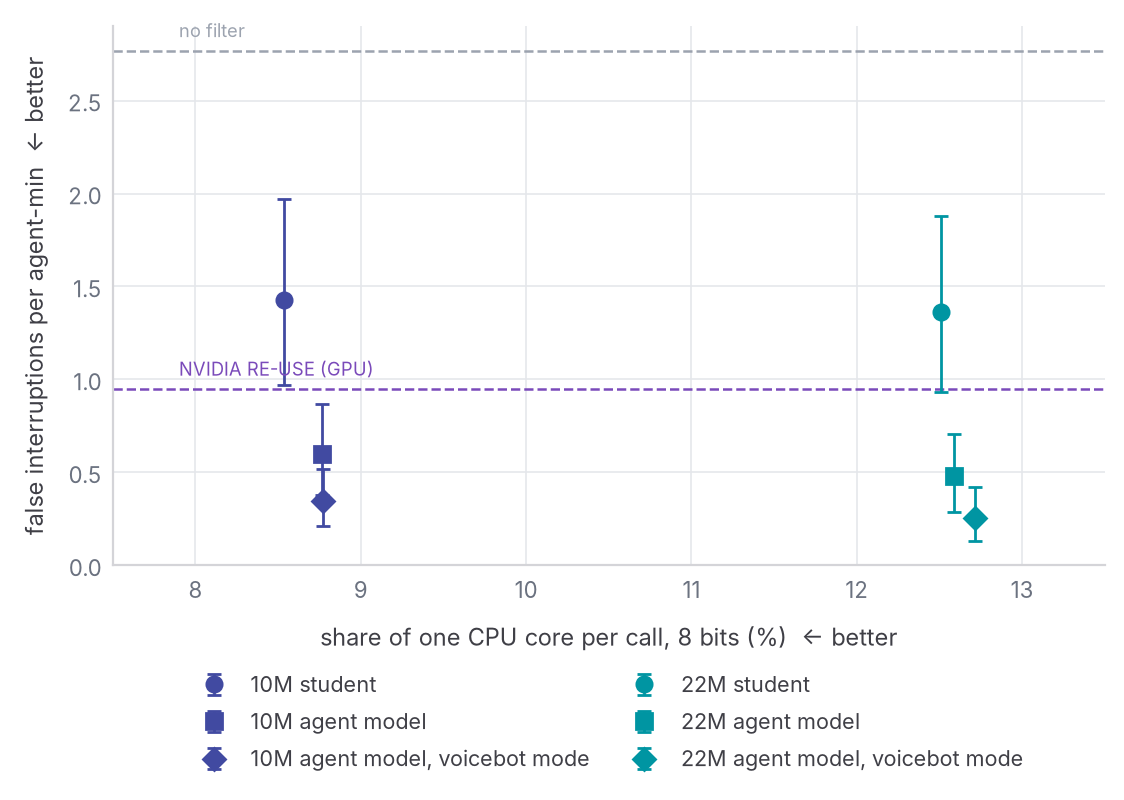}
\caption{The interruption bench, held out, through the LiveKit plugin in 8 bits: false interruptions per minute of the
agent's speech against the share of one CPU core per call, the 10M and 22M CPU models in each mode, with 90\% intervals
over resampled call plans. Dashed: RE-USE on a GPU and no filter.}
\label{fig:smallbench}
\end{figure}

\textbf{The 10M agent model} (Figure~\ref{fig:smallbench}), trained like the 22M one with its gate chosen by the rule
of Section~\ref{sec:clip}, paused the agent \val{vint/c10/c10ag8/all} times a minute held out (22M
\val{vint/c10/cpuag8/all}, RE-USE \val{vint/c10/reuse/all}) and \val{vint/c10/c10agv8/all} in the voicebot mode, and
keeps the caller as well (\val{vint/c10/silence/c10ag8/missed}\% of turns missed). After the first word it only ties
RE-USE: babble and home sounds get past the smaller head.

\section{Phone lines}
\label{sec:phone}

\begin{figure}[H]
\centering
\includegraphics[width=0.92\linewidth]{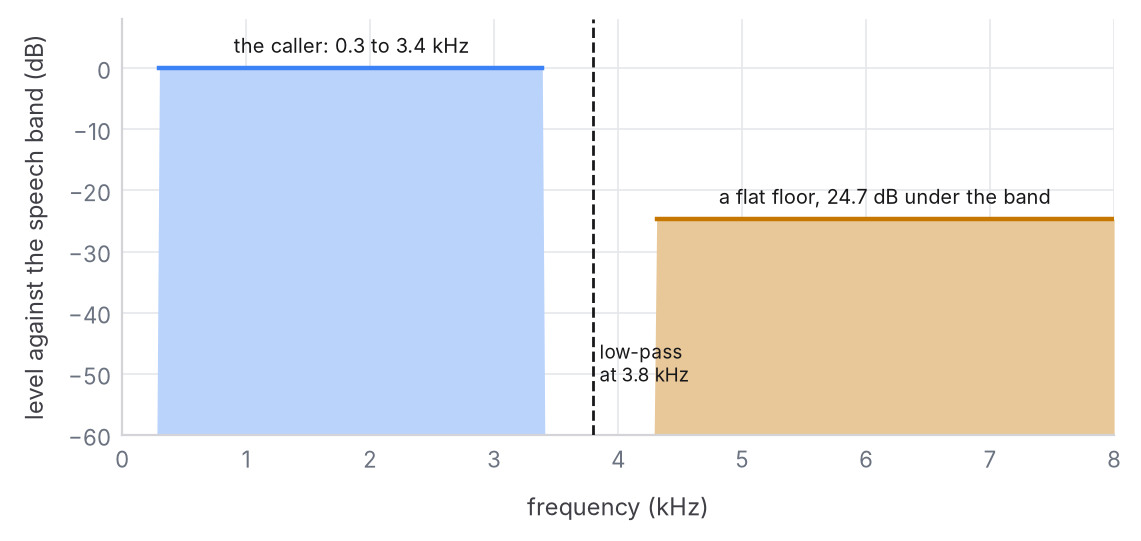}
\caption{The phone line as the agent receives it, drawn to the measured band edge and floor of the rebuilt calls
(median). Dashed: the low-pass we put in front of the isolator.}
\label{fig:phoneline}
\end{figure}

\textbf{The line agents receive.} A deployed agent receives a phone call as 8 kHz audio upsampled to 16 kHz, with a
flat noise floor above the speech band (Figure~\ref{fig:phoneline}). We rebuild that line from one recording and play
\val{phone/line/calls} held-out callers through it, over real noise and on a quiet line.

\begin{figure}[H]
\centering
\includegraphics[width=0.92\linewidth]{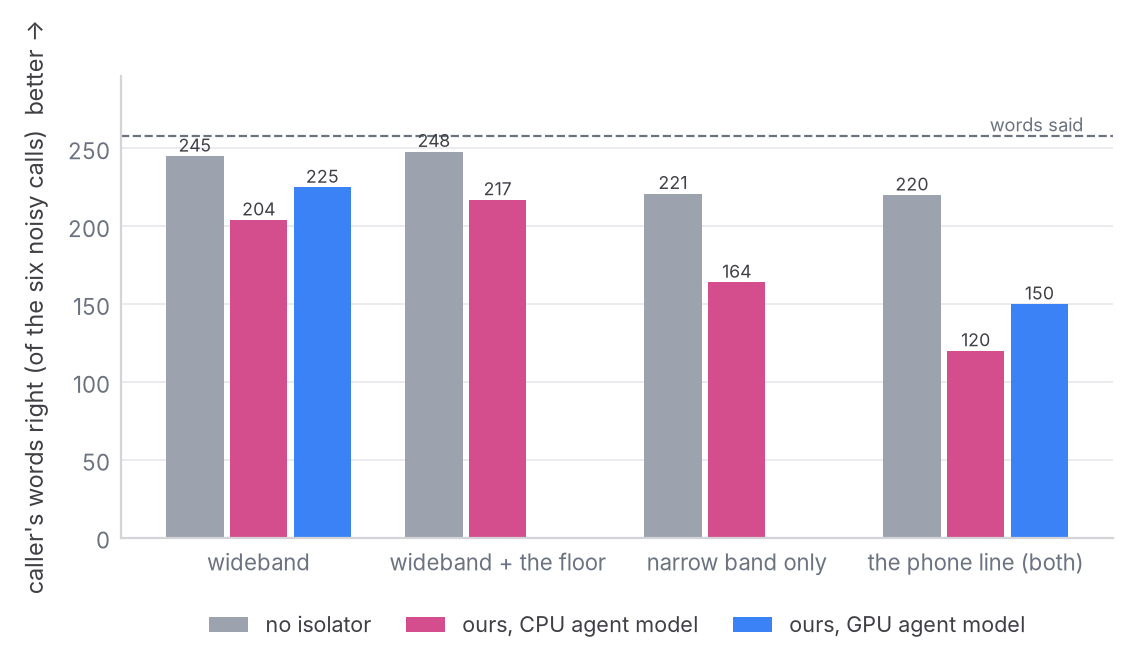}
\caption{The six noisy held-out calls: caller words Parakeet v3 gets right on the wide line, with only the floor, with
only the narrow band, and on the whole phone line. Agent models before the phone-line fine-tune, default mode; the GPU
model on the two lines it is served on.}
\label{fig:phoneparts}
\end{figure}

\textbf{Band and floor together hurt} (Figure~\ref{fig:phoneparts}). On noisy calls, behind the CPU agent model
Parakeet gets \val{phone/heard/cpu_default/noisy/wide} of the caller's words right on the wide line and
\val{phone/heard/cpu_default/noisy/phone} on the phone line, against \val{phone/heard/none/noisy/phone} with no
isolator. The narrow band alone costs most of that, the floor alone nothing, and both together more than either:
training calls are band-limited but never carry energy above the band. Quiet phone calls lose little.

\begin{figure}[H]
\centering
\includegraphics[width=0.92\linewidth]{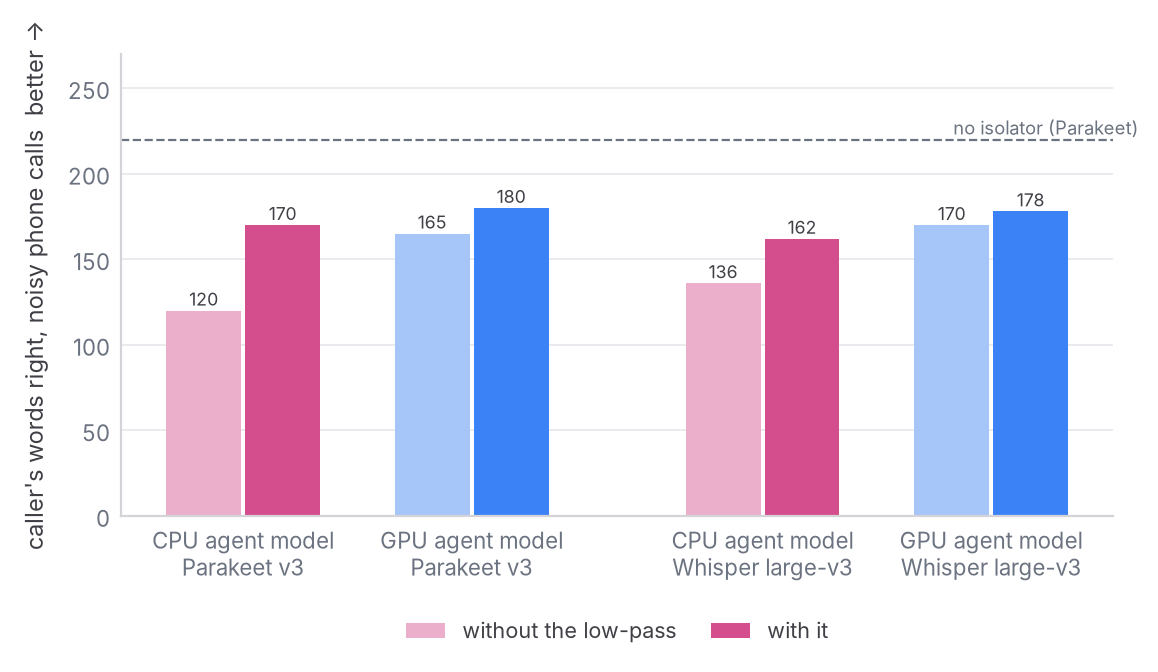}
\caption{The six noisy held-out phone calls: caller words right behind each agent model without and with a 3.8 kHz
low-pass, for both recognisers. Dashed: Parakeet with no isolator.}
\label{fig:phonelp}
\end{figure}

\textbf{A low-pass} at 3.8 kHz removes the floor (Figure~\ref{fig:phonelp}), raising caller words right from
\val{phone/lp/parakeet/cpu_dev/noisy} to \val{phone/lp/parakeet/cpu_lp/noisy} behind the CPU agent model and from
\val{phone/lp/parakeet/gpu_dev/noisy} to \val{phone/lp/parakeet/gpu_lp/noisy} behind the GPU one, short of no
isolator's \val{phone/heard/none/noisy/phone}.

\begin{figure}[H]
\centering
\includegraphics[width=0.92\linewidth]{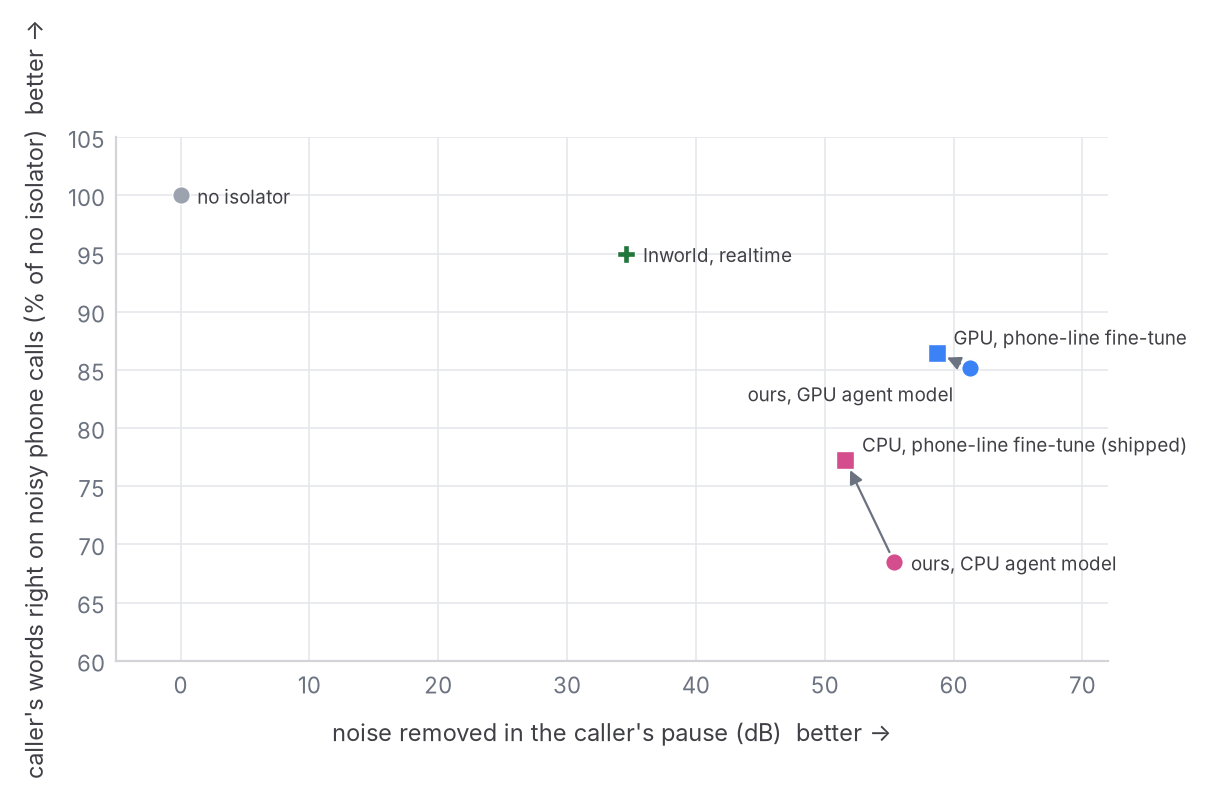}
\caption{The 108 phone-line development calls: noise removed in the caller's pause (median) against caller words right
on noisy phone calls, both recognisers, as a share of no isolator's. Arrows: each agent model to its phone-line pick.}
\label{fig:phonetrade}
\end{figure}

\begin{figure}[H]
\centering
\includegraphics[width=0.92\linewidth]{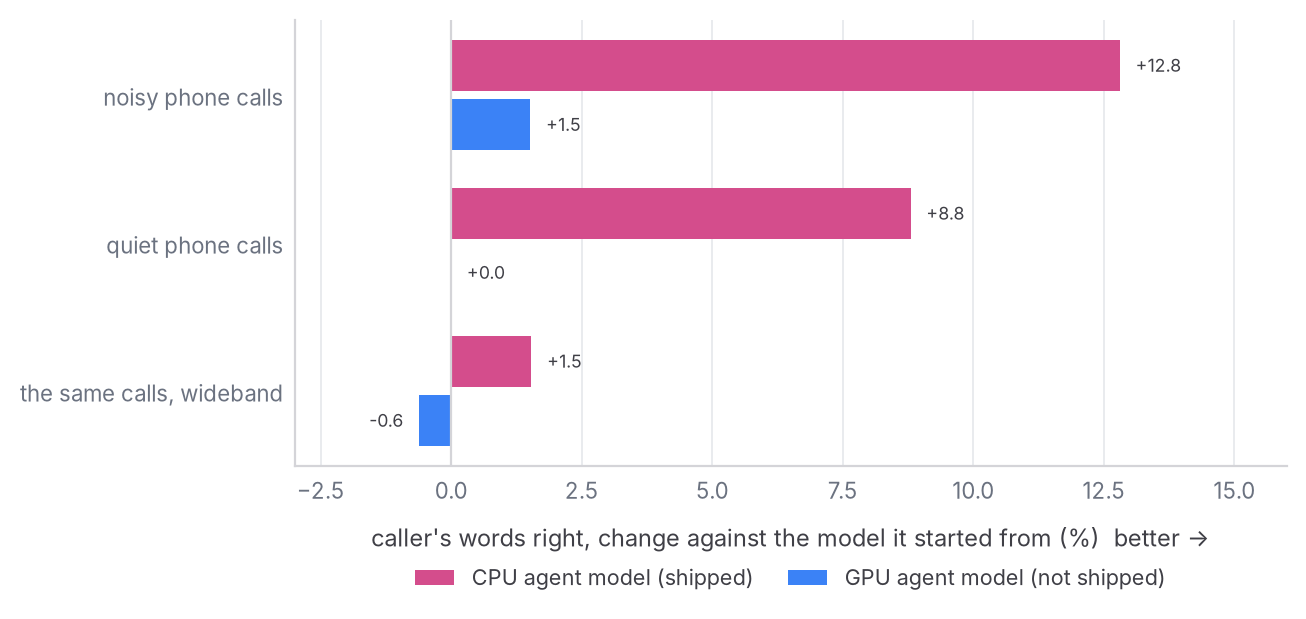}
\caption{The phone-line fine-tune on the development calls: change in caller words right against its starting model,
by kind of call.}
\label{fig:phoneft}
\end{figure}

\textbf{A phone-line fine-tune} (Figures~\ref{fig:phonetrade} and~\ref{fig:phoneft}). The simulator adds a flat floor
above a narrowband line, and both agent models train on. A rule written first, on 108 development calls, takes the
most words right on noisy phone calls while wideband and quiet calls, pause noise and other people's words hold. The
CPU pick gains \val{phone/dev/cpu_after-cpu_before/score-pct}\% on noisy phone calls and
\val{phone/dev/cpu_after-cpu_before/quiet_phone-pct}\% on quiet ones and is now the CPU agent model; the GPU pick
gains only \val{phone/dev/gpu_ft-gpu/score-pct}\% and is not shipped. Held out, the CPU pick is level on noisy calls
and ahead on quiet ones (\val{phone/test/cpu/cpu_pick/quiet} against \val{phone/test/cpu/cpu_cur/quiet} words right).

\textbf{What is left.} A noisy phone caller still loses words to every model of ours. Inworld keeps
\val{phone/dev/inworld-gpu/score-pct}\% more of them than our GPU agent model by removing less of everything
(Figure~\ref{fig:phonetrade}).

\section{Conclusion}
\label{sec:conclusion}

An isolator for transcription must remove every other voice however long it talks and leave the caller alone however
quietly they speak; short evaluation windows test neither. Long windows show that an attention window forgets the
caller, and lone real callers show that fixing it makes the model doubt them. WavLM's own weights, made causal and
trained inside the model, move that trade where width does not, and windows cut from longer calls remove most of a lone
caller's cold start. The release leads VoxLock live on long meetings and calls with a second voice, is level on Krisp,
and trails on the lone caller.

In an agent that talks back, noise paused it \val{vint/rel/fi} times a minute of its speech behind the release, against
\val{vint/none/fi} with no filter and \val{vint/voxlock/fi} behind VoxLock. A caller-presence gate that sends the
caller's pauses out as digital silence takes the agent model to \val{vint/gate/agent/all}, against RE-USE's
\val{vint/gate/reuse/all}, and the voicebot mode to \val{vint/r4/test/gpuv/all}. Distilled without WavLM, the recipe runs
on one CPU thread at 22M and 10M parameters (\val{vint/cpugate/cpuag8/all} and \val{vint/c10/c10ag8/all}), and the 22M
model serves \val{small/C22/cap-c2} calls on one GPU where the release serves \val{serve/MW/cap-c2}. On phone lines a
low-pass and a fine-tune win back part of what a noisy caller loses. The models and the LiveKit plugin are open at
\url{https://huggingface.co/collections/Scicom-intl/voice-isolator}.

\section{Limitations}

\textbf{Recognisers.} Every score comes from Parakeet or Whisper; other recognisers react differently to processing
\citep{iwamoto2024how}. \textbf{Calls.} Krisp's calls are scripted, and our call-centre windows carry a second voice
only as a mixture we made, with a recogniser's transcript as reference. \textbf{Enrollment.} The live rule takes the
loudest 3 s; when someone else is loudest, as in many AliMeeting windows, it keeps the wrong voice. \textbf{Selection.}
The meeting sets reward suppression and the clean calls restraint; another weighting picks another model. Two training
runs of one recipe at S land at \val{r/SB_ftlong_ws_cwlt2_selfl/dev4} and \val{r/SB_ftlong_ws_cwlt2r_selfl/dev4}\% on
dev-4, and at M at \val{r/pre_selfl/dev4} and \val{r/MB_ftlong_ws_cwlt2r_selfl/dev4}\%; they differ on the 60 s windows
and clean calls by more than our intervals, which resample callers, not runs. Dev-4 differences of a tenth or two rank
runs, not recipes; the release's last step held in two seeds at M, by less than that. \textbf{Cost.} The WavLM branch
adds \val{ssl/cwl/params}M parameters and steps in 20 ms frames: no 10 ms chunks, and fewer calls per GPU than M alone.
WavLM was pre-trained on English. \textbf{Phone lines.} Our phone-line calls are rebuilt from one recording's
measurements, and a noisy caller on them still loses words behind every model of ours. \textbf{Licences.} Krisp,
ai-coustics' calls and AppTek's recordings are evaluation-only; results are for internal study until a release is
cleared.

\section{Acknowledgement}

This work was carried out at Scicom (MSC) Berhad. The author thanks Scicom for the compute.

\bibliographystyle{plainnat}
\bibliography{neurips_2023}

\clearpage
\appendix
\section{Tables}
\label{app:tables}

The numbers behind the figures: every system on every test set (Table~\ref{tab:main}); the recipe
(Table~\ref{tab:recipe}) and its long-call stage (Tables~\ref{tab:caller}, \ref{tab:line} and~\ref{tab:longcall}); model size
(Table~\ref{tab:scale}); self-supervised models (Table~\ref{tab:ssl}); enrollment modes (Table~\ref{tab:enrollmodes});
and the stage-1 ablations (Tables~\ref{tab:loss}, \ref{tab:data}, \ref{tab:aug}, \ref{tab:real}, \ref{tab:keep}
and~\ref{tab:enrollarm}); the interruption bench, each system live (Table~\ref{tab:vintlive}) and the
voicebot mode through the shipped code (Table~\ref{tab:vintvb}); LiveKit's VAD behind each system on the test calls
(Table~\ref{tab:vad}); the distilled students on one CPU thread (Table~\ref{tab:cpu}); and the isolator inside running
LiveKit agents (Tables~\ref{tab:lkbench} and~\ref{tab:lkvad}).

\begin{table}[H]
\centering
\caption{Test sets: word error rate (AliMeeting: character error rate). Bold: the best live system per
column (VoxLock's loudest-3 s sample and its PyTorch script look up to 15 s ahead).}
\label{tab:main}
\scriptsize
\begin{adjustbox}{max width=\linewidth}
\begin{tabular}{lrrrrrrrrr}
\toprule
\schead \scth{System} & \scth{Krisp} & \scth{AMI (VoxLock)} & \scth{AMI test} & \scth{AMI 60 s} & \scth{NOTSOFAR} & \scth{AliMeeting} & \scth{Calls, clean} & \scth{Calls + voice} & \scth{ai-coustics} \\
\midrule
No processing & 23.19 & 33.00 & 42.59 & 121.79 & 34.70 & 29.40 & 0.04 & 69.79 & 66.36 \\
DeepFilterNet3 (noise only) & 13.15 & 21.56 & 26.00 & 41.77 & 15.38 & 18.48 & 2.63 & 69.65 & 35.35 \\
Krisp VI 2.5 Default & 5.30 &  &  &  &  &  &  &  &  \\
VoxLock, loudest 3 s & 5.19 & 11.26 & 9.70 & 19.52 & 7.61 & 17.11 & 3.37 & 47.05 & 20.31 \\
VoxLock, PyTorch script & 4.99 & 10.26 & 10.31 & 20.20 & 7.78 & 15.38 & 2.87 & 48.59 & 20.49 \\
VoxLock, live & 5.63 & 11.71 & 12.07 & 23.66 & 8.84 & 16.28 & 2.92 & 56.65 & 24.55 \\
Ours, Sidon architecture 188M & 29.49 & 32.45 & 37.51 & 101.80 & 30.01 & 42.94 &  &  &  \\
Ours, Sidon architecture lite & 18.27 & 17.45 & 24.06 & 66.15 & 13.40 & 19.31 &  &  &  \\
Ours, S, live & 6.12 & 10.23 & 10.47 & \textbf{15.48} & 7.78 & 18.09 & 8.55 & 48.86 & 18.46 \\
Ours, M, live & 5.79 & \textbf{9.95} & 10.51 & 16.84 & 7.58 & 18.57 & 8.80 & 48.22 & 16.57 \\
Ours, M, loudest 3 s & 5.76 & 11.23 & 10.62 & 16.44 & 7.70 & 17.85 & 7.60 & 47.99 & 15.47 \\
Ours, M, no enrollment & 6.55 & 10.26 & 10.27 & 18.39 & 7.64 & 16.02 & 6.84 & 50.85 & 16.86 \\
Ours, M + WavLM, live, before the window cut & 5.55 & 10.06 & \textbf{9.97} & 15.91 & 7.76 & 15.75 & 7.27 & 48.36 & 14.36 \\
Ours, M + WavLM, live & 5.74 & 10.19 & 10.18 & 16.13 & 7.67 & \textbf{15.69} & 5.58 & 47.52 & \textbf{14.26} \\
Ours, M + WavLM, loudest 3 s & 5.93 & 11.37 & 10.35 & 16.22 & 7.67 & 15.40 & 5.66 & 47.87 & 12.54 \\
Ours, M + WavLM, no enrollment & 6.28 & 10.02 & 10.09 & 16.19 & 7.74 & 15.74 & 5.90 & \textbf{47.43} & 14.47 \\
Ours, CPU student 22M, live & 5.66 & 11.23 & 10.29 & 15.55 & 7.58 & 15.82 & 5.32 & 49.90 & 18.03 \\
Ours, CPU student 10M, live & \textbf{5.44} & 11.06 & 10.73 & 15.94 & \textbf{7.52} & 15.73 & 4.35 & 52.30 & 16.11 \\
Inworld Voice Isolator v1, realtime & 22.24 & 21.39 & 30.06 & 102.51 & 22.41 & 18.94 & \textbf{2.07} & 69.02 & 60.69 \\
Inworld Voice Isolator v1, offline & 22.31 & 17.80 & 30.28 & 99.21 & 19.90 & 18.03 & 2.63 & 68.28 & 60.94 \\
\bottomrule
\end{tabular}
\end{adjustbox}

\end{table}
\begin{table}[H]
\centering
\caption{The recipe, S, live enrollment, with VoxLock. Word error rate on each development set (clean calls: against plain recognition, deleted words in brackets), the
mean of the three meeting sets, dev-4 (that and the clean calls), Krisp reported only.}
\label{tab:recipe}
\scriptsize
\begin{adjustbox}{max width=\linewidth}
\begin{tabular}{lrrrrrrrr}
\toprule
\schead \scth{System} & \scth{AMI dev} & \scth{NSF dev} & \scth{AMI 60 s} & \scth{Clean calls (del.)} & \scth{Calls + voice} & \scth{Meetings} & \scth{Dev-4} & \scth{Krisp} \\
\midrule
VoxLock, loudest 3 s & 13.48 & 8.11 & 23.66 & 2.39 (1.59) & 58.84 & 15.08 & 11.91 & 5.19 \\
VoxLock, live & 16.38 & 8.18 & 27.39 & 2.09 (1.53) & 64.82 & 17.32 & 13.51 & 5.63 \\
VoxLock, PyTorch & 14.34 & 8.28 & 26.97 & 1.89 (1.49) & 60.14 & 16.53 & 12.87 & 4.99 \\
S, stage 1 (6 s calls) & 15.69 & 8.71 & 33.87 & 7.16 (5.71) & 61.63 & 19.42 & 16.36 & 8.16 \\
\quad + real-crosstalk fine-tune & 13.66 & 7.86 & 44.29 & 3.85 (2.70) & 62.66 & 21.94 & 17.42 & 8.25 \\
\quad + long-call stage & 13.03 & 7.72 & 19.65 & 7.53 (5.90) & 52.71 & 13.47 & 11.98 & 6.06 \\
\quad + caller level drift, single talkers & 13.23 & 7.63 & 20.02 & 5.78 (4.56) & 51.42 & 13.63 & 11.67 & 6.12 \\
\bottomrule
\end{tabular}
\end{adjustbox}

\end{table}
\begin{table}[H]
\centering
\caption{Changes to the long-call stage, S, live unless marked. Word error rate on each development set (clean calls: against plain recognition, deleted words in brackets), the
mean of the three meeting sets, dev-4 (that and the clean calls), Krisp reported only.}
\label{tab:caller}
\scriptsize
\begin{adjustbox}{max width=\linewidth}
\begin{tabular}{lrrrrrrrr}
\toprule
\schead \scth{System} & \scth{AMI dev} & \scth{NSF dev} & \scth{AMI 60 s} & \scth{Clean calls (del.)} & \scth{Calls + voice} & \scth{Meetings} & \scth{Dev-4} & \scth{Krisp} \\
\midrule
long-call stage (reference) & 13.03 & 7.72 & 19.65 & 7.53 (5.90) & 52.71 & 13.47 & 11.98 & 6.06 \\
\quad 2,000 steps instead of 4,000 & 12.90 & 7.63 & 21.31 & 6.21 (4.86) &  & 13.95 & 12.01 & 5.52 \\
\quad caller level drift & 13.18 & 7.66 & 18.84 & 8.77 (7.00) & 52.30 & 13.23 & 12.11 & 6.63 \\
\quad more single-talker calls & 13.03 & 7.62 & 19.91 & 6.75 (5.19) & 51.24 & 13.52 & 11.83 & 6.68 \\
\quad both & 13.23 & 7.63 & 20.02 & 5.78 (4.56) & 51.42 & 13.63 & 11.67 & 6.12 \\
\quad enrollments taken as the live rule takes them & 13.25 & 7.85 & 18.40 & 8.82 (7.10) & 48.50 & 13.17 & 12.08 & 8.11 \\
\quad perceptual loss on the caller's frames & 13.64 & 7.94 & 21.34 & 5.85 (4.39) & 50.24 & 14.31 & 12.19 & 5.56 \\
\quad both + perceptual loss & 13.77 & 7.76 & 23.09 & 4.97 (3.67) & 57.78 & 14.87 & 12.40 & 5.07 \\
reference, no enrollment & 13.08 & 7.71 & 25.00 & 4.96 (3.97) & 51.20 & 15.26 & 12.69 & 7.01 \\
both, no enrollment & 13.65 & 7.63 & 26.59 & 3.66 (2.78) & 54.26 & 15.96 & 12.88 & 6.87 \\
\bottomrule
\end{tabular}
\end{adjustbox}

\end{table}
\begin{table}[H]
\centering
\caption{A lone caller on a recorded line: rules at inference on M + WavLM before the window cut, then changes to the
long-call stage (its pauses gated to digital silence, the enrollment through the call's own line, fewer held pauses,
windows cut from longer calls, which gives the release) and second runs with another seed. Columns as in
Table~\ref{tab:caller}.}
\label{tab:line}
\scriptsize
\begin{adjustbox}{max width=\linewidth}
\begin{tabular}{lrrrrrrrr}
\toprule
\schead \scth{System} & \scth{AMI dev} & \scth{NSF dev} & \scth{AMI 60 s} & \scth{Clean calls (del.)} & \scth{Calls + voice} & \scth{Meetings} & \scth{Dev-4} & \scth{Krisp} \\
\midrule
M + WavLM, live (from 3 s in) & 13.00 & 7.69 & 17.59 & 3.92 (2.82) & 51.05 & 12.76 & 10.55 & 5.55 \\
\quad no enrollment & 13.19 & 7.72 & 21.19 & 3.61 (2.62) & 49.77 & 14.04 & 11.43 & 5.85 \\
\quad live from 2 s in & 13.03 & 7.66 & 17.50 & 3.88 (2.77) & 51.17 & 12.73 & 10.52 & 5.58 \\
\quad live from 1 s in & 12.94 & 7.64 & 17.30 & 3.85 (2.68) & 51.14 & 12.63 & 10.43 & 5.63 \\
\quad the line's digital silence not stepped &  &  &  & 4.18 (3.01) &  & &  &  \\
\quad the input passed until the enrollment exists & 14.14 & 8.03 & 21.85 & 2.56 (1.67) & 51.46 & 14.67 & 11.64 & 5.51 \\
\quad the input passed for the first 5 s & 16.62 & 9.39 & 23.85 & 2.44 (1.58) & 51.50 & 16.62 & 13.07 & 5.47 \\
\quad the input mixed back at $-30$ dB & 30.84 & 9.62 & 82.13 & 1.96 (1.27) & 67.70 & 40.86 & 31.14 & 7.78 \\
\quad the input mixed back at $-20$ dB & 35.51 & 19.16 & 103.22 & 1.92 (1.18) & 70.53 & 52.63 & 39.95 & 12.79 \\
\quad a second of digital silence first & 13.48 & 7.73 & 18.66 & 2.74 (1.82) &  & 13.29 & 10.65 &  \\
\quad a second of digital silence first, no enrollment &  &  &  & 2.47 (1.58) &  & &  &  \\
\quad a second of digital silence first, the line's silence not stepped & 13.47 & 7.73 & 19.19 & 2.69 (1.57) &  & 13.47 & 10.77 &  \\
\quad trained with a gated line (30\% of calls) & 13.07 & 7.82 & 17.79 & 4.88 (3.80) & 50.90 & 12.89 & 10.89 & 5.54 \\
\quad trained with a gated line (50\% of calls) & 13.30 & 7.71 & 18.47 & 3.66 (2.62) & 54.22 & 13.16 & 10.78 & 5.49 \\
\quad trained on windows cut from longer calls (30\%): the release & 13.23 & 7.77 & 18.00 & 2.96 (1.96) & 51.32 & 13.00 & 10.49 & 5.74 \\
\quad\quad no enrollment & 13.29 & 7.77 & 22.57 & 2.62 (1.76) & 48.75 & 14.54 & 11.56 & 6.28 \\
\quad\quad live from 1 s in & 13.07 & 7.79 & 18.10 & 3.11 (2.02) & 51.87 & 12.99 & 10.52 & 5.79 \\
\quad M + WavLM again, another seed & 13.01 & 7.69 & 17.56 & 4.19 (3.26) &  & 12.75 & 10.61 & 5.60 \\
\quad windows cut, another seed & 13.08 & 7.73 & 18.66 & 2.68 (1.83) &  & 13.16 & 10.54 & 5.63 \\
\quad windows cut and a gated line & 13.48 & 7.75 & 19.60 & 2.32 (1.47) & 54.22 & 13.61 & 10.79 & 5.27 \\
\quad windows cut and fewer pauses held by another voice & 13.00 & 7.85 & 19.87 & 2.46 (1.64) & 50.13 & 13.57 & 10.79 & 5.30 \\
S + WavLM, live & 12.90 & 7.70 & 17.40 & 5.26 (3.91) & 51.63 & 12.67 & 10.81 & 5.37 \\
\quad gated line & 13.16 & 7.71 & 17.74 & 4.92 (3.27) & 52.03 & 12.87 & 10.89 & 5.50 \\
\quad enrollment through the call's line & 13.19 & 7.68 & 18.52 & 4.22 (3.02) & 50.88 & 13.13 & 10.90 & 5.79 \\
\quad both & 12.93 & 7.66 & 18.54 & 4.18 (2.96) & 50.29 & 13.05 & 10.83 & 5.24 \\
\quad fewer pauses held by another voice & 12.62 & 7.69 & 18.80 & 3.84 (2.59) & 54.20 & 13.04 & 10.74 & 5.15 \\
\quad windows cut from longer calls & 13.00 & 7.76 & 18.72 & 3.21 (2.10) & 52.73 & 13.16 & 10.67 & 5.50 \\
\quad gated line and windows cut from longer calls & 13.07 & 7.75 & 18.23 & 3.63 (2.38) & 52.75 & 13.01 & 10.67 & 5.35 \\
\quad the reference again, another seed & 12.71 & 7.72 & 18.91 & 4.50 (3.21) &  & 13.11 & 10.96 & 5.42 \\
\quad windows cut from longer calls, another seed & 13.14 & 7.69 & 19.04 & 3.63 (2.15) &  & 13.29 & 10.87 & 5.58 \\
\bottomrule
\end{tabular}
\end{adjustbox}

\end{table}
\begin{table}[H]
\centering
\caption{The long-call stage, S, 4,000 steps from one checkpoint, self-enrolled. Simulated development set: SI-SDR improvement, leak and keep (dB); real sets: word error rate / leaked words per
minute.}
\label{tab:longcall}
\scriptsize
\begin{adjustbox}{max width=\linewidth}
\begin{tabular}{lrrrrrrr}
\toprule
\schead \scth{Arm} & \scth{SI-SDRi} & \scth{Leak dB} & \scth{Keep dB} & \scth{AMI dev} & \scth{NOTSOFAR dev} & \scth{AMI dev, 60 s} & \scth{Krisp} \\
\midrule
6 s calls & 4.26 & $-27.0$ & $-1.14$ & 15.47 / 9.78 & 8.39 / 0.47 & 46.70 / 41.98 & 9.41 / 11.81 \\
6 s calls, 30\% with a long last stretch & 3.64 & $-28.0$ & $-1.36$ & 15.86 / 7.32 & 10.68 / 0.55 & 40.08 / 33.59 & 11.62 / 7.51 \\
24 s calls & 3.86 & $-19.4$ & $-1.19$ & 14.68 / 10.36 & 7.75 / 0.55 & 30.34 / 14.18 & 6.73 / 1.89 \\
24 s calls, 60\% with turns & 3.11 & $-26.0$ & $-2.45$ & 15.22 / 8.53 & 8.12 / 0.72 & 27.28 / 16.33 & 6.70 / 2.39 \\
\quad + 10 s attention window & 3.51 & $-25.4$ & $-1.91$ & 14.79 / 7.32 & 7.96 / 0.43 & 25.81 / 10.97 & 8.63 / 0.25 \\
\quad + one frame LSTM & 3.18 & $-25.7$ & $-2.24$ & 14.52 / 7.86 & 7.76 / 0.51 & 25.94 / 14.92 & 8.36 / 0.68 \\
\quad + four frame LSTMs & 3.12 & $-25.8$ & $-2.27$ & 14.84 / 8.39 & 7.83 / 0.47 & 26.73 / 14.94 & 6.89 / 0.76 \\
\quad + 10 s window and one frame LSTM & 3.54 & $-25.7$ & $-1.92$ & 14.45 / 7.14 & 7.93 / 0.38 & 23.91 / 9.89 & 6.64 / 0.43 \\
\bottomrule
\end{tabular}
\end{adjustbox}

\end{table}
\begin{table}[H]
\centering
\caption{Model size, live enrollment. Word error rate on each development set (clean calls: against plain recognition, deleted words in brackets), the
mean of the three meeting sets, dev-4 (that and the clean calls), Krisp reported only.}
\label{tab:scale}
\scriptsize
\begin{adjustbox}{max width=\linewidth}
\begin{tabular}{lrrrrrrrr}
\toprule
\schead \scth{System} & \scth{AMI dev} & \scth{NSF dev} & \scth{AMI 60 s} & \scth{Clean calls (del.)} & \scth{Calls + voice} & \scth{Meetings} & \scth{Dev-4} & \scth{Krisp} \\
\midrule
S, 140M & 13.23 & 7.63 & 20.02 & 5.78 (4.56) & 51.42 & 13.63 & 11.67 & 6.12 \\
M, 243M & 13.18 & 7.56 & 19.46 & 5.57 (4.30) & 52.19 & 13.40 & 11.44 & 5.79 \\
L, 374M & 13.32 & 7.58 & 21.65 & 6.83 (5.48) & 56.28 & 14.18 & 12.35 & 6.77 \\
L, gentler long-call stage & 13.36 & 7.61 & 20.35 & 6.83 (5.27) & 52.63 & 13.77 & 12.04 & 6.50 \\
S + perceptual loss & 13.77 & 7.76 & 23.09 & 4.97 (3.67) & 57.78 & 14.87 & 12.40 & 5.07 \\
M + perceptual loss & 14.62 & 7.98 & 23.88 & 3.43 (2.52) & 56.39 & 15.49 & 12.48 & 5.10 \\
S + WavLM, 452M & 12.90 & 7.70 & 17.40 & 5.26 (3.91) & 51.63 & 12.67 & 10.81 & 5.37 \\
M + WavLM, 555M & 13.00 & 7.69 & 17.59 & 3.92 (2.82) & 51.05 & 12.76 & 10.55 & 5.55 \\
L + WavLM, 687M & 12.82 & 7.76 & 17.89 & 4.16 (3.04) & 52.30 & 12.82 & 10.66 & 5.47 \\
M + WavLM, windows cut (release) & 13.23 & 7.77 & 18.00 & 2.96 (1.96) & 51.32 & 13.00 & 10.49 & 5.74 \\
\bottomrule
\end{tabular}
\end{adjustbox}

\end{table}
\begin{table}[H]
\centering
\caption{Self-supervised speech models with S. Word error rate on each development set (clean calls: against plain recognition, deleted words in brackets), the
mean of the three meeting sets, dev-4 (that and the clean calls), Krisp reported only.}
\label{tab:ssl}
\scriptsize
\begin{adjustbox}{max width=\linewidth}
\begin{tabular}{lrrrrrrrr}
\toprule
\schead \scth{System} & \scth{AMI dev} & \scth{NSF dev} & \scth{AMI 60 s} & \scth{Clean calls (del.)} & \scth{Calls + voice} & \scth{Meetings} & \scth{Dev-4} & \scth{Krisp} \\
\midrule
reference (stage 2, level range) & 13.30 & 8.06 & 20.17 & 9.99 (8.48) & 51.82 & 13.85 & 12.88 & 6.17 \\
\quad WavLM teacher: perceptual + distillation & 13.94 & 8.36 & 20.95 &  &  & 14.42 &  & 8.05 \\
\quad WavLM speaker encoder (self-enrolled) & 14.01 & 8.26 & 26.28 &  &  & 16.18 &  & 8.96 \\
reference (real-crosstalk recipe) & 13.03 & 7.72 & 19.65 & 7.53 (5.90) & 52.71 & 13.47 & 11.98 & 6.06 \\
\quad perceptual loss on the caller's frames & 13.64 & 7.94 & 21.34 & 5.85 (4.39) & 50.24 & 14.31 & 12.19 & 5.56 \\
\quad causal SSL branch, 20k-step student & 12.93 & 7.80 & 19.37 & 9.99 (8.59) & 49.38 & 13.37 & 12.52 & 8.18 \\
reference (+ level drift, single talkers) & 13.23 & 7.63 & 20.02 & 5.78 (4.56) & 51.42 & 13.63 & 11.67 & 6.12 \\
\quad causal SSL branch, 60k-step student & 13.61 & 7.73 & 19.97 & 6.36 (4.49) & 53.14 & 13.77 & 11.92 & 5.74 \\
M reference & 13.18 & 7.56 & 19.46 & 5.57 (4.30) & 52.19 & 13.40 & 11.44 & 5.79 \\
\quad causal SSL branch, 60k-step student & 13.08 & 7.63 & 20.64 & 6.66 (5.14) & 51.06 & 13.78 & 12.00 & 6.03 \\
\quad WavLM made causal, trained with the model & 12.64 & 7.62 & 18.79 & 3.62 (2.62) & 54.71 & 13.01 & 10.67 & 5.79 \\
\quad WavLM made causal and distilled, frozen & 13.00 & 7.58 & 19.71 & 4.76 (3.60) & 53.81 & 13.43 & 11.26 & 5.69 \\
\quad WavLM made causal and distilled, trained & 12.78 & 7.69 & 18.70 & 5.31 (4.22) & 51.02 & 13.05 & 11.12 & 5.69 \\
\quad the same from the fine-tune stage on (12k steps) & 13.00 & 7.69 & 17.59 & 3.92 (2.82) & 51.05 & 12.76 & 10.55 & 5.55 \\
\quad w2v-BERT 2.0 made causal, trained with the model & 12.89 & 7.57 & 19.51 & 5.42 (4.03) & 52.77 & 13.32 & 11.35 & 5.87 \\
\quad Whisper large-v3 encoder made causal, trained with the model & 13.03 & 7.68 & 19.50 & 6.75 (4.97) & 52.04 & 13.40 & 11.74 & 6.45 \\
\quad HuBERT-Large (316M) made causal, trained with the model & 13.00 & 7.64 & 20.20 & 5.93 (4.49) & 51.14 & 13.61 & 11.69 & 6.03 \\
\quad HuBERT-XLarge (964M) made causal, trained with the model & 13.08 & 7.65 & 21.22 & 4.52 (3.33) & 53.09 & 13.98 & 11.62 & 5.95 \\
\quad XLS-R 1B made causal, trained with the model & 12.69 & 7.70 & 18.75 & 4.13 (2.84) & 50.48 & 13.05 & 10.82 & 5.51 \\
\quad WavLM made causal, from the fine-tune stage on (12k steps) & 12.97 & 7.69 & 18.25 & 3.48 (2.57) & 52.66 & 12.97 & 10.60 & 5.54 \\
\quad w2v-BERT 2.0, from the fine-tune stage on (12k steps) & 12.98 & 7.70 & 19.17 & 5.75 (4.70) & 51.51 & 13.28 & 11.40 & 5.79 \\
\quad Whisper large-v3 encoder, from the fine-tune stage on (12k steps) & 13.04 & 7.70 & 19.43 & 4.79 (3.49) & 49.74 & 13.39 & 11.24 & 5.64 \\
\bottomrule
\end{tabular}
\end{adjustbox}

\end{table}
\begin{table}[H]
\centering
\caption{Enrollment modes of the S release candidate. Word error rate on each development set (clean calls: against plain recognition, deleted words in brackets), the
mean of the three meeting sets, dev-4 (that and the clean calls), Krisp reported only.}
\label{tab:enrollmodes}
\scriptsize
\begin{adjustbox}{max width=\linewidth}
\begin{tabular}{lrrrrrrrr}
\toprule
\schead \scth{System} & \scth{AMI dev} & \scth{NSF dev} & \scth{AMI 60 s} & \scth{Clean calls (del.)} & \scth{Calls + voice} & \scth{Meetings} & \scth{Dev-4} & \scth{Krisp} \\
\midrule
live: loudest 3 s heard so far & 13.23 & 7.63 & 20.02 & 5.78 (4.56) & 51.42 & 13.63 & 11.67 & 6.12 \\
live: from its own output & 13.46 & 7.57 & 21.52 & 5.33 (4.21) &  & 14.18 & 11.97 & 6.23 \\
offline: loudest 3 s of the first 15 s & 13.43 & 7.61 & 19.94 & 5.27 (4.03) &  & 13.66 & 11.56 & 6.08 \\
none & 13.65 & 7.63 & 26.59 & 3.66 (2.78) & 54.26 & 15.96 & 12.88 & 6.87 \\
oracle: another clip of the caller & 13.16 & 7.59 & 18.65 &  &  & 13.13 &  &  \\
\bottomrule
\end{tabular}
\end{adjustbox}

\end{table}
\begin{table}[H]
\centering
\caption{Training objective, stage 1 of S. Simulated development set: SI-SDR improvement, leak and keep (dB); real sets: word error rate / leaked words per
minute.}
\label{tab:loss}
\scriptsize
\begin{adjustbox}{max width=\linewidth}
\begin{tabular}{lrrrrrr}
\toprule
\schead \scth{Arm} & \scth{SI-SDRi} & \scth{Leak dB} & \scth{Keep dB} & \scth{AMI dev} & \scth{NOTSOFAR dev} & \scth{Krisp} \\
\midrule
SNR only & 6.06 & $-18.9$ & $-0.79$ & 21.10 / 32.37 & 14.42 / 21.56 & 7.50 / 10.16 \\
SNR + spectral 10 & 5.98 & $-19.7$ & $-0.85$ & 16.33 / 12.77 & 8.73 / 3.15 & 7.58 / 9.71 \\
SNR + silent 2 & 5.51 & $-32.9$ & $-1.14$ & 16.51 / 8.53 & 8.29 / 0.34 & 8.71 / 9.28 \\
SNR + spectral 10 + silent 2 & 5.41 & $-31.6$ & $-1.05$ & 16.29 / 9.24 & 8.97 / 0.38 & 7.51 / 5.88 \\
\bottomrule
\end{tabular}
\end{adjustbox}

\end{table}
\begin{table}[H]
\centering
\caption{Training speech, stage 1 of S. Simulated development set: SI-SDR improvement, leak and keep (dB); real sets: word error rate / leaked words per
minute.}
\label{tab:data}
\scriptsize
\begin{adjustbox}{max width=\linewidth}
\begin{tabular}{lrrrrrr}
\toprule
\schead \scth{Arm} & \scth{SI-SDRi} & \scth{Leak dB} & \scth{Keep dB} & \scth{AMI dev} & \scth{NOTSOFAR dev} & \scth{Krisp} \\
\midrule
LibriSpeech only (VoxLock's speech) & 4.91 & $-28.6$ & $-1.14$ & 16.26 / 7.81 & 8.43 / 0.51 & 8.23 / 6.28 \\
pool, English groups & 4.68 & $-28.0$ & $-1.18$ & 14.57 / 8.79 & 8.18 / 0.38 & 8.44 / 8.37 \\
pool, all languages & 5.41 & $-31.6$ & $-1.05$ & 16.29 / 9.24 & 8.97 / 0.38 & 7.51 / 5.88 \\
pool without restored groups & 5.50 & $-30.4$ & $-1.06$ & 14.52 / 7.68 & 7.99 / 0.38 & 8.01 / 7.56 \\
\bottomrule
\end{tabular}
\end{adjustbox}

\end{table}
\begin{table}[H]
\centering
\caption{Simulation, stage 1 of S, one part removed at a time. Simulated development set: SI-SDR improvement, leak and keep (dB); real sets: word error rate / leaked words per
minute.}
\label{tab:aug}
\scriptsize
\begin{adjustbox}{max width=\linewidth}
\begin{tabular}{lrrrrrr}
\toprule
\schead \scth{Arm} & \scth{SI-SDRi} & \scth{Leak dB} & \scth{Keep dB} & \scth{AMI dev} & \scth{NOTSOFAR dev} & \scth{Krisp} \\
\midrule
all augmentation & 5.41 & $-31.6$ & $-1.05$ & 16.29 / 9.24 & 8.97 / 0.38 & 7.51 / 5.88 \\
no synthetic crosstalk & 3.29 & $-10.7$ & $0.18$ & 29.84 / 58.62 & 7.85 / 0.47 & 12.75 / 27.04 \\
no background noise & 4.09 & $-26.5$ & $-1.05$ & 31.41 / 9.51 & 19.34 / 0.64 & 9.51 / 8.51 \\
no rooms & 4.75 & $-24.2$ & $-0.81$ & 21.83 / 10.00 & 8.71 / 0.26 & 11.45 / 19.81 \\
no phone line & 3.13 & $-34.6$ & $-4.49$ & 15.95 / 7.46 & 8.73 / 0.26 & 8.81 / 7.59 \\
no background-only stretches & 5.46 & $-31.2$ & $-0.94$ & 18.63 / 9.46 & 8.68 / 0.30 & 8.33 / 9.36 \\
\bottomrule
\end{tabular}
\end{adjustbox}

\end{table}
\begin{table}[H]
\centering
\caption{Real crosstalk, stage 1 of S. Simulated development set: SI-SDR improvement, leak and keep (dB); real sets: word error rate / leaked words per
minute.}
\label{tab:real}
\scriptsize
\begin{adjustbox}{max width=\linewidth}
\begin{tabular}{lrrrrrr}
\toprule
\schead \scth{Arm} & \scth{SI-SDRi} & \scth{Leak dB} & \scth{Keep dB} & \scth{AMI dev} & \scth{NOTSOFAR dev} & \scth{Krisp} \\
\midrule
reference (SNR + spectral 10 + silent 2) & 5.41 & $-31.6$ & $-1.05$ & 16.29 / 9.24 & 8.97 / 0.38 & 7.51 / 5.88 \\
+ real AMI, NOTSOFAR, AliMeeting crosstalk (30\%) & 4.60 & $-29.5$ & $-1.09$ & 14.58 / 11.56 & 8.05 / 0.85 & 8.48 / 10.78 \\
\bottomrule
\end{tabular}
\end{adjustbox}

\end{table}
\begin{table}[H]
\centering
\caption{A keep term, stage 1 of S. Simulated development set: SI-SDR improvement, leak and keep (dB); real sets: word error rate / leaked words per
minute.}
\label{tab:keep}
\scriptsize
\begin{adjustbox}{max width=\linewidth}
\begin{tabular}{lrrrrrr}
\toprule
\schead \scth{Arm} & \scth{SI-SDRi} & \scth{Leak dB} & \scth{Keep dB} & \scth{AMI dev} & \scth{NOTSOFAR dev} & \scth{Krisp} \\
\midrule
reference (SNR + spectral 10 + silent 2) & 5.41 & $-31.6$ & $-1.05$ & 16.29 / 9.24 & 8.97 / 0.38 & 7.51 / 5.88 \\
+ keep term (hinge on turning the caller down), weight 2 & 5.55 & $-12.5$ & $-0.24$ & 15.47 / 13.08 & 11.93 / 13.33 & 8.63 / 13.28 \\
\bottomrule
\end{tabular}
\end{adjustbox}

\end{table}
\begin{table}[H]
\centering
\caption{The enrollment branch, stage 1 of S. Simulated development set: SI-SDR improvement, leak and keep (dB); real sets: word error rate / leaked words per
minute.}
\label{tab:enrollarm}
\scriptsize
\begin{adjustbox}{max width=\linewidth}
\begin{tabular}{lrrrrrr}
\toprule
\schead \scth{Arm} & \scth{SI-SDRi} & \scth{Leak dB} & \scth{Keep dB} & \scth{AMI dev} & \scth{NOTSOFAR dev} & \scth{Krisp} \\
\midrule
reference (SNR + spectral 10 + silent 2) & 5.41 & $-31.6$ & $-1.05$ & 16.29 / 9.24 & 8.97 / 0.38 & 7.51 / 5.88 \\
no enrollment branch & 4.28 & $-26.2$ & $-1.18$ & 14.59 / 8.04 & 8.06 / 0.38 & 11.93 / 19.03 \\
\bottomrule
\end{tabular}
\end{adjustbox}

\end{table}

\begin{table}[H]
\centering
\caption{The interruption bench (Section~\ref{sec:interrupt}): each system live, one session each on the same 48 call plans, as the bench scored the agent's own events.
Late-transcript pauses: the agent paused when the recogniser's answer for the caller's last words came after the turn
had ended. End of turn: the caller's last word to the agent's commit. No reply: turns the agent never answered.
Frame: the filter's time per 50 ms frame on the agent's audio loop.}
\label{tab:vintlive}
\begin{adjustbox}{max width=\linewidth}
\begin{tabular}{lrrrrrrrr}
\toprule
 & VAD false & late-transcript & false starts & cut-ins & end of turn & no & unfiltered & frame ms \\
System & int. / agent-min & pauses / agent-min & per silent min & caught \% & p50 / p90 (s) & reply \% & \% & p50 / p99 \\
\midrule
No filter & 2.61 & 5.30 & 3.97 & 93 & 0.79 / 2.07 & 7.6 & 0.0 & 0.0 / 0.0 \\
GTCRN, noise only, in the agent & 5.09 & 5.05 & 6.44 & 95 & 0.85 / 2.00 & 9.1 & 0.0 & 3.2 / 6.2 \\
VoxLock, live, in the agent & 0.89 & 3.37 & 1.48 & 78 & 0.97 / 1.97 & 15.4 & 0.0 & 39.2 / 55.1 \\
VoxLock, live, fastest 8-bit model, in the agent & 0.96 & 4.52 & 1.55 & 86 & 0.82 / 1.59 & 12.5 & 0.0 & 26.6 / 46.0 \\
NVIDIA Real-time RE-USE, GPU server & 0.87 & 4.05 & 1.40 & 89 & 0.90 / 1.67 & 7.3 & 0.1 & 0.1 / 27.3 \\
Ours, CPU student, in the agent & 1.42 & 4.22 & 1.93 & 90 & 0.76 / 1.55 & 3.4 & 0.0 & 10.9 / 19.8 \\
Ours, release, GPU server & 0.91 & 3.44 & 1.36 & 95 & 0.87 / 1.64 & 6.2 & 0.0 & 0.2 / 25.7 \\
Ours, release fine-tuned for agents & 0.92 & 3.85 & 1.17 & 86 & 0.82 / 1.62 & 2.9 & 0.1 & 0.2 / 25.6 \\
\bottomrule
\end{tabular}
\end{adjustbox}

\end{table}

\begin{table}[H]
\centering
\caption{The voicebot mode (Section~\ref{sec:interrupt}) on the interruption bench's held-out sessions, through the shipped code: our release with its GPU servers'
enrollment timing and our CPU student in the agent, through the LiveKit plugin in 50 ms frames, each in its default
mode and in the voicebot mode; RE-USE on its GPU server. Change: the voicebot mode against the same system's default, paired over resampled call plans. Before
and after the first word: false interruptions per minute of the agent's speech before the caller's first word (the
greeting) and after it. Whisper: large-v3 on the caller's turns after the filter (two sessions).}
\label{tab:vintvb}
\begin{adjustbox}{max width=\linewidth}
\begin{tabular}{llrrrrrrr}
\toprule
 & & false interruptions & change, & before the & after the & false starts & turns & Whisper \\
System & mode & per agent-min (90\% int.) & paired (90\% int.) & first word & first word & per silent min & missed \% & WER \% \\
\midrule
No filter &  & 2.77 (1.91 to 3.73) &  & 5.52 & 2.42 & 3.04 & 0.4 & 7.4 \\
NVIDIA Real-time RE-USE (GPU) &  & 0.95 (0.63 to 1.30) &  & 6.54 & 0.24 & 1.01 & 2.6 & 16.9 \\
\addlinespace
Ours, release (GPU) & default & 1.02 (0.66 to 1.45) &  & 5.17 & 0.50 & 1.22 & 0.6 & 13.4 \\
 & voicebot & 0.78 (0.49 to 1.13) & -0.24 (-0.37 to -0.13) & 3.02 & 0.50 & 1.05 & 0.6 & 13.4 \\
\addlinespace
Ours, CPU student, 8-bit & default & 1.36 (0.93 to 1.88) &  & 5.77 & 0.80 & 1.58 & 0.3 & 11.8 \\
 & voicebot & 1.13 (0.75 to 1.59) & -0.23 (-0.34 to -0.14) & 3.61 & 0.82 & 1.40 & 0.3 & 11.7 \\
\addlinespace
Ours, CPU student, full precision & default & 1.36 (0.92 to 1.87) &  & 5.74 & 0.80 & 1.57 & 0.3 & 11.9 \\
 & voicebot & 1.13 (0.75 to 1.58) & -0.23 (-0.34 to -0.13) & 3.69 & 0.80 & 1.42 & 0.3 & 11.8 \\
\bottomrule
\end{tabular}
\end{adjustbox}

\end{table}

\begin{table}[H]
\centering
\caption{LiveKit's VAD behind each system on the test calls, against the same VAD on the caller's own channel. False
starts: speech starts while the caller is silent. Turn ends held: the caller's turn ends the VAD does not see before
the caller talks again. Turn end p90: how much later than the reference it sees the others, 90th percentile.}
\label{tab:vad}
\begin{adjustbox}{max width=\linewidth}
\begin{tabular}{lrrrrrr}
\toprule
 & \multicolumn{5}{c}{Calls + another voice} & Caller alone \\ \cmidrule(lr){2-6} \cmidrule(lr){7-7}
 & speech in the & false starts & turn ends & turn end & caller & caller \\
System & caller's pauses \% & per min & held \% & p90 (s) & missed \% & missed \% \\
\midrule
No processing & 47.9 & 3.33 & 28 & 2.43 & 0.1 & 0.0 \\
GTCRN (noise only; STT-API's plugin) & 48.3 & 3.58 & 26 & 2.40 & 0.3 & 0.6 \\
DeepFilterNet3 (noise only) & 45.2 & 3.14 & 27 & 2.28 & 0.6 & 0.9 \\
VoxLock, live & 27.5 & 2.28 & 24 & 1.51 & 11.0 & 2.6 \\
Ours, M, live & 24.2 & 2.27 & 19 & 1.41 & 8.4 & 8.4 \\
Ours, release, live & 22.8 & 1.94 & 18 & 1.00 & 8.4 & 5.2 \\
Ours, CPU student, live & 26.2 & 2.27 & 20 & 1.34 & 6.7 & 4.9 \\
\bottomrule
\end{tabular}
\end{adjustbox}

\end{table}
\begin{table}[H]
\centering
\caption{The distilled students through the plugin on one thread: share of one core per live call in 20 ms and 50 ms
frames, float and 8-bit weights, and dev-4. VoxLock's package steps 20 ms at a time; its 8-bit column is its default
model.}
\label{tab:cpu}
\begin{adjustbox}{max width=\linewidth}
\begin{tabular}{lrrrrrr}
\toprule
 & \multicolumn{2}{c}{\% of a core, 20 ms} & \multicolumn{2}{c}{\% of a core, 50 ms} & \multicolumn{2}{c}{dev-4 WER} \\ \cmidrule(lr){2-3} \cmidrule(lr){4-5} \cmidrule(lr){6-7}
Model & float & 8-bit & float & 8-bit & float & 8-bit \\
\midrule
VoxLock (its ONNX package, 27M) & 158 & 132 & 88 & 65 & 13.51 & \\
GTCRN (noise only, STT-API's plugin) & 5 & & 5 & & & \\
{Ours, C10, 9.8M} & 18 & 14 & 10 & 8 & 11.06 & 11.10 \\
\textbf{Ours, C22, 22.1M} & 33 & 22 & 17 & 12 & 10.66 & 10.52 \\
{Ours, C51, 50.9M} & 64 & 39 & 31 & 21 & 10.78 & 10.87 \\
\bottomrule
\end{tabular}
\end{adjustbox}

\end{table}
\begin{table}[H]
\centering
\caption{What the isolator costs a LiveKit agent: \val{lk/final/none/calls} test calls with another voice streamed at
once (the last row \val{lk/final64/remote/calls}). Frame: its time per frame on the agent's audio loop (on the GPU rows
mostly a wait for the server's reply); core: its share of one CPU core per call, where it computes; delay: how far its
output stood behind the call; unfiltered: audio let through because the
server's reply came later than the plugin's cap on its delay (40 ms while frames wait for their replies, 250 ms when
they do not).}
\label{tab:lkbench}
\begin{adjustbox}{max width=\linewidth}
\begin{tabular}{lrrrr}
\toprule
Filter & frame p50 / p99 (ms) & core per call & delay p50 / p99 (ms) & unfiltered \% \\
\midrule
No filter & 0.00 / 0.00 & 0.000 & 0 &  \\
GTCRN (noise only) & 3.05 / 5.58 & 0.066 &  &  \\
Ours, CPU student & 15.21 / 24.88 & 0.323 & 10 / 10 &  \\
Ours, CPU student, 8-bit & 9.49 / 16.66 & 0.200 & 10 / 10 &  \\
Ours, release on a GPU, 50 ms frames & 14.46 / 30.25 &  & 40 / 50 & 0.1 \\
Ours, release on a GPU, 50 ms frames, no waiting & 0.11 / 0.31 &  & 90 / 90 & 0.0 \\
Ours, release on a GPU, 20 ms frames & 0.12 / 11.83 &  & 30 / 50 & 0.0 \\
Ours, release on a GPU, 50 ms frames, 64 calls & 18.23 / 30.30 &  & 50 / 50 & 1.0 \\
\bottomrule
\end{tabular}
\end{adjustbox}

\end{table}
\begin{table}[H]
\centering
\caption{What the agent's VAD did in the same runs, against the same rooms with each caller's own channel; the columns
are those of Table~\ref{tab:vad}.}
\label{tab:lkvad}
\begin{adjustbox}{max width=\linewidth}
\begin{tabular}{lrrrrr}
\toprule
 & false starts & speech in the & turn end & turn ends & caller \\
Filter & per min & caller's pauses \% & p90 (s) & held \% & missed \% \\
\midrule
No filter & 2.84 & 47.5 & 2.04 & 23 & 0.3 \\
GTCRN (noise only) & 2.88 & 46.8 & 2.03 & 22 & 0.7 \\
Ours, CPU student & 2.12 & 27.7 & 1.03 & 18 & 3.0 \\
Ours, CPU student, 8-bit & 2.25 & 28.2 & 1.04 & 16 & 2.9 \\
Ours, release on a GPU, 50 ms frames & 1.75 & 24.9 & 1.06 & 18 & 6.9 \\
Ours, release on a GPU, 50 ms frames, no waiting & 1.59 & 25.4 & 1.15 & 19 & 7.0 \\
Ours, release on a GPU, 20 ms frames & 1.59 & 24.1 & 1.03 & 18 & 6.8 \\
Ours, release on a GPU, 50 ms frames, 64 calls & 1.61 & 22.8 & 1.02 & 21 & 8.5 \\
\bottomrule
\end{tabular}
\end{adjustbox}

\end{table}

\section{Full results}
\label{app:full}

Word error rate, missed and inserted words as shares of the reference words, leaked words per minute of
background-only time, and the difference in word error rate to VoxLock's PyTorch script with a 95\% interval from
2,000 resamples of callers.

\begin{table}[H]
\centering
\caption{Krisp.}
\scriptsize
\begin{adjustbox}{max width=\linewidth}
\begin{tabular}{lrrrrrrl}
\toprule
\schead \scth{System} & \scth{WER} & \scth{Missed} & \scth{Inserted} & \scth{Leaked/min} & \scth{DNSMOS-C} & \scth{$\Delta$WER vs VoxLock} \\
\midrule
No processing & 23.19 & 4.27 & 18.92 & 66.44 & 3.59 & $+18.20$ [$+15.50$, $+20.70$] \\
DeepFilterNet3 (noise only) & 13.15 & 6.96 & 6.18 & 20.34 & 4.06 & $+8.15$ [$+5.73$, $+10.23$] \\
Krisp VI 2.5 Default & 5.30 & 4.21 & 1.09 & 1.52 & 3.92 & $+0.31$ [$-0.27$, $+0.93$] \\
Krisp VI 2.5 Balanced & 5.32 & 4.06 & 1.26 & 2.02 &  & $+0.32$ [$-0.31$, $+0.90$] \\
Krisp VI 2.5 HD & 5.79 & 4.36 & 1.42 & 2.74 & 4.00 & $+0.79$ [$-0.09$, $+1.77$] \\
Krisp VI 2.5 Lite & 6.21 & 4.26 & 1.94 & 4.52 &  & $+1.21$ [$+0.23$, $+2.03$] \\
VoxLock 1.1.1, live & 5.63 & 4.78 & 0.84 & 0.29 & 3.91 & $+0.63$ [$+0.21$, $+1.31$] \\
VoxLock 1.1.1, loudest 3 s & 5.19 & 4.46 & 0.72 & 0.07 & 3.95 & $+0.20$ [$+0.01$, $+0.39$] \\
VoxLock, its PyTorch script & 4.99 & 4.25 & 0.74 & 0.09 & 3.91 &  \\
Sidon CallEnhancer v2, as released & 33.50 & 4.09 & 29.41 & 103.40 & 4.00 & $+28.51$ [$+24.24$, $+31.76$] \\
Ours, M + WavLM, live, loudest 3 s so far (causal) & 5.74 & 5.08 & 0.67 & 0.04 &  & $+0.75$ [$+0.12$, $+1.71$] \\
Ours, M + WavLM, loudest 3 s of first 15 s & 5.93 & 5.26 & 0.67 & 0.05 & 3.75 & $+0.93$ [$+0.27$, $+1.88$] \\
Ours, M + WavLM, no enrollment & 6.28 & 5.14 & 1.14 & 1.80 &  & $+1.28$ [$+0.41$, $+2.38$] \\
Ours, transformer M, live, loudest 3 s so far (causal) & 5.79 & 5.05 & 0.73 & 0.21 & 3.77 & $+0.79$ [$+0.31$, $+1.37$] \\
Ours, transformer M, loudest 3 s of first 15 s & 5.76 & 4.92 & 0.84 & 0.48 & 3.77 & $+0.77$ [$+0.36$, $+1.23$] \\
Ours, transformer M, no enrollment & 6.55 & 5.12 & 1.43 & 2.72 & 3.73 & $+1.56$ [$+0.95$, $+2.13$] \\
Ours, transformer S, live, loudest 3 s so far (causal) & 6.12 & 5.40 & 0.73 & 0.12 & 3.72 & $+1.13$ [$+0.57$, $+1.91$] \\
Ours, CPU student C22, live, loudest 3 s so far (causal) & 5.66 & 4.98 & 0.69 & 0.04 &  & $+0.67$ [$+0.12$, $+1.27$] \\
Ours, Sidon architecture 188M & 29.49 & 4.04 & 25.45 & 91.02 & 4.08 & $+24.50$ [$+20.93$, $+27.17$] \\
Ours, Sidon architecture lite & 18.27 & 6.22 & 12.05 & 42.50 & 4.00 & $+13.28$ [$+10.77$, $+15.36$] \\
\bottomrule
\end{tabular}
\end{adjustbox}

\end{table}

\begin{table}[H]
\centering
\caption{AMI, VoxLock's 134 windows.}
\scriptsize
\begin{adjustbox}{max width=\linewidth}
\begin{tabular}{lrrrrrl}
\toprule
\schead \scth{System} & \scth{WER} & \scth{Missed} & \scth{Inserted} & \scth{Leaked/min} & \scth{$\Delta$WER vs VoxLock} \\
\midrule
No processing & 33.00 & 8.05 & 24.95 & 78.20 & $+22.74$ [$+17.85$, $+29.23$] \\
DeepFilterNet3 (noise only) & 21.56 & 12.68 & 8.88 & 23.33 & $+11.30$ [$+5.57$, $+19.20$] \\
VoxLock 1.1.1, live & 11.71 & 7.98 & 3.73 & 3.05 & $+1.45$ [$-0.22$, $+3.96$] \\
VoxLock 1.1.1, loudest 3 s & 11.26 & 8.12 & 3.14 & 2.65 & $+1.00$ [$-0.49$, $+3.71$] \\
VoxLock, its PyTorch script & 10.26 & 6.88 & 3.39 & 2.12 &  \\
Sidon CallEnhancer v2, as released & 44.54 & 9.09 & 35.45 & 121.14 & $+34.28$ [$+31.14$, $+38.66$] \\
Ours, M + WavLM, live, loudest 3 s so far (causal) & 10.19 & 6.74 & 3.46 & 2.39 & $-0.07$ [$-0.41$, $+0.40$] \\
Ours, M + WavLM, loudest 3 s of first 15 s & 11.37 & 7.98 & 3.39 & 2.25 & $+1.11$ [$-0.45$, $+3.90$] \\
Ours, M + WavLM, no enrollment & 10.02 & 6.70 & 3.32 & 2.25 & $-0.24$ [$-0.63$, $+0.26$] \\
Ours, transformer M, live, loudest 3 s so far (causal) & 9.95 & 6.67 & 3.28 & 2.25 & $-0.31$ [$-0.64$, $+0.10$] \\
Ours, transformer M, loudest 3 s of first 15 s & 11.23 & 8.09 & 3.14 & 2.25 & $+0.97$ [$-0.55$, $+3.57$] \\
Ours, transformer M, no enrollment & 10.26 & 6.84 & 3.42 & 2.92 & $+0.00$ [$-0.53$, $+0.74$] \\
Ours, transformer S, live, loudest 3 s so far (causal) & 10.23 & 6.88 & 3.35 & 2.12 & $-0.03$ [$-0.30$, $+0.24$] \\
Ours, CPU student C22, live, loudest 3 s so far (causal) & 11.23 & 7.91 & 3.32 & 2.52 & $+0.97$ [$-0.58$, $+3.53$] \\
Ours, Sidon architecture 188M & 32.45 & 9.02 & 23.43 & 79.39 & $+22.18$ [$+17.94$, $+28.37$] \\
Ours, Sidon architecture lite & 17.45 & 9.02 & 8.43 & 21.07 & $+7.19$ [$+3.29$, $+12.35$] \\
\bottomrule
\end{tabular}
\end{adjustbox}

\end{table}

\begin{table}[H]
\centering
\caption{AMI, evaluation meetings.}
\scriptsize
\begin{adjustbox}{max width=\linewidth}
\begin{tabular}{lrrrrrl}
\toprule
\schead \scth{System} & \scth{WER} & \scth{Missed} & \scth{Inserted} & \scth{Leaked/min} & \scth{$\Delta$WER vs VoxLock} \\
\midrule
No processing & 42.59 & 9.75 & 32.84 & 101.40 & $+32.28$ [$+28.55$, $+36.23$] \\
DeepFilterNet3 (noise only) & 26.00 & 14.40 & 11.60 & 32.79 & $+15.69$ [$+12.53$, $+19.51$] \\
VoxLock 1.1.1, live & 12.07 & 8.39 & 3.69 & 3.80 & $+1.76$ [$+0.76$, $+2.83$] \\
VoxLock 1.1.1, loudest 3 s & 9.70 & 7.53 & 2.17 & 1.20 & $-0.62$ [$-1.34$, $-0.12$] \\
VoxLock, its PyTorch script & 10.31 & 7.87 & 2.45 & 2.14 &  \\
Sidon CallEnhancer v2, as released & 49.92 & 9.30 & 40.62 & 130.80 & $+39.61$ [$+35.51$, $+43.48$] \\
Ours, M + WavLM, live, loudest 3 s so far (causal) & 10.18 & 7.51 & 2.67 & 1.90 & $-0.13$ [$-0.97$, $+0.52$] \\
Ours, M + WavLM, loudest 3 s of first 15 s & 10.35 & 7.83 & 2.51 & 1.90 & $+0.03$ [$-0.79$, $+0.70$] \\
Ours, M + WavLM, no enrollment & 10.09 & 7.54 & 2.55 & 1.90 & $-0.22$ [$-0.94$, $+0.30$] \\
Ours, transformer M, live, loudest 3 s so far (causal) & 10.51 & 7.95 & 2.56 & 2.02 & $+0.19$ [$-0.76$, $+1.17$] \\
Ours, transformer M, loudest 3 s of first 15 s & 10.62 & 8.13 & 2.49 & 1.87 & $+0.31$ [$-0.60$, $+1.12$] \\
Ours, transformer M, no enrollment & 10.27 & 7.83 & 2.44 & 1.64 & $-0.04$ [$-0.83$, $+0.61$] \\
Ours, transformer S, live, loudest 3 s so far (causal) & 10.47 & 8.05 & 2.41 & 1.55 & $+0.15$ [$-0.74$, $+0.87$] \\
Ours, CPU student C22, live, loudest 3 s so far (causal) & 10.29 & 7.89 & 2.40 & 1.29 & $-0.02$ [$-0.85$, $+0.65$] \\
Ours, Sidon architecture 188M & 37.51 & 8.47 & 29.04 & 98.03 & $+27.20$ [$+23.91$, $+30.11$] \\
Ours, Sidon architecture lite & 24.06 & 8.26 & 15.81 & 49.26 & $+13.75$ [$+10.90$, $+16.67$] \\
\bottomrule
\end{tabular}
\end{adjustbox}

\end{table}

\begin{table}[H]
\centering
\caption{AMI, 60 s windows, evaluation meetings.}
\scriptsize
\begin{adjustbox}{max width=\linewidth}
\begin{tabular}{lrrrrrl}
\toprule
\schead \scth{System} & \scth{WER} & \scth{Missed} & \scth{Inserted} & \scth{Leaked/min} & \scth{$\Delta$WER vs VoxLock} \\
\midrule
No processing & 121.79 & 13.08 & 108.71 & 137.17 & $+101.59$ [$+82.24$, $+122.96$] \\
DeepFilterNet3 (noise only) & 41.77 & 22.17 & 19.60 & 24.37 & $+21.57$ [$+14.69$, $+31.79$] \\
VoxLock 1.1.1, live & 23.66 & 16.00 & 7.66 & 9.11 & $+3.46$ [$+1.10$, $+6.01$] \\
VoxLock 1.1.1, loudest 3 s & 19.52 & 15.95 & 3.57 & 3.73 & $-0.67$ [$-3.01$, $+0.98$] \\
VoxLock, its PyTorch script & 20.20 & 14.78 & 5.41 & 5.47 &  \\
Ours, M + WavLM, live, loudest 3 s so far (causal) & 16.13 & 11.55 & 4.58 & 4.11 & $-4.07$ [$-7.02$, $-1.68$] \\
Ours, M + WavLM, loudest 3 s of first 15 s & 16.22 & 11.82 & 4.39 & 3.91 & $-3.98$ [$-6.98$, $-1.57$] \\
Ours, M + WavLM, no enrollment & 16.19 & 11.10 & 5.09 & 5.34 & $-4.01$ [$-8.02$, $-0.43$] \\
Ours, transformer M, live, loudest 3 s so far (causal) & 16.84 & 12.94 & 3.90 & 3.59 & $-3.36$ [$-6.41$, $-0.63$] \\
Ours, transformer M, loudest 3 s of first 15 s & 16.44 & 12.98 & 3.46 & 3.08 & $-3.75$ [$-7.03$, $-0.98$] \\
Ours, transformer M, no enrollment & 18.39 & 11.99 & 6.40 & 6.72 & $-1.80$ [$-5.64$, $+2.53$] \\
Ours, transformer S, live, loudest 3 s so far (causal) & 15.48 & 11.74 & 3.74 & 3.10 & $-4.72$ [$-7.81$, $-2.29$] \\
Ours, CPU student C22, live, loudest 3 s so far (causal) & 15.55 & 12.04 & 3.51 & 3.08 & $-4.64$ [$-7.62$, $-2.23$] \\
Ours, Sidon architecture 188M & 101.80 & 10.04 & 91.76 & 118.92 & $+81.61$ [$+65.07$, $+98.84$] \\
Ours, Sidon architecture lite & 66.15 & 12.69 & 53.46 & 70.88 & $+45.95$ [$+34.42$, $+60.76$] \\
\bottomrule
\end{tabular}
\end{adjustbox}

\end{table}

\begin{table}[H]
\centering
\caption{NOTSOFAR, evaluation split.}
\scriptsize
\begin{adjustbox}{max width=\linewidth}
\begin{tabular}{lrrrrrl}
\toprule
\schead \scth{System} & \scth{WER} & \scth{Missed} & \scth{Inserted} & \scth{Leaked/min} & \scth{$\Delta$WER vs VoxLock} \\
\midrule
No processing & 34.70 & 6.84 & 27.86 & 90.90 & $+26.92$ [$+24.75$, $+29.28$] \\
DeepFilterNet3 (noise only) & 15.38 & 6.23 & 9.14 & 26.09 & $+7.59$ [$+6.32$, $+9.02$] \\
VoxLock 1.1.1, live & 8.84 & 6.53 & 2.31 & 2.99 & $+1.05$ [$+0.52$, $+1.66$] \\
VoxLock 1.1.1, loudest 3 s & 7.61 & 5.77 & 1.84 & 0.68 & $-0.17$ [$-0.33$, $-0.02$] \\
VoxLock, its PyTorch script & 7.78 & 5.84 & 1.94 & 1.02 &  \\
Sidon CallEnhancer v2, as released & 47.35 & 8.47 & 38.89 & 131.19 & $+39.57$ [$+37.17$, $+42.12$] \\
Ours, M + WavLM, live, loudest 3 s so far (causal) & 7.67 & 5.76 & 1.91 & 0.38 & $-0.11$ [$-0.35$, $+0.11$] \\
Ours, M + WavLM, loudest 3 s of first 15 s & 7.67 & 5.79 & 1.89 & 0.47 & $-0.11$ [$-0.36$, $+0.11$] \\
Ours, M + WavLM, no enrollment & 7.74 & 5.82 & 1.91 & 0.49 & $-0.05$ [$-0.30$, $+0.17$] \\
Ours, transformer M, live, loudest 3 s so far (causal) & 7.58 & 5.74 & 1.84 & 0.42 & $-0.21$ [$-0.46$, $+0.03$] \\
Ours, transformer M, loudest 3 s of first 15 s & 7.70 & 5.83 & 1.87 & 0.38 & $-0.08$ [$-0.36$, $+0.20$] \\
Ours, transformer M, no enrollment & 7.64 & 5.81 & 1.83 & 0.47 & $-0.14$ [$-0.42$, $+0.16$] \\
Ours, transformer S, live, loudest 3 s so far (causal) & 7.78 & 5.89 & 1.90 & 0.53 & $+0.00$ [$-0.28$, $+0.29$] \\
Ours, CPU student C22, live, loudest 3 s so far (causal) & 7.58 & 5.74 & 1.84 & 0.55 & $-0.21$ [$-0.46$, $+0.03$] \\
Ours, Sidon architecture 188M & 30.01 & 6.81 & 23.20 & 76.82 & $+22.23$ [$+20.25$, $+24.26$] \\
Ours, Sidon architecture lite & 13.40 & 6.04 & 7.36 & 20.65 & $+5.61$ [$+4.48$, $+6.80$] \\
\bottomrule
\end{tabular}
\end{adjustbox}

\end{table}

\begin{table}[H]
\centering
\caption{AliMeeting, evaluation split (character error rate).}
\scriptsize
\begin{adjustbox}{max width=\linewidth}
\begin{tabular}{lrrrrrl}
\toprule
\schead \scth{System} & \scth{WER} & \scth{Missed} & \scth{Inserted} & \scth{Leaked/min} & \scth{$\Delta$WER vs VoxLock} \\
\midrule
No processing & 29.40 & 15.12 & 14.28 & 61.12 & $+14.02$ [$+10.52$, $+18.03$] \\
DeepFilterNet3 (noise only) & 18.48 & 16.92 & 1.56 & 4.46 & $+3.10$ [$+1.90$, $+4.58$] \\
VoxLock 1.1.1, live & 16.28 & 15.27 & 1.01 & 1.37 & $+0.90$ [$+0.41$, $+1.45$] \\
VoxLock 1.1.1, loudest 3 s & 17.11 & 15.15 & 1.97 & 1.21 & $+1.74$ [$+0.26$, $+4.39$] \\
VoxLock, its PyTorch script & 15.38 & 14.38 & 1.00 & 1.01 &  \\
Ours, M + WavLM, live, loudest 3 s so far (causal) & 15.69 & 14.47 & 1.21 & 1.79 & $+0.31$ [$-0.16$, $+0.80$] \\
Ours, M + WavLM, loudest 3 s of first 15 s & 15.40 & 14.25 & 1.15 & 1.88 & $+0.02$ [$-0.47$, $+0.53$] \\
Ours, M + WavLM, no enrollment & 15.74 & 14.49 & 1.25 & 1.93 & $+0.36$ [$-0.17$, $+0.93$] \\
Ours, transformer M, live, loudest 3 s so far (causal) & 18.57 & 15.28 & 3.30 & 12.83 & $+3.20$ [$+0.57$, $+7.98$] \\
Ours, transformer M, loudest 3 s of first 15 s & 17.85 & 14.56 & 3.29 & 13.03 & $+2.48$ [$-0.10$, $+7.38$] \\
Ours, transformer M, no enrollment & 16.02 & 14.94 & 1.08 & 1.46 & $+0.64$ [$+0.11$, $+1.18$] \\
Ours, transformer S, live, loudest 3 s so far (causal) & 18.09 & 14.70 & 3.39 & 12.97 & $+2.72$ [$+0.13$, $+7.63$] \\
Ours, CPU student C22, live, loudest 3 s so far (causal) & 15.82 & 14.66 & 1.17 & 1.96 & $+0.45$ [$-0.09$, $+1.02$] \\
Ours, Sidon architecture 188M & 42.94 & 16.82 & 26.12 & 124.03 & $+27.57$ [$+24.18$, $+31.48$] \\
Ours, Sidon architecture lite & 19.31 & 15.75 & 3.56 & 9.25 & $+3.93$ [$+1.87$, $+6.61$] \\
\bottomrule
\end{tabular}
\end{adjustbox}

\end{table}

\section{Simulator parameters}
\label{app:sim}

Table~\ref{tab:simparams} lists every random choice the simulator makes for one training example, at the defaults the
ablations start from.

\begin{table}[H]
\centering
\caption{The simulator's default draws (\texttt{vi/simulate.py}). Percentages are probabilities per example (per
background talker for the talker rows).}
\label{tab:simparams}
\scriptsize
\begin{adjustbox}{max width=\linewidth}
\begin{tabular}{ll}
\toprule
\schead \scth{Parameter} & \scth{Value} \\
\midrule
Example length & 6 s \\
Primary speech pieces (1 / 2 / 3) & 45\% / 35\% / 20\% \\
Pause between pieces & 0.3--3.0 s \\
Primary starts late (1.5--4 s) & 10\% \\
Background talkers (0 / 1 / 2 / 3) & 15\% / 50\% / 25\% / 10\% \\
SIR per talker & normal(8, 7) dB, clipped to [$-6$, 30] \\
Talker placed in a primary pause & 60\% \\
Talker dry (no room) & 10\% \\
Noises (0 / 1 / 2) & 20\% / 60\% / 20\% \\
SNR per noise & normal(15, 10) dB, clipped to [$-5$, 40] \\
No room at all & 8\% \\
Target reflections kept & 50 ms \\
High-pass / peaking EQ & 50\% / 60\% \\
Narrowband (of which $\mu$-law) / 7 kHz & 25\% (50\%) / 15\% \\
Real codec on the mixture & 8\%: gsm, g726, opus8, opus16, mp3, speex, g722 \\
Held-out codecs (tests only) & g723\_1, codec2 \\
Mixture level & $-45$ to $-14$ dBFS \\
Clipping & 2\% \\
Enrollment given (length) & 70\% (2--5 s), noisy in 60\% \\
\bottomrule
\end{tabular}
\end{adjustbox}

\end{table}

\end{document}